\documentclass[11pt]{article}
\usepackage[a4paper,total={16cm,24cm},centering]{geometry}
\usepackage[french,english]{babel}% Main language : english, auxilliary. french
\usepackage[sc]{mathpazo}
\usepackage{amsmath}
\usepackage{amssymb,amsthm}		%needed for \mathcal and fancy therorems
\usepackage{mathtools}
\usepackage{enumerate}
\usepackage{epigraph}
\usepackage{empheq} 
\usepackage{booktabs}
\usepackage{cite} %citation [1-10]

\usepackage{comment}
\usepackage{graphicx}  %scalebox
\usepackage{subcaption}
\usepackage[title,toc,titletoc,page]{appendix}
\usepackage{titling}

\usepackage{pstricks}
\usepackage{tikz}
\usepackage{pgfplots}
\pgfplotsset{compat=newest}
\usepackage{pgfplotstable}
\usepgfplotslibrary{groupplots}

\pgfplotsset{select coords between index/.style 2 args={
    x filter/.code={
        \ifnum\coordindex<#1\fi
        \ifnum\coordindex>#2\fi
    }
}}

\usetikzlibrary{decorations.pathmorphing,patterns,decorations.pathreplacing,decorations.markings}
\usetikzlibrary{shapes.geometric}
\usepackage{hyperref}
\usepackage[capitalise,noabbrev]{cleveref}
\newtheorem{theorem}{Theorem}[section]
\newtheorem{lemma}[theorem]{Lemma}
\newtheorem{proposition}[theorem]{Proposition}

\renewcommand{\i}{\mathrm{i}}		%imaginary unit
\newcommand{\ee}{\mathrm{e}} 	%exponential
\newcommand{\lr}[1]% grande parenthèse
{\left( #1 \right)
}

\newcommand{\vertical}[1]% grande barre verticale
{\left. #1\right\vert 
}

\newcommand{\arrowup}[2]% coordonnées (abscisse, ordonnée) de l'extrémité de la flèche up
{\draw[thick,color=black] (#1-0.15,#2-0.15) -- (#1,#2);
\draw[thick,color=black] (#1+0.15,#2-0.15) -- (#1,#2);
}
\newcommand{\arrowdown}[2]% coordonnées (abscisse, ordonnée) de l'extrémité de la flèche down
{\draw[thick,color=black] (#1-0.15,#2+0.15) -- (#1,#2);
\draw[thick,color=black] (#1+0.15,#2+0.15) -- (#1,#2);
}
\newcommand{\arrowright}[2]% coordonnées (abscisse, ordonnée) de l'extrémité de la flèche right
{\draw[thick,color=black] (#1-0.15,#2+0.15) -- (#1,#2);
\draw[thick,color=black] (#1-0.15,#2-0.15) -- (#1,#2);
}
\newcommand{\arrowleft}[2]% coordonnées (abscisse, ordonnée) de l'extrémité de la flèche left
{\draw[thick,color=black] (#1+0.15,#2+0.15) -- (#1,#2);
\draw[thick,color=black] (#1+0.15,#2-0.15) -- (#1,#2);
}

\newcommand{\fr}[2]% fraction
{\frac{#1}{#2}
}				% input just copy paste the contents of this file here 

\renewcommand{\i}{\mathrm{i}}
\newcommand{\pa}{\partial}

\renewcommand{\le}{\leqslant}

\def\be{\begin{equation}}
\def\ee{\end{equation}}
\newcommand\bea{\begin{eqnarray}}
\newcommand\eea{\end{eqnarray}}

\renewcommand{\le}{\leqslant}
\def\benn{\begin{eqnarray*}}
\def\eenn{\end{eqnarray*}}

\def\a{\alpha}

\newcommand{\JFadd}[1]{\textcolor{black}{#1}}
\newcommand{\rev}[1]{\textcolor{black}{#1}}
\numberwithin{equation}{section}

\begin{document}

\thispagestyle{plain}
\begin{center}
	\huge
	\textbf{\Large{Arctic curves of the $6$V model with partial DWBC and\\ double Aztec rectangles\\}}
	\vspace{0.4cm}
	\large
	Jean-François de Kemmeter$^{1}$, Bryan Debin$^{2}$, Philippe Ruelle$^{2}$\\ % Your name
	\vspace{0.4cm}
	\normalsize
	$^{1}$ Department of Mathematics and Namur Institute for Complex Systems (naXys), University of Namur, 61 rue de Bruxelles, Namur, B-5000, Belgium\\
	$^{2}$ Institut de Recherche en Mathématique et Physique, Université catholique de Louvain, Louvain-la-Neuve, B-1348, Belgium  % Your institution
\end{center}

\begin{abstract}

Previous numerical studies have shown that in the disordered and anti-ferroelectric phases the six-vertex ($6$V) model with partial domain wall boundary conditions (DWBC) exhibits an arctic curve whose exact shape is unknown. The model is defined on a $s\times n$ square lattice \JFadd{($s\leq n$)}. In this paper, we derive the analytic expression of the arctic curve, for $a=b=1$ and $c=\sqrt{2}$ ($\Delta=0$), while keeping the ratio $s/n \JFadd{\,\in [0,1]}$ as a free parameter. The computation relies on the tangent method. We also consider domino tilings of double Aztec rectangles and show via the tangent method that\JFadd{, for particular parameters,} the arctic curve is identical to that of the $6$V model with partial DWBC. Our results are confirmed by extensive numerical simulations.   
\end{abstract}

\tableofcontents

\section{Introduction}

In statistical physics, random domino tilings of Aztec diamonds \rev{(AD)} provide one of the most famous examples of systems whose bulk properties are strongly influenced by their boundary conditons. In those models, the constraints induce by the boundary propagate deep inside the domain, leading to an arctic phenomenon, which is characterised by a phase separation between ordered (frozen) regions adjacent to the boundary and a central disordered (entropic) region. In the thermodynamic limit, obtained by sending the mesh size to $0$ while keeping the size of the domain unchanged, this separation is the so-called \textit{arctic curve} of the model. Without being exhaustive, such an arctic phenomenon is observed in lozenge tilings of hexagons \cite{gorin2019lectures}, in domino tilings of Aztec rectangles with defects \cite{di2019tangent4}, in bounded lecture hall tableaux \cite{corteel2019Hall} or in configuations of the six-vertex ($6$V) with various boundary conditions \cite{lyberg20186V_varietyBC}.

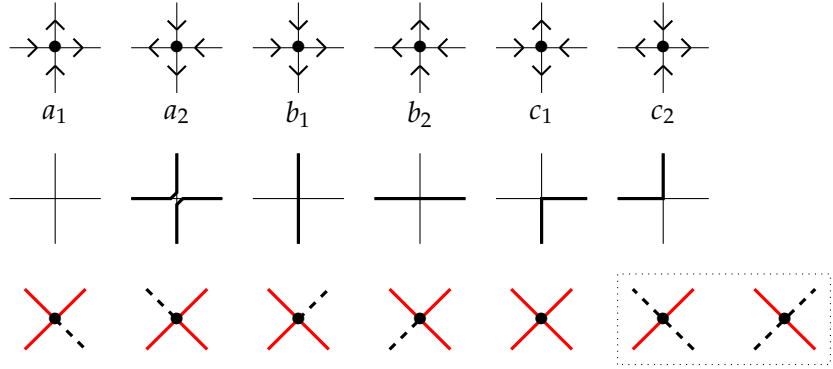
\begin{figure}
\begin{center}
	\begin{tikzpicture}[scale=0.8]
\begin{scope}[yshift=-2.5cm]
	%\draw [thick,<-,>=latex] (0.75, 1.75) -- (0.75, 1.25);
	\draw[color=black] (0,0) -- (1.5,0);
	\draw[color=black] (0.75,-0.75) -- (0.75,0.75);
\end{scope}
\begin{scope}[yshift=-7cm]
	\draw[very thick, color=red] (0.25,2) -- (1.25,3);
	\draw[very thick, color=red] (0.75,2.5) -- (0.25,3);
	%\draw[very thick, dashed, color=black] (0.25,2) -- (1.25,3);
	\draw[very thick, dashed, color=black] (0.75,2.5) -- (1.25,2);
	\draw[black] (0.75,2.5) node[below]{} node{$\bullet$};
\end{scope}
	%\draw [thick,<-,>=latex] (0.75, 1.75) -- (0.75, 1.25);
	\draw[color=black] (0,0) -- (1.5,0);
	\draw[color=black] (0.75,-0.75) -- (0.75,0.75);
	\draw[thick,color=black] (0.3,-0.15) -- (0.45,0);
	\draw[thick,color=black] (0.3,0.15) -- (0.45,0);
	\draw[thick,color=black] (1.05,-0.15) -- (1.2,0);
	\draw[thick,color=black] (1.05,0.15) -- (1.2,0);
	\draw[thick,color=black] (0.6,-0.45) -- (0.75,-0.3);
	\draw[thick,color=black] (0.9,-0.45) -- (0.75,-0.3);
	\draw[thick,color=black] (0.6,0.3) -- (0.75,0.45);
	\draw[thick,color=black] (0.9,0.3) -- (0.75,0.45);
	\draw[black] (0.75,0) node[below]{} node{$\bullet$};
	\node[] at (0.75,-1.1) {$a_1$};
\begin{scope}[xshift=2cm]
\begin{scope}[yshift=-2.5cm]
	%\draw [thick,<-,>=latex] (0.75, 1.75) -- (0.75, 1.25);
	\draw[color=black] (0,0) -- (1.5,0);
	\draw[color=black] (0.75,-0.75) -- (0.75,0.75);
	\draw[very thick] (0,0)--(0.75-0.1,0)--(0.75,0+0.1)--(0.75,0.75);
	\draw[very thick] (0.75,-0.75)--(0.75,0-0.1)--(0.75+0.1,0)--(1.5,0);
\end{scope}
\begin{scope}[yshift=-7cm]
    \draw[very thick, color=red] (0.25,2) -- (1.25,3);
	\draw[very thick, color=red] (0.75,2.5) -- (1.25,2);
	%\draw[very thick, dashed, color=black] (0.25,2) -- (1.25,3);
	\draw[very thick, dashed, color=black] (0.25,3) -- (0.75,2.5);
	\draw[black] (0.75,2.5) node[below]{} node{$\bullet$};
\end{scope}
	\draw[color=black] (0,0) -- (1.5,0);
	\draw[color=black] (0.75,-0.75) -- (0.75,0.75);
	%\draw [thick,<-,>=latex] (0.75, 1.75) -- (0.75, 1.25);
	\draw[thick,color=black] (0.3,0) -- (0.45,0.15);
	\draw[thick,color=black] (0.3,0) -- (0.45,-0.15);
	\draw[thick,color=black] (1.05,0) -- (1.2,0.15);
	\draw[thick,color=black] (1.05,0) -- (1.2,-0.15);
	\draw[thick,color=black] (0.6,-0.3) -- (0.75,-0.45);
	\draw[thick,color=black] (0.9,-0.3) -- (0.75,-0.45);
	\draw[thick,color=black] (0.6,0.45) -- (0.75,0.3);
	\draw[thick,color=black] (0.9,0.45) -- (0.75,0.3);
	\draw[black] (0.75,0) node[below]{} node{$\bullet$};
	\node[] at (0.75,-1.1) {$a_2$};
\end{scope}
\begin{scope}[xshift=4cm]
\begin{scope}[yshift=-2.5cm]
	%\draw [thick,<-,>=latex] (0.75, 1.75) -- (0.75, 1.25);
	\draw[color=black] (0,0) -- (1.5,0);
	\draw[color=black] (0.75,-0.75) -- (0.75,0.75);
	\draw[very thick] (0.75,-0.75)--(0.75,0.75);
\end{scope}
\begin{scope}[yshift=-7cm]
	\draw[very thick, color=red] (0.25,2) -- (0.75,2.5);
	\draw[very thick, color=red] (1.25,2) -- (0.25,3);
	\draw[very thick, dashed, color=black] (0.75,2.5) -- (1.25,3);
	%\draw[very thick, dashed, color=black] (0.25,3) -- (1.25,2);
	\draw[black] (0.75,2.5) node[below]{} node{$\bullet$};
\end{scope}
	\draw[color=black] (0,0) -- (1.5,0);
	\draw[color=black] (0.75,-0.75) -- (0.75,0.75);
	%\draw [thick,<-,>=latex] (0.75, 1.75) -- (0.75, 1.25);
	\draw[thick,color=black] (0.3,-0.15) -- (0.45,0);
	\draw[thick,color=black] (0.3,0.15) -- (0.45,0);
	\draw[thick,color=black] (1.05,-0.15) -- (1.2,0);
	\draw[thick,color=black] (1.05,0.15) -- (1.2,0);
	\draw[thick,color=black] (0.6,-0.3) -- (0.75,-0.45);
	\draw[thick,color=black] (0.9,-0.3) -- (0.75,-0.45);
	\draw[thick,color=black] (0.6,0.45) -- (0.75,0.3);
	\draw[thick,color=black] (0.9,0.45) -- (0.75,0.3);
	\draw[black] (0.75,0) node[below]{} node{$\bullet$};
	\node[] at (0.75,-1.1) {$b_1$};
\end{scope}
\begin{scope}[xshift=6cm]
\begin{scope}[yshift=-2.5cm]
	%\draw [thick,<-,>=latex] (0.75, 1.75) -- (0.75, 1.25);
	\draw[color=black] (0,0) -- (1.5,0);
	\draw[color=black] (0.75,-0.75) -- (0.75,0.75);
	\draw[very thick] (0,0)--(1.5,0);
\end{scope}
\begin{scope}[yshift=-7cm]
	\draw[very thick, color=red] (0.75,2.5) -- (1.25,3);
	\draw[very thick, color=red] (1.25,2) -- (0.25,3);
	\draw[very thick, dashed, color=black] (0.25,2) -- (0.75,2.5);
	%\draw[very thick, dashed, color=black] (0.25,3) -- (1.25,2);
	\draw[black] (0.75,2.5) node[below]{} node{$\bullet$};
\end{scope}
	\draw[color=black] (0,0) -- (1.5,0);
	\draw[color=black] (0.75,-0.75) -- (0.75,0.75);
	%\draw [thick,<-,>=latex] (0.75, 1.75) -- (0.75, 1.25);
	\draw[thick,color=black] (0.3,0) -- (0.45,0.15);
	\draw[thick,color=black] (0.3,0) -- (0.45,-0.15);
	\draw[thick,color=black] (1.05,0) -- (1.2,0.15);
	\draw[thick,color=black] (1.05,0) -- (1.2,-0.15);
	\draw[thick,color=black] (0.6,-0.45) -- (0.75,-0.3);
	\draw[thick,color=black] (0.9,-0.45) -- (0.75,-0.3);
	\draw[thick,color=black] (0.6,0.3) -- (0.75,0.45);
	\draw[thick,color=black] (0.9,0.3) -- (0.75,0.45);
	\draw[black] (0.75,0) node[below]{} node{$\bullet$};
	\node[] at (0.75,-1.1) {$b_2$};
\end{scope}
\begin{scope}[xshift=8cm]
\begin{scope}[yshift=-2.5cm]
	%\draw [thick,<-,>=latex] (0.75, 1.75) -- (0.75, 1.25);
	\draw[color=black] (0,0) -- (1.5,0);
	\draw[color=black] (0.75,-0.75) -- (0.75,0.75);
	\draw[very thick] (0.75,-0.75)--(0.75,0)--(1.5,0);
\end{scope}
\begin{scope}[yshift=-7cm]
	\draw[very thick, color=red] (0.25,2) -- (1.25,3);
	\draw[very thick, color=red] (0.25,3) -- (1.25,2);
	%\draw[very thick, dashed, color=black] (0.25,2) -- (1.25,3);
	%\draw[very thick, dashed, color=black] (0.25,3) -- (1.25,2);
	\draw[black] (0.75,2.5) node[below]{} node{$\bullet$};
\end{scope}
	\draw[color=black] (0,0) -- (1.5,0);
	\draw[color=black] (0.75,-0.75) -- (0.75,0.75);
	%\draw [thick,<-,>=latex] (0.75, 1.75) -- (0.75, 1.25);
	\draw[thick,color=black] (0.3,-0.15) -- (0.45,0);
	\draw[thick,color=black] (0.3,0.15) -- (0.45,0);
	\draw[thick,color=black] (1.05,0) -- (1.2,0.15);
	\draw[thick,color=black] (1.05,0) -- (1.2,-0.15);
	\draw[thick,color=black] (0.6,-0.3) -- (0.75,-0.45);
	\draw[thick,color=black] (0.9,-0.3) -- (0.75,-0.45);
	\draw[thick,color=black] (0.6,0.3) -- (0.75,0.45);
	\draw[thick,color=black] (0.9,0.3) -- (0.75,0.45);
	\draw[black] (0.75,0) node[below]{} node{$\bullet$};
	\node[] at (0.75,-1.1) {$c_1$};
\end{scope}
\begin{scope}[xshift=9cm]
\begin{scope}[yshift=-2.5cm,xshift=1cm]
	%\draw [thick,<-,>=latex] (0.75, 1.75) -- (0.75, 1.25);
	\draw[color=black] (0,0) -- (1.5,0);
	\draw[color=black] (0.75,-0.75) -- (0.75,0.75);
	\draw[very thick] (0,0)--(0.75,0)--(0.75,0.75);
\end{scope}
\begin{scope}[xshift=1cm, yshift=-7cm]
    \draw[dotted] (0,1.75) rectangle ++(3.5,1.5);
	\draw[very thick, color=red] (0.25,2) -- (1.25,3);
	\draw[very thick, color=red] (3.25,2) -- (2.25,3);
	\draw[very thick, dashed, color=black] (1.25,2) -- (0.25,3);
	%\draw[very thick, dashed, color=black] (1.25,3) -- (0.25,2);
	\draw[very thick, dashed, color=black] (2.25,2) -- (3.25,3);
	%\draw[very thick, dashed, color=black] (2.25,3) -- (3.25,2);
	\draw[black] (2.75,2.5) node[below]{} node{$\bullet$};
	\draw[black] (0.75,2.5) node[below]{} node{$\bullet$};
\end{scope}
	\draw[color=black] (1,0) -- (2.5,0);
	\draw[color=black] (1.75,-0.75) -- (1.75,0.75);
	%\draw [thick,<-,>=latex] (0.75, 1.75) -- (1.25, 1.25);
	%\draw [thick,<-,>=latex] (2.75, 1.75) -- (2.25, 1.25);
	\draw[thick,color=black] (1.3,0) -- (1.45,0.15);
	\draw[thick,color=black] (1.3,0) -- (1.45,-0.15);
	\draw[thick,color=black] (2.05,-0.15) -- (2.2,0);
	\draw[thick,color=black] (2.05,0.15) -- (2.2,0);
	\draw[thick,color=black] (1.6,-0.45) -- (1.75,-0.3);
	\draw[thick,color=black] (1.9,-0.45) -- (1.75,-0.3);
	\draw[thick,color=black] (1.6,0.45) -- (1.75,0.3);
	\draw[thick,color=black] (1.9,0.45) -- (1.75,0.3);
	\draw[black] (1.75,0) node[below]{} node{$\bullet$};
	\node[] at (1.75,-1.1) {$c_2$};
\end{scope}
	\end{tikzpicture}
\end{center}
\vspace{-0.5cm}
\caption{\small \textit{Top}: The six possible arrow orientations around each vertex. \textit{Middle}: Osculating paths representation of those six vertices. \textit{Bottom}: Rules used in Figure \ref{correspondance_AD_6vertex} in the one-to-many correspondence between $6$V configurations and domino tilings \cite{zinnjustin20006V_onematrix}. \JFadd{Red segments indicate separations between dominoes.}}
\label{fig_6V_weights_non_crossing}
\end{figure}

As we shall be interested in the $6$V model with particular boundary conditions, we briefly review the main features of this model. The latter is defined on a two-dimensional square lattice. A configuration of this model is obtained by drawing an arrow on each edge with the restriction that the ice-rule and the boundary conditions be satisfied: at each vertex, there must be exactly two ingoing and two outgoing edges, leading to $6 = {4 \choose 2}$ possible arrow configurations around a vertex, hence the name of the model. Each local configuration around a vertex is assigned a weight among the set $\{a_1, a_2, b_1, b_2, c_1, c_2\}$, see Figure \ref{fig_6V_weights_non_crossing}. The partition function is the weighted sum over configurations $\mathcal{C}$ allowed by the ice rule and the boundary conditions 
\begin{equation}
Z=\sum_{\mathcal{C}} \prod_{i=1}^2 a_i^{N_{a_i}} b_i^{N_{b_i}} c_i^{N_{c_i}},
\end{equation}
where $N_{a_i}$, $N_{b_i}$ and $N_{c_i}$ are the number of vertices of each type in configuration $\mathcal{C}$. In the following, we restrict our attention to the subset of weights that are symmetric under arrow reversal, namely $a_1=a_2=a$, $b_1=b_2=b$ and $c_1=c_2=c$. 

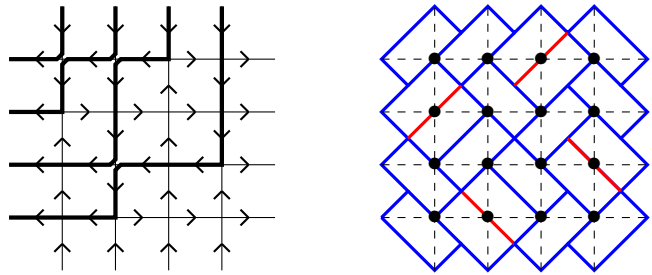
\begin{figure}[h!]
\begin{center}
\begin{tikzpicture}[scale=0.7]
\begin{scope}[xshift=7cm]
	\draw[very thick,color=blue] (0,0) -- (1,-1)--(1.5,-0.5)--(2,-1)--(2.5,-0.5)--(3,-1)--(3.5,-0.5)--(4,-1)--(5,0)--(4.5,0.5)--(5,1)--(4.5,1.5)--(5,2)--(4.5,2.5)--(5,3)--(4,4)--(3.5,3.5)--(3,4)--(2.5,3.5)--(2,4)--(1.5,3.5)--(1,4)--(0,3)--(0.5,2.5)--(0,2)--(0.5,1.5)--(0,1)--(0.5,0.5)--(0,0);
	\draw[very thick,color=blue] (0.5,0.5) -- (1.5,-0.5);
	\draw[very thick,color=blue] (0.5,1.5) -- (1.5,0.5);
	\draw[very thick,color=red] (1.5,0.5) -- (2.5,-0.5);
	\draw[very thick,color=blue] (1,0) -- (4,3);
	\draw[very thick,color=blue] (3.5,3.5) -- (4.5,2.5);
	\draw[very thick,color=blue] (2,1) -- (3,0);
	\draw[very thick,color=blue] (2.5,-0.5) -- (3.5,0.5);
	\draw[very thick,color=blue] (3.5,-0.5) -- (4.5,0.5);
	\draw[very thick,color=blue] (4,0) -- (1,3);
	\draw[very thick,color=blue] (4.5,0.5) -- (3.5,1.5);
	\draw[very thick,color=blue] (4.5,1.5) -- (3.5,2.5);
	\draw[very thick,color=blue] (3,1) -- (4,2);
	\draw[very thick,color=blue] (0.5,2.5) -- (1.5,3.5);
	\draw[very thick,color=red] (0.5,1.5) -- (1.5,2.5);
	\draw[very thick,color=blue] (1.5,2.5) -- (2.5,3.5);
	\draw[very thick,color=blue] (1,1) -- (2,2);
	\draw[very thick,color=blue] (2,3) -- (3,2);
	\draw[very thick,color=red] (2.5,2.5) -- (3.5,3.5);
	\draw[very thick,color=red] (3.5,1.5) -- (4.5,0.5);
	%\draw[very thick,dashed,color=black] (0.5,-0.5) -- (4.5,3.5);
	%\draw[very thick,dashed,color=black] (1.5,-0.5) -- (4.5,2.5);
	%\draw[very thick,dashed,color=black] (2.5,-0.5) -- (4.5,1.5);
	%\draw[very thick,dashed,color=black] (0.5,0.5) -- (3.5,3.5);
	%\draw[very thick,dashed,color=black] (0.5,2.5) -- (3.5,-0.5);
	%\draw[very thick,dashed,color=black] (0.5,3.5) -- (4.5,-0.5);
	%\draw[very thick,dashed,color=black] (1.5,3.5) -- (4.5,0.5);
	%\draw[very thick,dashed,color=black] (2.5,3.5) -- (4.5,1.5);
	%\draw[very thick,dashed,color=black] (3.5,3.5) -- (4.5,2.5);
	%\draw[very thick,dashed,color=black] (0.5,1.5) -- (2.5,-0.5);
	%\draw[very thick,dashed,color=black] (0.5,0.5) -- (1.5,-0.5);
	\draw[dashed,color=black] (1,-1) -- (1,4);
	\draw[dashed,color=black] (2,-1) -- (2,4);
	\draw[dashed,color=black] (3,-1) -- (3,4);
	\draw[dashed,color=black] (4,-1) -- (4,4);
	\draw[dashed,color=black] (0,0) -- (5,0);
	\draw[dashed,color=black] (0,1) -- (5,1);
	\draw[dashed,color=black] (0,2) -- (5,2);
	\draw[dashed,color=black] (0,3) -- (5,3);
	\draw[black] (1,0) node[below]{} node{$\bullet$};
	\draw[black] (2,0) node[below]{} node{$\bullet$};
	\draw[black] (3,0) node[below]{} node{$\bullet$};
	\draw[black] (4,0) node[below]{} node{$\bullet$};
	\draw[black] (1,1) node[below]{} node{$\bullet$};
	\draw[black] (2,1) node[below]{} node{$\bullet$};
	\draw[black] (3,1) node[below]{} node{$\bullet$};
	\draw[black] (4,1) node[below]{} node{$\bullet$};
	\draw[black] (1,2) node[below]{} node{$\bullet$};
	\draw[black] (2,2) node[below]{} node{$\bullet$};
	\draw[black] (3,2) node[below]{} node{$\bullet$};
	\draw[black] (4,2) node[below]{} node{$\bullet$};
	\draw[black] (1,3) node[below]{} node{$\bullet$};
	\draw[black] (2,3) node[below]{} node{$\bullet$};
	\draw[black] (3,3) node[below]{} node{$\bullet$};
	\draw[black] (4,3) node[below]{} node{$\bullet$};
\end{scope}
\begin{scope}
    \draw[color=black] (1,-1) -- (1,4);
	\draw[color=black] (2,-1) -- (2,4);
	\draw[color=black] (3,-1) -- (3,4);
	\draw[color=black] (4,-1) -- (4,4);
	\draw[color=black] (0,0) -- (5,0);
	\draw[color=black] (0,1) -- (5,1);
	\draw[color=black] (0,2) -- (5,2);
	\draw[color=black] (0,3) -- (5,3);
	\arrowdown{1}{3.5};
	\arrowdown{2}{3.5};
	\arrowdown{3}{3.5};
	\arrowdown{4}{3.5};
	\arrowleft{0.5}{0};
	\arrowleft{0.5}{1};
	\arrowleft{0.5}{2};
	\arrowleft{0.5}{3};
	\arrowleft{1.5}{3};
	\arrowleft{2.5}{3};
	\arrowright{3.5}{3};
	\arrowright{4.5}{3};
	\arrowright{4.5}{2};
	\arrowright{4.5}{1};
	\arrowright{4.5}{0};
	\arrowright{1.5}{2};
	\arrowright{2.5}{2};
	\arrowright{3.5}{2};
	\arrowleft{1.5}{1};
	\arrowleft{2.5}{1};
	\arrowleft{3.5}{1};
	\arrowleft{1.5}{0};
	\arrowright{2.5}{0};
	\arrowright{3.5}{0};
	\arrowup{1}{-0.5};
	\arrowup{2}{-0.5};
	\arrowup{3}{-0.5};
	\arrowup{4}{-0.5};
	\arrowdown{1}{2.5};
	\arrowdown{2}{2.5};
	\arrowup{3}{2.5};
	\arrowdown{4}{2.5};
	\arrowup{3}{1.5};
	\arrowdown{4}{1.5};
	\arrowup{1}{1.5};
	\arrowdown{2}{1.5};
	\arrowup{1}{0.5};
	\arrowdown{2}{0.5};
	\arrowup{3}{0.5};
	\arrowup{4}{0.5};
	\draw[ultra thick] (0,0)--++(2,0)--++(0,1-0.1)--++(0.1,0.1)--++(2-0.1,0)--++(0,3);
	\draw[ultra thick] (0,1)--++(2-0.1,0)--++(0.1,0.1)--++(0,2-0.2)--++(0.1,0.1)--++(1-0.1,0)--++(0,1);
	\draw[ultra thick] (0,2)--++(1,0)--++(0,1-0.1)--++(0.1,0.1)--++(1-0.2,0)--++(0.1,0.1)--++(0,1-0.1);
	\draw[ultra thick] (0,3)--++(0.9,0)--++(0.1,0.1)--++(0,0.9);
\end{scope}
\end{tikzpicture}
\end{center}
\vspace{-0.4cm}
\caption{\small \textit{Left}: Configuration of the $6$V model with DWBC, along with its osculating paths description. \textit{Right}: corresponding tilings of the AD, obtained using the rules of Figure \ref{fig_6V_weights_non_crossing}. \JFadd{The correspondence is one-to-many as red edges may be rotated by $\frac{\pi}{2}$.}}
\label{correspondance_AD_6vertex}
\end{figure}

The $6$V model is strongly affected by boundary conditions. This property was not \textit{a priori} obvious and was first substantiated in \cite{korepin2000thermo_6V_DWBC_toda}, \JFadd{where the free energy per site was computed for the domain wall boundary conditions (DWBC). These boundary conditions are presented in Figure \ref{correspondance_AD_6vertex} (see \cite{slavnov2018ABA} for a review and Figure 6.10 therein for an explanation of the name ``domain wall''). The phase diagrams of the $6$V model with DWBC and periodic boundary conditions are identical, although the nature of the phases are different. The computation shows that depending on the value of $\Delta\equiv\frac{a^2+b^2-c^2}{2ab}$, the model can be in three distinct phases: ferroelectric (FE) when $\Delta > 1$, disordered (D) for $-1 < \Delta < 1$, and anti-ferroelectric if $\Delta < -1$. In the disordered and anti-ferroelectric phases, the $6$V model with DWBC exhibits an arctic phenomenon, namely a separation between ordered or frozen regions (in which all the arrows have the same orientation) and an inner disordered or liquid region (where the $6$ types of vertices are found without any apparent pattern) \cite{Syljuasen2004directed_loop, allison20056V_MonteCarlo,cugliandolo2015Bethe_Peierls}.}

The arctic curve of the $6$V model with DWBC was derived for fully general weights in \cite{colomo2010ac_6V_disord,colomo2010ac_6V_AF,colomo2016tangent}. The first successful approach \cite{colomo2010ac_6V_disord,colomo2010ac_6V_AF} was based on a non-local observable called the emptiness formation probability (EFP), whose asymptotic was extracted using the technical assumption that the roots of some saddle point equations condensate. The EFP technique yields an arctic curve whose shape is the solution of $F(x,y,z)=0$ and $\pa_z F(x,y,z)=0$, which exactly describes the envelope of a family of curves parametrized by $z$, which moreover turned out to be straight lines. This observation was then elevated to a universal geometric principle by the development of the tangent method \cite{colomo2016tangent}, which provides an alternative derivation of the arctic curve and which has been applied to a large class of models  \cite{di2018tangent1,di2018tangent2,di2019tangent3,di2019tangent4,debin2020_20V,colomo2019Lshaped,aggarwal2019three_bundle,corteel2019Hall,passos2019ac_6V_refl_end,passos2020PhD,di2021tangent5,di2021twenty}.\newline

Interestingly, configurations of vertex models with DWBC are related to domino tilings of the AD, using the correspondence shown in Figure \ref{fig_6V_weights_non_crossing} \cite{zinnjustin20006V_onematrix}. The  correspondence is not one-to-one, due to the ambiguity in the $c_2$ vertex, but relates partition functions of the two models, when $a=b=1$ and $c=\sqrt{2}$ ($\Delta=0$). Unsurprisingly, the arctic curve is found to be a circle at this special point called the free-fermion point \cite{colomo2008arctic_circle}. For other weights, the $6$V model with DWBC  may be seen as an interacting version of the AD tiling model. There exist other boundary conditions (all resembling DWBC  in some way) for which the $6$V model exhibits an arctic curve \cite{lyberg20186V_varietyBC}. The reflecting boundary conditions have been studied analytically in \cite{passos2019ac_6V_refl_end, passos2020PhD,di2021tangent5}, \JFadd{but the so-called partial domain-wall boundary conditions (pDWBC) have not been considered to this day.}

This paper is organised as follows. In Section \ref{sec1}, we compute, by means of the tangent method, the arctic curve of the $6$V model with pDWBC at the free fermion point and for particular values of the weights ($a=1, b=1, c=\sqrt{2}$). The size of the rectangular domain yields an extra free parameter. The core of the computation relies in the evaluation of the asymptotics of a ratio of determinants, that is addressed by means of a LU decomposition.  \JFadd{In Section \ref{sec3}, we argue that the analytic continuation of this arctic curve should be the arctic curve of an other model, namely a particular case of domino tilings of the double Aztec rectangle. We verify that this is the case by computing the arctic curve via the tangent method, for the double Aztec rectangle.}. Our results are confirmed by extensive numerical simulations.

\section{Six-vertex model with pDWBC}\label{sec1}

\begin{figure}[h!]
\centering
\begin{tikzpicture}[scale=0.8]
\draw[step=0.75cm,color=black] (0,0) grid (3.75,2.25);
\draw[step=0.75cm,color=black] (-0.75,0) grid (0,0);
\draw[step=0.75cm,color=black] (-0.75,0.75) grid (0,0.75);
\draw[step=0.75cm,color=black] (-0.75,1.5) grid (0,1.5);
\draw[step=0.75cm,color=black] (-0.75,2.25) grid (0,2.25);
\draw[step=0.75cm,color=black] (3.75,0) grid (4.5,0);
\draw[step=0.75cm,color=black] (3.75,0.75) grid (4.5,0.75);
\draw[step=0.75cm,color=black] (3.75,1.5) grid (4.5,1.5);
\draw[step=0.75cm,color=black] (3.75,2.25) grid (4.5,2.25);
\draw[step=0.75cm,color=black] (0,-0.75) -- (0,0);
\draw[step=0.75cm,color=black] (0.75,-0.75) -- (0.75,0);
\draw[step=0.75cm,color=black] (1.5,-0.75) -- (1.5,0);
\draw[step=0.75cm,color=black] (2.25,-0.75) -- (2.25,0);
\draw[step=0.75cm,color=black] (3,-0.75) -- (3,0);
\draw[step=0.75cm,color=black] (3.75,-0.75) -- (3.75,0);
\draw[step=0.75cm,color=black] (0,2.25) -- (0,3);
\draw[step=0.75cm,color=black] (0.75,2.25) -- (0.75,3);
\draw[step=0.75cm,color=black] (1.5,2.25) -- (1.5,3);
\draw[step=0.75cm,color=black] (2.25,2.25) -- (2.25,3);
\draw[step=0.75cm,color=black] (3,2.25) -- (3,3);
\draw[step=0.75cm,color=black] (3.75,2.25) -- (3.75,3);
\draw[thick,color=red] (-0.45,0) -- (-0.3,-0.15);
\draw[thick,color=red] (-0.45,0) -- (-0.3,0.15);
\draw[thick,color=red] (-0.45,0.75) -- (-0.3,0.6);
\draw[thick,color=red] (-0.45,0.75) -- (-0.3,0.9);
\draw[thick,color=red] (-0.45,1.5) -- (-0.3,1.35);
\draw[thick,color=red] (-0.45,1.5) -- (-0.3,1.65);
\draw[thick,color=red] (-0.45,2.25) -- (-0.3,2.1);
\draw[thick,color=red] (-0.45,2.25) -- (-0.3,2.4);

\draw[thick,color=blue] (-0.15,2.70) -- (0,2.55);
\draw[thick,color=blue] (+0.15,2.70) -- (0,2.55);

\draw[thick,color=blue] (-0.15+1.5,2.70) -- (0+1.5,2.55);
\draw[thick,color=blue] (+0.15+1.5,2.70) -- (0+1.5,2.55);

\draw[thick,color=blue] (-0.15+1.5+0.75,2.70) -- (0+1.5+0.75,2.55);
\draw[thick,color=blue] (+0.15+1.5+0.75,2.70) -- (0+1.5+0.75,2.55);

\draw[thick,color=blue] (-0.15+3.75,2.70) -- (0+3.75,2.55);
\draw[thick,color=blue] (+0.15+3.75,2.70) -- (0+3.75,2.55);

\draw[thick,color=red] (-0.45,0) -- (-0.3,-0.15);
\draw[thick,color=red] (-0.45,0) -- (-0.3,0.15);
\draw[thick,color=red] (4.05,-0.15) -- (4.2,0);
\draw[thick,color=red] (4.05,0.15) -- (4.2,0);
\draw[thick,color=red] (4.05,0.6) -- (4.2,0.75);
\draw[thick,color=red] (4.05,0.9) -- (4.2,0.75);
\draw[thick,color=red] (4.05,1.35) -- (4.2,1.5);
\draw[thick,color=red] (4.05,1.65) -- (4.2,1.5);
\draw[thick,color=red] (4.05,2.1) -- (4.2,2.25);
\draw[thick,color=red] (4.05,2.4) -- (4.2,2.25);
\draw[thick,color=red] (-0.15,-0.45) -- (0,-0.3);
\draw[thick,color=red] (0.15,-0.45) -- (0,-0.3);
\draw[thick,color=red] (0.6,-0.45) -- (0.75,-0.3);
\draw[thick,color=red] (0.9,-0.45) -- (0.75,-0.3);
\draw[thick,color=red] (1.35,-0.45) -- (1.5,-0.3);
\draw[thick,color=red] (1.65,-0.45) -- (1.5,-0.3);
\draw[thick,color=red] (2.1,-0.45) -- (2.25,-0.3);
\draw[thick,color=red] (2.4,-0.45) -- (2.25,-0.3);
\draw[thick,color=red] (2.85,-0.45) -- (3,-0.3);
\draw[thick,color=red] (3.15,-0.45) -- (3,-0.3);
\draw[thick,color=red] (3.6,-0.45) -- (3.75,-0.3);
\draw[thick,color=red] (3.9,-0.45) -- (3.75,-0.3);
\draw[<->,very thick, >=latex] (0,-1)--(3.75,-1); 
\draw[<->,very thick, >=latex] (-1,0)--(-1,2.25);
\draw (1.875,-1.5) node(N) {$n$};
\draw (-1.5,1.125) node(N) {$s$};%
\arrowup{0}{0.45};
\arrowup{0.75}{0.45};
\arrowup{1.5}{0.45};
\arrowdown{2.25}{0.3};
\arrowup{3}{0.45};
\arrowup{3.75}{0.45};%
\arrowup{0}{1.2};
\arrowup{0.75}{1.2};
\arrowdown{1.5}{1.05};
\arrowdown{2.25}{1.05};
\arrowdown{3}{1.05};
\arrowup{3.75}{1.2};%
\arrowup{0}{1.95};
\arrowdown{0.75}{1.8};
\arrowup{1.5}{1.95};
\arrowdown{2.25}{1.8};
\arrowup{3}{1.95};
\arrowdown{3.75}{1.8};%
%
%\arrowdown{0}{2.55};
\arrowup{0.75}{2.7};
%\arrowdown{1.5}{2.55};
%\arrowdown{2.25}{2.55};
\arrowup{3}{2.7};
%\arrowdown{3.75}{2.55};%
%
\arrowleft{0.3}{0};
\arrowleft{1.05}{0};
\arrowleft{1.75}{0};
\arrowright{2.7}{0};
\arrowright{3.45}{0};%
\arrowleft{0.3}{0.75};
\arrowleft{1.05}{0.75};
\arrowleft{1.8}{0.75};
\arrowleft{2.55}{0.75};
\arrowright{3.45}{0.75};%
\arrowleft{0.3}{1.5};
\arrowright{1.2}{1.5};
\arrowleft{1.8}{1.5};
\arrowright{2.7}{1.5};
\arrowleft{3.3}{1.5};%
\arrowright{0.45}{2.25};
\arrowleft{1.05}{2.25};
\arrowright{1.95}{2.25};
\arrowright{2.7}{2.25};
\arrowright{3.45}{2.25};%
\draw (0,3.25);
\draw (0.75,3.25);
\draw (1.5,3.25);
\draw (2.25,3.25);
\draw (3,3.25);
\draw (3.75,3.25);%
\draw (4.75,2.25);
\draw (4.75,1.5);
\draw (4.75,0.75);
\draw (4.75,0);
\end{tikzpicture}\hspace{1cm}
\begin{tikzpicture}[scale=0.8]
\clip (-0.8,-1.7) rectangle (5.1,3);
\draw[step=0.75cm,color=black] (0,0) grid (3.75,2.25);
\draw[step=0.75cm,color=black] (-0.75,0) grid (0,0);
\draw[step=0.75cm,color=black] (-0.75,0.75) grid (0,0.75);
\draw[step=0.75cm,color=black] (-0.75,1.5) grid (0,1.5);
\draw[step=0.75cm,color=black] (-0.75,2.25) grid (0,2.25);
\draw[step=0.75cm,color=black] (3.75,0) grid (4.5,0);
\draw[step=0.75cm,color=black] (3.75,0.75) grid (4.5,0.75);
\draw[step=0.75cm,color=black] (3.75,1.5) grid (4.5,1.5);
\draw[step=0.75cm,color=black] (3.75,2.25) grid (4.5,2.25);
\draw[step=0.75cm,color=black] (0,-0.75) -- (0,0);
\draw[step=0.75cm,color=black] (0.75,-0.75) -- (0.75,0);
\draw[step=0.75cm,color=black] (1.5,-0.75) -- (1.5,0);
\draw[step=0.75cm,color=black] (2.25,-0.75) -- (2.25,0);
\draw[step=0.75cm,color=black] (3,-0.75) -- (3,0);
\draw[step=0.75cm,color=black] (3.75,-0.75) -- (3.75,0);
\draw[step=0.75cm,color=black] (0,2.25) -- (0,3);
\draw[step=0.75cm,color=black] (0.75,2.25) -- (0.75,3);
\draw[step=0.75cm,color=black] (1.5,2.25) -- (1.5,3);
\draw[step=0.75cm,color=black] (2.25,2.25) -- (2.25,3);
\draw[step=0.75cm,color=black] (3,2.25) -- (3,3);
\draw[step=0.75cm,color=black] (3.75,2.25) -- (3.75,3);
\draw[line width=0.5mm,color=black] (-0.75,0) -- (2.25,0)--(2.25,0.75-0.1)--(2.25+0.1,0.75)--(3,0.75)--(3,1.5)--(3.75,1.5)--(3.75,3);
\draw[line width=0.5mm,color=black] (-0.75,0.75) -- (1.5+0.65,0.75)--(2.25,1.5-0.65)--(2.25,3);
\draw[line width=0.5mm,color=black] (-0.75,1.5) --(0.75,1.5)--(0.75,2.25)--(1.5,2.25)--(1.5,3);
\draw[line width=0.5mm,color=black] (-0.75,2.25)--(0,2.25)--(0,3);
\end{tikzpicture}
\vspace{-0.5cm}
\caption{\small \textit{Left}: Configuration of the $6$V with pDWBC for $s=4$ and $n=6$. The arrows in red are fixed. The upper boundary remains free and must contain $s$ down arrows (in blue) due to the ice rule. \textit{Right}: Osculating paths description of this configuration.}
\label{graphe_6V_pDWBC}
\end{figure}
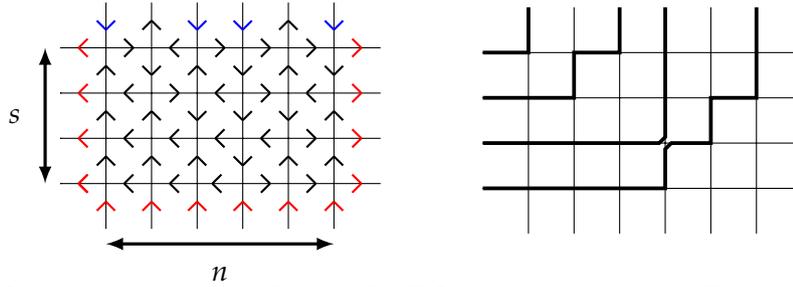

In this section, we study the $6$V model on a rectangular $s \times n$ (with $s\leq n$) domain with pDWBC. These boundary conditions are a mixture of fixed and free boundary conditions. In the case of pDWBC, arrows on the upper boundary of the rectangle are free, \JFadd{while they are fixed as in DWBC on the other boundaries}, see Figure \ref{graphe_6V_pDWBC}. Because of the ice rule, there are exactly $s$ down arrows on the upper boundary. Hence, for $s=n$, pDWBC exactly reproduce DWBC. A given configuration can equally well be described by osculating paths instead of arrows using the dictionary shown in Figure \ref{fig_6V_weights_non_crossing}. \JFadd{These osculating paths start from the $s$ vertices on the left boundary and end at $s$ vertices on the upper boundary, whose positions are not imposed.} An arctic phenomenon occurs in the D and AF phases \cite{cugliandolo2015Bethe_Peierls, lyberg20186V_varietyBC}. In the FE case, the partition function was studied in detail \JFadd{and as for DWBC, there is no arctic phenomenon.}

For DWBC, one may impose that $a_1=a_2$, $b_1=b_2$ and $c_1=c_2$, without loss of generality. Indeed, for fixed boundary conditions the number of horizontal and vertical steps is fixed, and so is the difference between the number of left-turns and right-turns. This leads to four \textit{conservations laws}, namely it fixes $N_{a_1}+N_{a_2}+N_{b_1}+N_{b_2}+N_{c_1}+N_{c_2}$, $2N_{a_2}+2N_{b_2}+N_{c_1}+N_{c_2}$, $2N_{a_2}+2N_{b_1}+N_{c_1}+N_{c_2}$ and $N_{c_2}-N_{c_1}$, which allow to assume the aforementioned conditions. In the pDWBC case, the number of horizontal steps is not fixed and imposing $a_1=a_2$, $b_1=b_2$ and $c_1=c_2$ is a \textit{restriction}. Asymmetric weights also lead to interesting arctic phenomena for other boundary conditions \cite{aggarwal2020stochast, dGKW18}. For pDWBC it was shown numerically that this asymmetry deforms the arctic curve \cite{lyberg20186V_varietyBC}. In this section, we use the tangent method \cite{di2018tangent1} to compute the arctic curve in the symmetric (free-fermion) case $a=b=1$ and $c=\sqrt{2}$.

\subsection{Family of tangent lines}\label{sec:famille_tan}
The tangent method applies to configurations that are in bijection with a set of directed non-crossing random paths. This is the case of domino tilings and configurations of vertex models. The tangent method relies on the reasonable assumption that the arctic curve is not influenced by the displacement of a single random path. In the scaling limit, this displaced path is a straight line that hits tangentially the arctic curve \cite{debin2019concavity}. This property is at the core of the tangent method. \JFadd{In its most usual form, the tangent method prescribes us to move outside the original} domain the starting point of a well-chosen path and to compute (in the scaling limit) the equation of the corresponding straight line on the basis of the most likely entry point of the displaced path in the domain. Varying the position of the starting point leads to a family of straight lines whose envelope is a portion of the arctic curve. Recently a reformulation of the tangent method \JFadd{that does not call for an extension of the domain} was also proposed in \cite{debin2021multi}.\newline  

\begin{figure}[h!]
	\centering
	\includegraphics[scale=0.6, trim = 0 0 100 0]{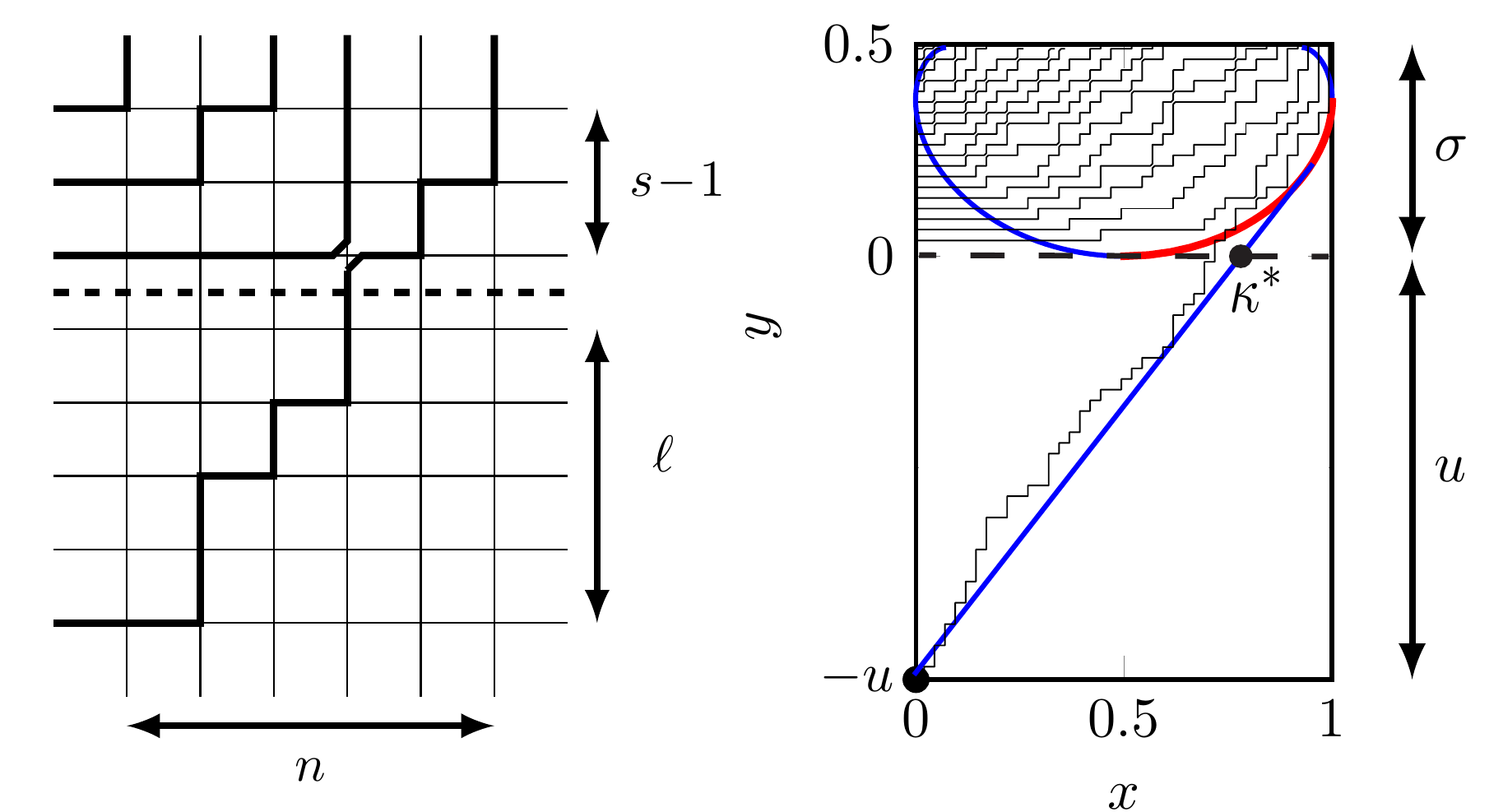}
	\vspace{-0.5cm}
	\caption{\small \textit{Left}: Tangent method set-up, on the lattice. \textit{Right}: Rescaled domain picture \JFadd{superposed with a finite size configuration}. The red portion is the only one accessible using this extension.}
	\label{domaine_meth_6V}
\end{figure}

The geometrical setting for the tangent method is identical to that of the DWBC  case, see Figure \ref{domaine_meth_6V}. The two derivations differ in the only non-trivial part of the tangent method, namely the computation of the asymptotics of the boundary one-point function $H_{n,s}^{(k)}$, encompassed in a function called $r_\sigma(z)$. For completeness and to introduce the notations, we briefly recall the determination of the family of tangent lines $F(x,y,z)$ in terms of $r_\sigma(z)$, which is identical to that of \cite{colomo2016tangent}.

The one-point function $H_{n,s}^{(k)}$ is the probability that the only type $c$ vertex on the bottom row is in the $k$-th column (with $1\leq k \leq n$), see Figure \ref{graphe_6V_pDWBC}. To its left there are $k-1$ type $b$ vertices and to its right $(n-k)$ type $a$ ones. In the osculating paths description, $H_{n,s}^{(k)}$ is the probability that the lowest path takes its first step upwards at position $k$. It is given by
\begin{equation}
    H_{n,s}^{(k)}=b^{k-1}\,c\,a^{n-k}\frac{Z_{n,s-1}^{(k)}}{Z_{n,s}},
    \label{1point_f}
\end{equation}
where $Z_{n,s-1}^{(k)}$ is the  partition function of a $6$V model on a $(s-1) \times n$ grid with slightly modified pDWBC, in the sense that the bottom line has a down arrow at position $k$. A natural quantity to consider  when dealing with  \textit{inhomogeneous} models is---as will become clear with \eqref{eq_onept_gen}---the generating function of the one-point function
\begin{equation}
    h_{n,s}(z)\equiv \sum_{k=1}^n H_{n,s}^{(k)}\; z^{k-1}.
    \label{eq_def_hn}
\end{equation}
All we need to know about its asymptotics is captured in 
\begin{equation}
    r_{\sigma}(z)\equiv \lim_{n\rightarrow+\infty} \frac{1}{n}z\frac{d}{dz}\log h_{n,s}(z),
    \label{resolvent}
\end{equation}
where the aspect ratio $\sigma=\frac{s}{n}$ is kept fixed. Indeed, $r_{\sigma}(z)$  can be used to compute $S_\sigma (\zeta) \equiv \lim_{n\to \infty} \frac{1}{n} \log H_{n,s}^{(k)}$, for $k=n\zeta$. The link is established by studying the asymptotics of \eqref{eq_def_hn}, 
\begin{equation}
    h_{n,s}(z)\approx \int_0^1 \text{d}\zeta \;{\rm e}^{n[S_{\sigma}(\zeta)+\zeta\log{z}]}\approx {\rm e}^{n[S_{\sigma}(\zeta^*)+\zeta^*\log{z}]},
    \label{eq_saddle_pt_hz}
\end{equation}
\JFadd{with the notation $a_n \approx b_n \Leftrightarrow \lim_{n\rightarrow \infty} \frac{1}{n} \log(\frac{a_n}{b_n})=0$} and with $\zeta^*$ solution of the saddle point equation. We then have
\begin{equation}
\left\lbrace
    \begin{aligned}
    &S_{\sigma}'(\zeta^*)=-\log z,\\
    &r_{\sigma}(z)={\zeta}^*, 
    \end{aligned}
    \right. 
    \label{eq_saddle_rz}
\end{equation}
so that $S_{\sigma}(\zeta^*)$ is related to $r_\sigma(z)$ through some kind of Legendre transform. The two relations \eqref{eq_saddle_rz} may also be seen as being the inverse of each other. It will turn out to be sufficient to only have an explicit expression for $r_\sigma(z)$; computing $S_{\sigma}(\zeta)$ is not required for the present purpose. Notice that in \eqref{eq_saddle_pt_hz}, \JFadd{we have assumed that the saddle point solution $\zeta^*$ is in $[0,1]$, which can be verified a posteriori using} the explicit formula $r_\sigma(z)$ derived below. 

To apply the tangent method we extend the domain as shown in Figure \ref{domaine_meth_6V} \JFadd{and let the lowest path start on the left side of the extension, at the point $(0,-\ell) \equiv (0,-u n)$}. The partition function of the extended domain $Z^\text{ext}_n$ can be expressed as a sum over the entry point $k$ of two independent partition functions. If we parametrize the weights as $t\equiv \frac{b}{a}$ and $\frac{c}{a}=\sqrt{1-2\Delta t +t^2}$, we get 
\begin{equation}
    \frac{Z^\text{ext}_n}{Z_{n,s}}
    = \frac {1}{Z_{n,s}} \sum_k Z^{(k)}_{n,s-1} Y_{k,l}
    =\sum_{k=1}^n\frac{1}{a^n}\frac{1}{t^{k-1}\sqrt{1-2\Delta t + t^2}}\,H_{n,s}^{(k)}\,Y_{k,\ell},
    \label{Z_D}
\end{equation}
where $Y_{k,\ell}$ is the partition function in the $\ell\times n$ extension  that includes all vertices below the dashed line in Figure \ref{domaine_meth_6V}, with a prescribed exit point at column $k$. To evaluate $Y_{k,\ell}$, we simply notice the identity $N_{c_2}=N_{c_1}+1$ in this extension. The enumeration for a fixed $m=N_{c_2}$ is straightforward, and we find
\begin{equation}
    Y_{k,\ell}={\left(\frac{b}{a}\right)}^{k+\ell-1}a^{n\ell}\sum_{m=0}^{k-1}\binom{k-1}{m}\binom{\ell-1}{m}{\left(\frac{c}{b}\right)}^{2m+1},
\end{equation}
Injecting this last expression into \eqref{Z_D}, we obtain:
\begin{equation}
    \frac{Z^\text{ext}_n}{Z_{n,s}}=a^{n(\ell-1)}t^{\ell-1}\sum_{k=1}^{n}H_{n,s}^{(k)}\sum_{m=0}^{k-1}\binom{k-1}{m}\binom{\ell-1}{m}{\left(\frac{1-2\Delta t+t^2}{t^2}\right)}^{m}.
\end{equation}
With $k = n \kappa$ and $m = n \eta$, the asymptotics is given by 
\begin{equation}
    \frac{Z^\text{ext}_n}{Z_{n,s}}\approx a^{n(\ell-1)}t^\ell 
    \int_0^1\mbox{d}\kappa \int_0^{\kappa}\mbox{d}\eta \, {\rm e}^{n S(\kappa,\eta,u)},
\end{equation}
where, \JFadd{using the Stirling formula}, 
\be
S(\kappa,\eta,u)=S_{\sigma}(\kappa) +\mathcal{L}(\kappa)-\mathcal{L}(\eta)-\mathcal{L}(\kappa-\eta)+\mathcal{L}(u)-\mathcal{L}(\eta)
-\mathcal{L}(u-\eta)+\eta \log \lr{\frac{t^2-2\Delta t +1}{t^2}},
\ee
with $\mathcal{L}(x)\equiv x\log(x)$. To find the most likely entry point $\kappa^*$ in the rescaled domain, we perform a saddle point analysis. The saddle point equations are
\begin{empheq}{align}
    &S_{\sigma}'({\kappa}^*)+\log({\kappa}^*)-\log({\kappa}^*-{\eta}^*) = 0, 
    \label{eq_pt_selle1}
    \\
    &-2\log({\eta}^*)+\log({\kappa}^*-{\eta}^*)+\log(u-{\eta}^*)+\log\lr{\frac{t^2-2\Delta t+1}{t^2}} = 0.
    \label{eq_pt_selle2}
\end{empheq}
Using \eqref{eq_saddle_rz}, we deduce from \eqref{eq_pt_selle1} that
\begin{equation}
    S_{\sigma}'({\kappa}^*)=-\log\lr{\frac{{\kappa}^*}{{\kappa}^*-{\eta}^*}} \quad \Rightarrow \quad z=\frac{\kappa^*}{\kappa^*-\eta^*}, \quad \kappa^*(z)=r_\sigma(z).
    \label{eq_S_0}
\end{equation}
We now have all the ingredients---with the exception of a formula for $r_\sigma(z)$---to compute the family of tangent lines.  We could solve the implicit equation \eqref{eq_S_0} to find $\eta^*(\kappa^*)$ and then inject this solution into \eqref{eq_pt_selle2} to find $\kappa^*(u)$. Instead of doing that, we choose to parametrize the family of tangent lines by $z$ instead of $u$. In those terms the family of lines is 
\begin{equation}
    F(x,y,z)=x-\frac{{\kappa}^*(u(z))}{u(z)}y-r_{\sigma}(z)=0.
    \label{famille_droites}
\end{equation}
\JFadd{In order to express $\frac{{\kappa}^*(u(z))}{u(z)}$ as an explicit function of $z$, we solve the quadratic equation \eqref{eq_pt_selle2}. Using $z^{-1}=1-\frac{\eta^*}{\kappa^*}$}, one gets after elementary manipulations
\begin{equation}
    \frac{{\kappa}^*(u(z))}{u(z)}=\frac{z(t^2-2\Delta t+1)}{(z-1)(t^2z-2\Delta t +1)}.
\end{equation}
Hence, we conclude that
\begin{equation}
    F(x,y,z)=x-\frac{z(t^2-2\Delta t+1)}{(z-1)(t^2z-2\Delta t +1)}y-r_{\sigma}(z)=0.
    \label{famille_Dz}
\end{equation}
\JFadd{In what follows, we compute $r_{\sigma}(z)$, which strongly depends on the specific model at hand.}

\subsection{Almost homogeneous limit}\label{sec:Limites_det}

\JFadd{The computation of $r_{\sigma}(z)$, defined in (\ref{resolvent}), will be achieved by considering} 
the vertically \textit{inhomogeneous} $6$V model with pDWBC where vertices on line $j$ (numbered from $1$ to $s$ from top to bottom) are assigned the weights
\begin{equation}
a_j=1, \quad b_j=t_j, \quad c_j=\sqrt{1-2\Delta t_j +t_j^2},
\end{equation}
in such a way that $\Delta_j=\Delta$. \JFadd{The original partition function $Z_{n,s}$ is recovered by taking the limit $t_j \rightarrow t$. In the present case, an almost homogeneous limit, by which the inhomogeneity is kept in the last row only, is enough to make contact with the one-point function,}
\begin{equation}
    \frac{Z_{n,s}(t,\cdots,t, t_s)}{Z_{n,s}}=\frac{c_s}{c}\, \sum_{k=1}^n H_{n,s}^{(k)} \,{\left(\frac{b_s}{b}\right)}^{k-1}=\frac{c_s}{c} \,h_{n,s}(t_s/t).
    \label{eq_onept_gen}
\end{equation}

\JFadd{From now on we consider the free-fermion point $\Delta=0$. In that case, the fully inhomogeneous  partition function has been computed in \cite{pronko2019pronko_pronko}, with the following result,}
\begin{equation}
    Z_{n,s}(t_1,\cdots,t_s)=\prod_{i=1}^s\frac{c_i}{1-t_i}\prod_{1\leq i < j \leq s}\frac{1+t_i t_j}{(1-t_it_j)(t_j-t_i)}\det_{1\leq i,j \leq s}\left[t_j^{i-1}-{t_j}^{n+s-i}\right].
    \label{det_Pronko}
\end{equation}

Let us first evaluate the homogeneous limit where $t_j \to t=1$. For the DWBC, the analogue of \eqref{det_Pronko} is the Izergin-Korepin determinantal formula, for which the homogeneous limit can be taken stepwise, by evaluating the limit over one inhomogeneity at a time. In this case, zeros coming from Taylor expanding the determinant are exactly cancelled by poles contained in the prefactor \cite{IKdet}. In the case at hand, the homogeneous limit of formula \eqref{det_Pronko} can be evaluated in a similar way, \JFadd{with only poles of odd order appearing in the prefactor}. The result of the homogeneous limit is 
\begin{equation}
    Z_{n,s}(1,\cdots,1)= 
    -c^s2^{s(s-1)/2}\lr{\prod_{k=1}^s\frac{1}{(2k-1)!}}
    \det_{1\leq i,j \leq s}(A_{ij}),
    \label{Z_hom}
\end{equation}
where the entries
\begin{equation}
    A_{ij}\equiv f_i^{(2j-1)}(1)=\frac{(i-1)!}{(i-1-(2j-1))!}-\frac{(n+s-i)!}{(n+s-i-(2j-1))!},
    \label{elements_A}
\end{equation}
are the derivatives of $f_i(t)\equiv t^{i-1}-t^{n+s-i}$. We use the (standard) convention that if the factorial of a negative number appears in a denominator, the corresponding term is set to zero. A rigorous derivation of \eqref{Z_hom} is presented in Appendix \ref{sec:App_Homogeneous}. The almost homogeneous limit is taken in a similar fashion and yields 
\begin{equation}
    Z_{n,s}(1,\cdots,1,t_s)= c^{s-1}c_s2^{\frac{(s-1)(s-2)}{2}}\frac{{(1+t_s)}^{s-1}}{{(1-t_s)}^{2s-1}} \lr{\prod_{k=1}^{s-1}\frac{1}{(2k-1)!}}\det_{1\leq i,j \leq s}(\tilde{A}_{ij}), \\
    \label{Z_inh}
\end{equation}
where the matrix $\tilde A$ only differs from $A$ in the last column, with $\tilde{A}_{is}=t_s^{i-1}-{t_s}^{n+s-i}$. We define $t_s=1+\xi$ and permute the lines of $A$ and $\tilde{A}$, therefore working with
\be
B_{i,j}=A_{s+1-i,j} \quad , \quad \Tilde{B}_{i,j}=\Tilde{A}_{s+1-i,j}.
\ee
Putting \eqref{Z_hom} and \eqref{Z_inh} into \eqref{eq_onept_gen} one gets
\begin{equation}
    h_{n,s}(1+\xi)=2^{1-s}(2s-1)!\,\lr{\frac{1}{\xi}}{\lr{\frac{2+\xi}{{\xi}^2}}}^{s-1}\frac{\det \Tilde{B}}{\det B}
    \label{fct_1pt_det}.
\end{equation}
We have to evaluate the asymptotics of a ratio of determinants that only differ in their last column, a typical situation in the application of the tangent method. In  \cite{colomo2016tangent,colomo2010ac_6V_AF}, the asymptotics of a ratio of determinants is evaluated directly by using random matrix model techniques \cite{zinnjustin20006V_onematrix}. \JFadd{An alternative is to look for} the LU decomposition of $B$ and use techniques of \cite{di2018tangent1} to find a finite size expression for $\frac{\det \Tilde{B}}{\det B}$, whose asymptotics is then straightforward to evaluate. 
%The partially homogeneous limit we just performed seems particularly appropriate when applying the tangent method, even if other formulae for $h_{n,s}(z)$ might also be obtained using the inverse scattering method (also called the algebraic Bethe ansatz). ??

\subsection{LU decomposition}\label{sec:decompositionLU}

The general strategy is that if one can find a \JFadd{lower triangular matrix $L$ with unit diagonal entries} and an upper triangular matrix $U$ such that $B=LU$, then since $\tilde{B}$ differs from $B$ only by its last column we have $\tilde{B}=L\tilde{U}$, with the same matrix $L$. The matrix $\tilde{U}$ differs from $U$ by its last column and can be computed from $\tilde{B}$ using $L^{-1} \tilde{B}=\tilde{U}$. We then obtain 
\begin{equation}
    \frac{\det \Tilde{B}}{\det B}=\frac{\Tilde{U}_{ss}}{U_{ss}},
    \label{ratioU_B}
\end{equation}
with
\begin{equation}
    \Tilde{U}_{ss}=\sum_{p=1}^{s}L^{-1}_{sp}{\Tilde{B}}_{ps}
    =\sum_{p=1}^{s}L^{-1}_{sp}\left[{(1+\xi)}^{s-p}-{(1+\xi)}^{n+p-1}\right].
    \label{U_ss}
\end{equation}
\JFadd{The LU decomposition of $B$ is given below, and proved in Appendix \ref{sec:App_LUdec}.}
\begin{proposition}\label{LUdecomposition}
The matrices $L^{-1}$ and $U$ that appear in the LU decomposition of $B$ have matrix elements
\begin{equation}
    \begin{split}
    &L^{-1}_{ij}={(-1)}^{i+j}\binom{i-1}{j-1}\prod_{l=j}^{i-1}\frac{n-s+i+l}{n-s+l}\quad \text{for }i\geq j,\\
    &U_{ii}=-(i-1)!\prod_{l=i}^{2i-1}(n+l-s),
    \end{split}
    \label{decomp_LU}.
\end{equation}
\end{proposition}
Using \eqref{U_ss} we now have a (very explicit) finite size formula for $h_{n,s}(1+\xi)$. It takes the form of an alternating sum, not well-suited for an asymptotic evaluation by the saddle point method, but can be turned into a non-alternating sum; the technical details of this computation are again gathered in Appendix \ref{sec:App_alternating}. One finds that
\begin{equation}
    \Tilde{U}_{ss}=-{\binom{n-1}{s-1}}^{-1}\sum_{k=2s-1}^{n+s-1}\binom{n+s-1}{k}\binom{k-s}{s-1}{\xi}^k,
\end{equation}
so that \eqref{fct_1pt_det} and \eqref{ratioU_B} yield
\begin{equation}
    h_{n,s}(1+\xi)=C
    \lr{\frac{1}{\xi}} {\lr{\frac{2+\xi}{{\xi}^2}}}^{s-1}
    \sum_{k=2s-1}^{n+s-1}\binom{n+s-1}{k}\binom{k-s}{s-1}{\xi}^k,
    \label{fct_gen_fin}
\end{equation}
with
\begin{equation}
    C=
    \frac{2^{1-s}(2s-1)!
    {\binom{n-1}{s-1}}^{-1}
    }
    {(s-1)!\prod_{l=s}^{2s-1}(n+l-s)
    }.
\end{equation}
The precise value of $C$ does not matter, since it is independent on $\xi$ and disappears in the logarithmic derivative in \eqref{resolvent}. A similar observation shows that the only important input of the LU decomposition is actually $L^{-1}_{sj}$.

\subsection{Asymptotics and arctic curve}\label{sec:eq_CA}
\JFadd{The last step is to compute the asymptotics of (\ref{fct_gen_fin}) and evaluate the following sum}
\begin{equation}
    \sum_{k=2s-1}^{n+s-1}\binom{n+s-1}{k}\binom{k-s}{s-1}{\xi}^k\approx  \int_{2\sigma}^{1+\sigma}\mbox{d}u\,{\rm e}^{nR(\sigma,u,\xi)},
\end{equation}
where
\begin{equation}
    R(\sigma,u,\xi)=\mathcal{L}(1+\sigma)-\mathcal{L}(u)-\mathcal{L}(1+\sigma-u)+\mathcal{L}(u-\sigma)-\mathcal{L}(u-2\sigma)-\mathcal{L}(\sigma)+u\log \xi.
    \label{FR}
\end{equation}
The saddle point equation 
\begin{equation}
    \frac{\partial R}{\partial u}=\log \lr{\frac{\xi(1+\sigma-u)(\sigma-u)}{u(2\sigma-u)}}=0
\end{equation}
yields two solutions
\begin{equation}
    u_{\pm}(\sigma,\xi)=\frac{
    \xi+2\sigma(1+\xi)\pm \sqrt{{\xi}^2+4\sigma^2(1+\xi)}
    }
    {
    2(1+\xi)
    },
\end{equation}
of which we keep the solution in $[2\sigma,1+\sigma]$. We notice that $\xi \in [0,+\infty[$, since $z=1+\xi=\frac{\kappa^*}{\kappa^*-\eta^*}$ and $0<\eta^*<\kappa^*$. Hence, we exclude the solution $u_{-}(\sigma,\xi)<\sigma$. We also check that $u_+(\sigma, \xi)$ spans the whole range $[2\sigma,1+\sigma]$ when $\xi \in [0, \infty[$.

Using \eqref{FR} and $\vertical{\frac{\partial R}{\partial u}}_{u_{+}}=0$, we obtain $\frac{\partial}{\partial\xi}R(\sigma,u_+,\xi)=\frac{u_{+}}{\xi}$. We hence have
\begin{equation}
    r_\sigma(1+\xi)=(1+\xi)\left[\frac{-\sigma(4+\xi)}{\xi(2+\xi)}+\frac{u_{+}}{\xi}\right],
\end{equation}
which upon substitution in \eqref{famille_Dz} specialized at $t=1$ and $\Delta=0$ finally leads to the equation 
\begin{equation}
\begin{split}
    &F(x,y,\xi)\\
    &=x-\frac{2(1+\xi)}{\xi(2+\xi)}y-(1+\xi)\left[
    \frac{-\sigma(4+\xi)}{\xi(2+\xi)}+\frac{
    \xi+2\sigma(1+\xi)+\sqrt{{\xi}^2+4{\sigma}^2(1+\xi)}
    }
    {
    2\xi(1+\xi)
    }
    \right].
    \label{famille_F}
\end{split}
\end{equation}

\begin{figure}
		\centering
		\includegraphics[scale = 1.2, trim = 0 0 0 0]{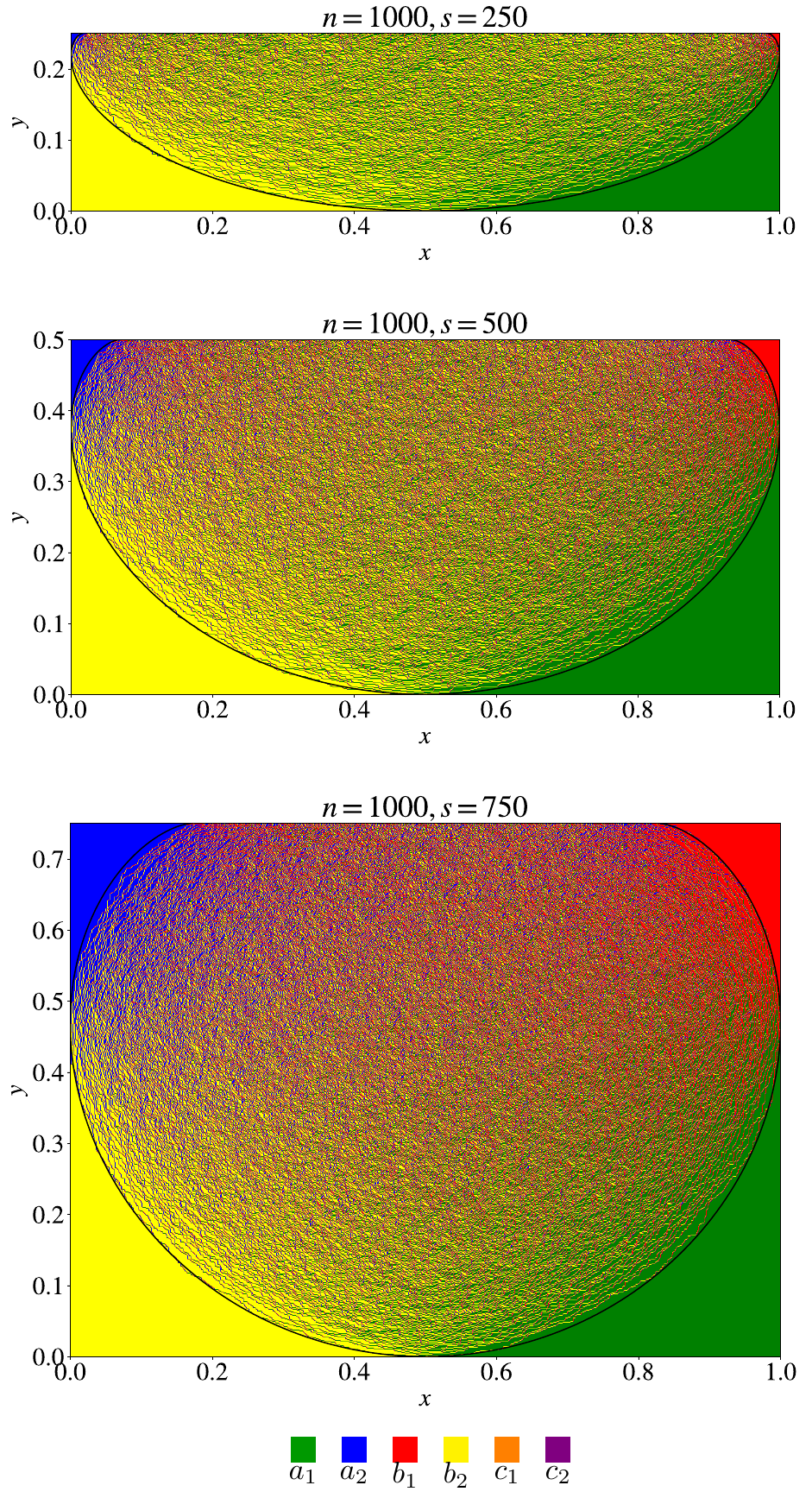}
		\caption{\small Configurations of the $6$V model with pDWBC for $a=b=1$ and  $c=\sqrt{2}$ ($\Delta=0$ and $t=1$) for $n=1000$ and $s=750, 500, 250$ from top to bottom. \JFadd{The colors distinguish the $6$ types of vertices defined in Figure \ref{fig_6V_weights_non_crossing}.} The black curve is the analytic continuation of the arctic curve obtained from the parametrisation \eqref{param_AC}.}
		\label{config_pDWBC_3}
\end{figure}

The arctic curve is the \JFadd{envelope of this family of straight lines and satisfies} $F(x,y,\xi)=0$ and $\frac{\partial}{\partial \xi}F(x,y,\xi)=0$. Solving these equations leads to a parametrization of the arctic curve
\begin{equation}
\begin{aligned}
x(\xi)&=\frac{1}{2}+\frac{
	2{\sigma}^2\xi(\xi+1)+\xi(2+\xi(2+\xi))
}
{
	2(2+\xi(2+\xi))\sqrt{{\xi}^2+4{\sigma}^2(1+\xi)}
},\\
y(\xi)&=\sigma -\frac{
	{\sigma}^2{(2+\xi)}^3
}
{
	2(2+\xi(2+\xi))\sqrt{{\xi}^2+4{\sigma}^2(1+\xi)}
},
\end{aligned}
\label{param_AC}
\end{equation}
where $\xi\in[0,\infty[$. The curve \eqref{param_AC} can be shown to be a degree six algebraic curve, whose exact expression is not very enlightening. \JFadd{As an example, for $\sigma=1/2$, the curve is given by: 
\begin{equation}
\begin{split}
&-16y^6+48y^5-\frac{143}{2} y^4+63y^3 -\frac{8161}{256} y^2 + \frac{2145}{256} y - \frac{225}{256}  \\
&+ x\big(44y^4 -88 y^3+\frac{311}{4} y^2 -\frac{135}{4} y + \frac{373}{64}\big)
+ x^2\big(-44y^4+88y^3-\frac{471}{4}y^2+\frac{295}{4}y-\frac{1157}{64}\big)\\
&+x^3\big(80y^2-80y+\frac{73}{2}\big) + x^4\big(-40y^2+40y-\frac{193}{4}\big)+36x^5-12x^6 = 0.
\end{split}
\end{equation}
}
\rev{The algebraic character of the arctic curve turns out to be a typical property of free-fermion models\cite{KO2,astala2020conformal}. Although our computation is only valid for the south-east portion of the arctic curve, we conjecture that the remaining portions of the arctic curve also satisfy the same algebraic equation.}
Figure \ref{config_pDWBC_3} shows that this conjecture is well supported by numerical simulations. The latter were generated by a Metropolis algorithm similar to that of \cite{lyberg20186V_varietyBC} and exploits parallelization techniques inspired from \cite{keating2018sim_tiling_TW}. A description of the algorithm, adapted to deal with pDWBC, is provided in Appendix \ref{App:simu}. When $\sigma=1$ an arctic circle is recovered from (\ref{famille_F}), which is expected since the boundary conditions then reduce to DWBC  \cite{colomo2008arctic_circle}. Finally, let us mention that the south-west and north-west portions are also recovered from the parametrisation (\ref{param_AC}) but the the north-east portion, although it can be obtained from the north-west portion by symmetry, cannot be recovered from (\ref{param_AC}) (even for complex values of $\xi$)\rev{\footnote{\rev{
The north-east portion can be recovered from the following slightly different parametrization:
\begin{equation}
\begin{aligned}
x(\xi)&=\frac{1}{2}-\frac{
	2{\sigma}^2\xi(\xi+1)+\xi(2+\xi(2+\xi))
}
{
	2(2+\xi(2+\xi))\sqrt{{\xi}^2+4{\sigma}^2(1+\xi)}
},\\
y(\xi)&=\sigma +\frac{
	{\sigma}^2{(2+\xi)}^3
}
{
	2(2+\xi(2+\xi))\sqrt{{\xi}^2+4{\sigma}^2(1+\xi)}
},
\end{aligned}
\label{param_AC_NE}
\end{equation}
which satisfies the same algebraic equation as the other portions of the arctic curve.
}
}}.

 % use \input instead of \include to avoid starting the section in a new page
\usetikzlibrary{shapes.geometric}
\section{Arctic curves of double Aztec rectangles}\label{sec3}

\begin{figure}
	\centering
	\includegraphics[width = \textwidth]{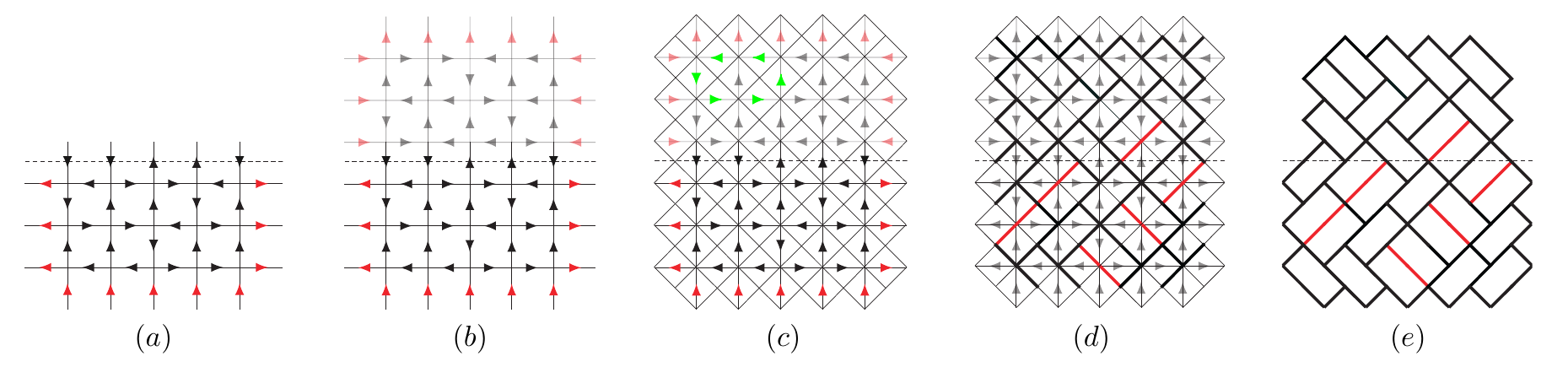}
	\caption{The steps establishing the correspondence between configurations of the six-vertex model with partial DWBC and domino tilings of double Aztec rectangles, illustrated for $m_1=2, m_2=3, n_1=4, n_2=5$ and $k=1$. The first step is obtained by reflecting the $s\times n$ configuration with respect to the dashed line and then reversing all the arrows in the upper copy. The second step is the relaxation of the strict relation between the lower and upper parts: we allow all the arrows of the upper part, except those belonging to the boundary, to be reversed, as long as the ice rule remains satisfied, see green edges in panel (c). The third step shows the one-to-many mapping between $6$V configurations and domino tilings, based in Figure \ref{fig_6V_weights_non_crossing}. Red edges may be rotated by $\frac{\pi}{2}$. In the last step, we remove the forced unit squares.}
	\label{fig:from6VpDWBCtoDAD}
\end{figure}

The real section of the algebraic curve of degree 6 discussed in the previous section covers the entire arctic curve but can in fact be shown to be symmetric under a reflection with respect to the horizontal axis $y=\sigma$. This means that it has a second lobe which is the reflection symmetric of the arctic curve of the 6V with pDWBC at $t=1$ and $\Delta=0$. The whole algebraic curve, with its two symmetric lobes, can be viewed as the arctic curve of a larger but equivalent model, defined on a $2s \times n$ grid, and obtained by reflecting the $s \times  n$ configurations with respect to the dashed line, see Figure \ref{fig:from6VpDWBCtoDAD}, and then reversing all arrows in the upper copy. Although the upper and lower halves are deterministically related to each other, they actually contain different vertices because the two transformations, the reflection and the arrow reversal, exchange the types of vertices ($a_1 \leftrightarrow b_{2}, a_2 \leftrightarrow b_{1}$ and $c_1 \leftrightarrow c_2$). However the frozen and entropic regions of the two halves are reflection symmetric. Relaxing the strict relation between the upper and lower parts defines a new model, which may or may not have the same arctic curve. It is however tempting to conjecture that the new model does have in fact the same arctic curve, namely the algebraic curve found in the previous section for the 6V model and pDWB\rev{C}. In this section, we check that this is indeed the case, and in fact compute the arctic curve for a generalisation of the doubled domain.
	
Using the one-to-many mapping between $6$V configurations and domino tilings \cite{zinnjustin20006V_onematrix} given at Figure \ref{fig_6V_weights_non_crossing},  this new model can in turn be put in correspondence with a particular case of domino tilings of double Aztec rectangles defined in \cite{lai2016double_aztec_rect}. In the notation of \cite{lai2016double_aztec_rect}, this domain corresponds to $k=1, m_1=s-1, n_1=n-1, m_2=s$ and $n_2=n$. This domain is obtained by matching the southeast side of an Aztec rectangle of order $(m_1,n_1)$ with the northwest side of an Aztec rectangle of order $(m_2,n_2)$, with an offset of $k$ unit squares, see Figure \ref{fig:bijectionNILP}. This double Aztec rectangle is denoted by $\mathcal{DR}_{m_1,n_1,k}^{m_2,n_2}$ for which we assume $m_1 \le n_1$ and $m_2 \le n_2$. Although these domains are reminiscent of double Aztec diamonds and skew-Aztec rectangles investigated in \cite{adler2014double,adler2019singular}; they give rise to a completely different arctic phenomenon, as illustrated by the numerical simulations shown in Figure \ref{fig:config_DAR}.

\begin{figure}
	\centering
	\includegraphics[scale=1]{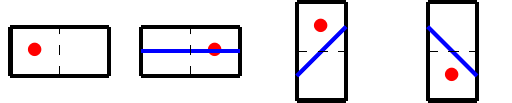}
	\caption{Bijection between dominoes and elementary steps of lattice paths.}
	\label{fig:bij_NILP}
\end{figure}

\begin{figure}
	\centering
	\includegraphics[width=\textwidth]{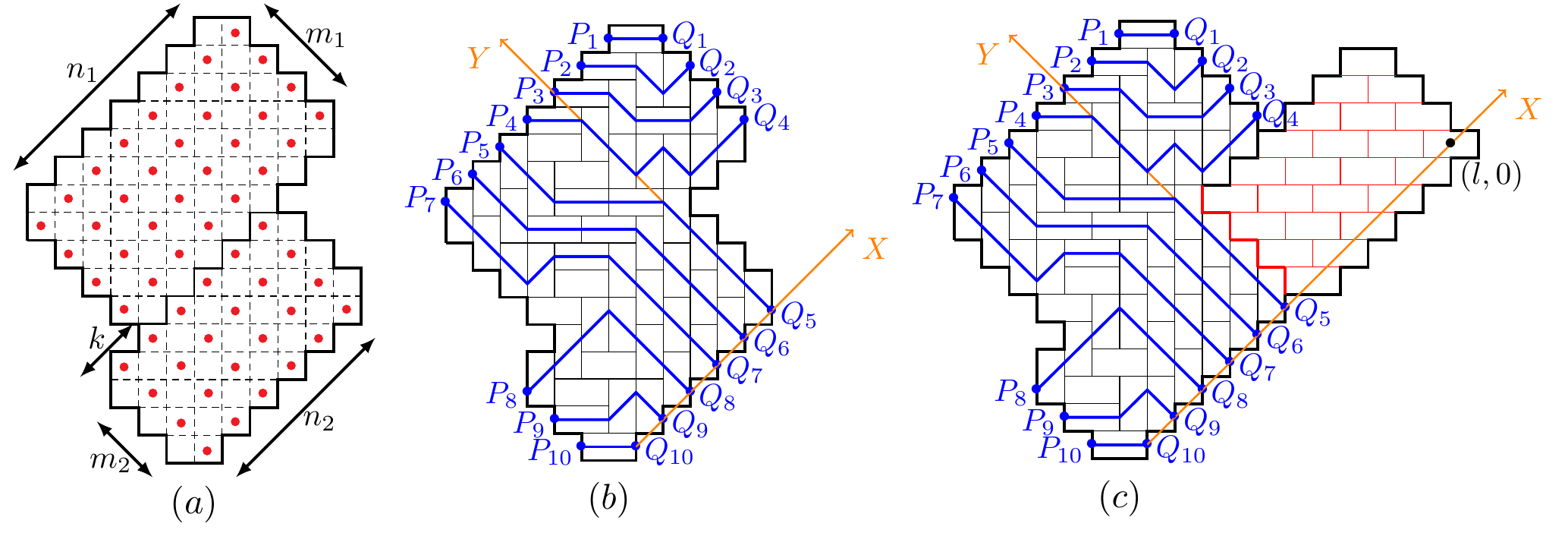}
	\caption{(a) The double Aztec rectangle $\mathcal{DR}_{4,7,2}^{3,6}$ corresponding to $m_1=4, n_1=7, m_2=3, n_2=6$ and $k=2$. (b) A domino tiling of  $\mathcal{DR}_{4,7,2}^{3,6}$ with its equivalent description in terms of non-intersecting lattice paths. (c) The extended double Aztec rectangle $\mathcal{DR}_{4,7,2,6}^{3,6}$. Dominoes in red are necessarily frozen due to the shape of the boundary at the point $(l,0)$.}
	\label{fig:bijectionNILP}
\end{figure}

The aim of the following analysis is to compute the arctic curve of double Aztec rectangles, for general values of the parameters $m_1, n_1, m_2, n_2$ and $k$, with the constraints that $n_1-m_1 = n_2-m_2$ and $k\leq \min (m_2,n_2-1)$. Horizontal and vertical dominoes will be assigned respectively weights $\alpha$ and $\beta$. However since the probability measure associated to the set of tilings only depends on the ratio $\frac{\beta}{\alpha}$, we will without loss of generality consider $\alpha=1$. In the special case discussed above ($k=1, m_1=s-1, n_1=n-1, m_2=s$, $n_2=n$ and $\beta=1$), which came from the 6V model with pDWBC, our results show that the arctic curve is identical whether or not we relax the symmetry between the two halves. 
In order to compute the arctic curve, it will be useful to describe domino tilings in terms of non-intersecting lattice paths. To this end, let us consider a checkerboard coloring of the double Aztec rectangle, as indicated by the red dots, see Figure \ref{fig:bijectionNILP}(a). We use the convention that the right topmost square contains a red dot. A domino being the union of two unit squares, this enables to distinguish four types of dominoes, as depicted in Figure \ref{fig:bij_NILP}, each of them but one being assigned an elementary step, $(2,0)$, $(1,1)$ or $(1,-1)$.
	
Applying this construction produces a set of $n_1+m_2$ paths that may be seen as starting from $\{Q_i\}$ and ending at $\{P_i\}$ for $i=1\cdots,n_1+m_2$, see Figure \ref{fig:bijectionNILP}(b). These paths are in bijection with the underlying tiling. They can neither intersect nor have kissing points because dominoes cannot overlap.

Throughout the following analysis, we will use an orthonormal coordinate system whose origin is the point $Q_{m_1+1}$ and such that a diagonal elementary step is of length one, see Figure \ref{fig:bijectionNILP}$(b)$. 
Let us denote by $Z_{m_1,n_1,k}^{m_2,n_2}(\beta)$ the partition function of the double Aztec rectangle $\mathcal{DR}_{m_1,n_1,k}^{m_2,n_2}$ with $\beta$ the weight attributed to vertical dominoes (without loss of generality, horizontal dominoes are assigned a weight $1$). Let us observe that $Z_{m_1,n_1,k}^{m_2,n_2}(\beta)$ remains unchanged if we replace the bottom Aztec rectangle of size $(m_2,n_2)$ by an extended Aztec rectangle of size $(m_2,n_2+l)$ ($l\in \mathbb{N}_0$) whose lower boundary is slightly modified so that the part outside the original domain  $\mathcal{DR}_{m_1,n_1,k}^{m_2,n_2}$ can only be filled by horizontal dominoes void of paths, see Figure \ref{fig:bijectionNILP}(c). We denote by $\mathcal{DR}_{m_1,n_1,k,l}^{m_2,n_2}$ this extended double Aztec rectangle. Let us stress that after this extension procedure, the boundary of the initial double Aztec rectangle was kept except for the northeast side of the original bottom Aztec rectangle which was deleted and delimits partly the brickwall of the extended double Aztec rectangle (thick red curve).

\subsection{Tangent method}

As for the $6$V model, we will take advantage of the description in terms of non-intersecting lattice paths to compute the arctic curve of the model using the tangent method. Let us first recall that the displacement of one random path is not expected to change the arctic curve of the corresponding model. Moreover we know that an isolated random path travelling between two lattice points converges, in the scaling limit, to the straight line between these two points\cite{debin2019concavity}. When combined, these properties are the set-up of the tangent method and allow in principle to obtain the arctic curve as the envelope of a family of straight lines.  

\begin{figure}
	\centering
	\includegraphics[width=\textwidth]{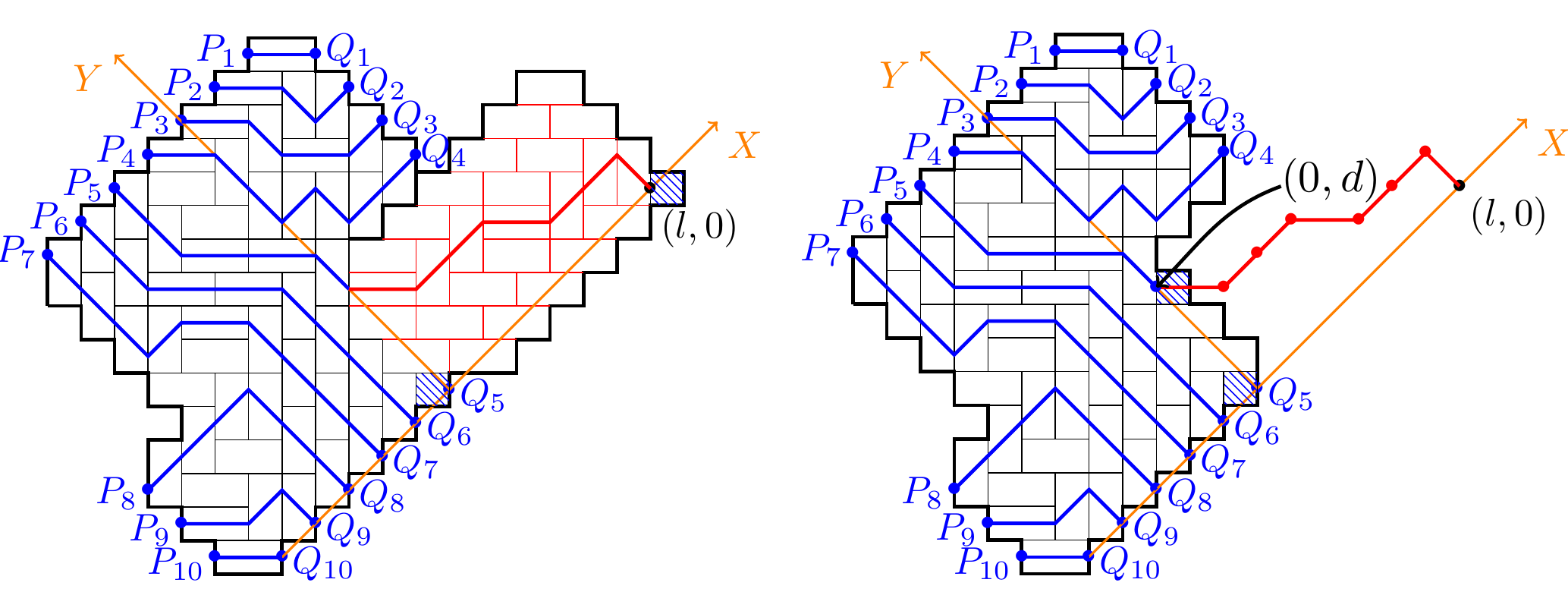}
	\caption{Left: A $1$-refined configuration $Z_{m_1,n_1,k,l}^{m_2,n_2}(\beta)$ corresponding to $m_1=4, n_1=7, m_2=3, n_2=6$, $k=2$ and $l=6$. Right: The $1$-refined configuration $Z_{m_1,n_1,k,l}^{m_2,n_2}(\beta)$ can be decomposed into a path from $(l,0)$ to $(0,d)$ and the $1$-refined configuration $Z_{m_1,n_1,k}^{m_2,n_2,d}(\beta)$ of the double Aztec rectangle (here $d=3$).}
	\label{fig:tangent_method}
\end{figure}

For the case of interest, we consider the $1$-refined partition function $Z_{m_1,n_1,k,l}^{m_2,n_2}(\beta)$ obtained by inserting two monomers centered at $(-1/4,1/4)$ and $(l+1/4,-1/4)$, see Figure \ref{fig:tangent_method}. As a consequence, the starting point of the $(m_1+1)-$th path is now $(l,0)$ instead of $(0,0)$. The displaced path can only enter the double Aztec rectangle by crossing the thick red curve at some point whose coordinates are $(1/2,d-1/2)$ with $1\leq d\leq m_2$, see Figure \ref{fig:tangent_method}. 

\begin{figure}[htb!]
	\centering
	\includegraphics[width=\textwidth]{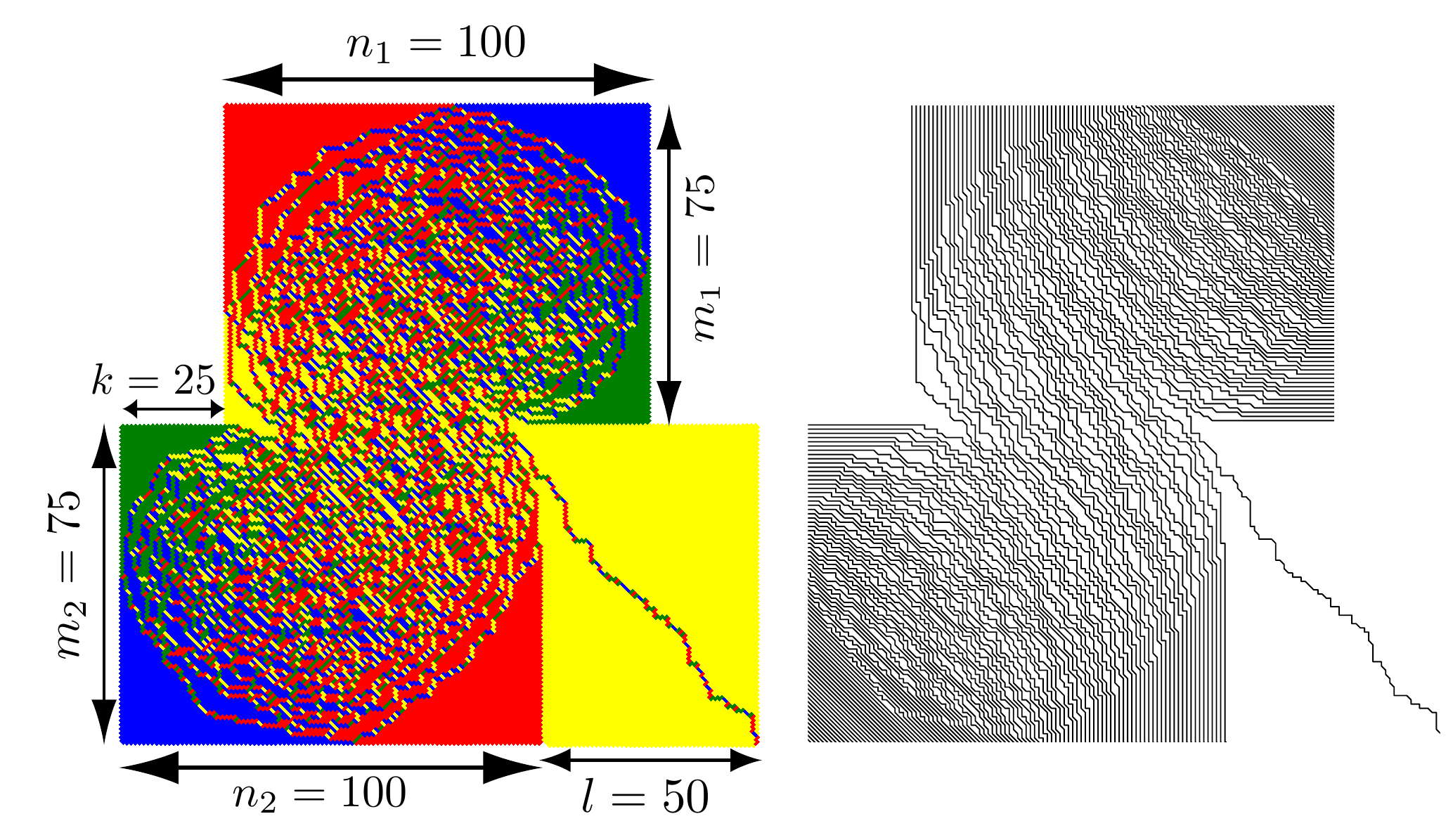}
	\caption{(Left) $1$-refined configuration of the extended double Aztec rectangle $\mathcal{DR}_{m_1,n_1,k,l}^{m_2,n_2}$ and its bijection in terms of non-intersecting lattice paths (Right). Configurations were rotated by $-\frac{\pi}{4}$. The four types of dominoes are distinguished by distinct colors. The displaced path, starting in the original domain at $(l,0)$, becomes, in the scaling limit, a straight line that hits tangentially the arctic curve. For small enough values of $l$, the entry point within the (non rescaled) domain has coordinates $(0,d)$. These configurations were obtained using the Janvresse algorithm which is a generalization of the shuffling algorithm that allows for vanishing weights and hence provides a way to generate tilings on domains that can be embedded in an Aztec diamond \cite{janvresse2006shuffling}.}
	\label{fig:entry_point}
\end{figure}

Let us introduce the following rescaled coordinates: 
	\begin{equation}
	\phi:=\frac{d}{n_2}~,~\lambda=\frac{l}{n_2}~,~ \kappa=\frac{k}{n_2}~,~ \sigma_1=\frac{m_1}{n_2}~,~ \sigma_2=\frac{m_2}{n_2}~,~x=\frac{X}{n_2}~,~y=\frac{Y}{n_2}
	\end{equation}	
In the scaling limit obtained by dividing all the lengths by $n_2$ and sending $n_2\rightarrow + \infty$, the displaced path is a straight line segment that crosses the domain at some point $(0,\phi^*(\lambda))$ and reaches the arctic curve tangentially \cite{debin2019concavity}, see Figure \ref{fig:entry_point}.

By varying $\lambda$, we obtain a family of tangent lines parametrized by $\lambda$ whose equation is given by:
\begin{equation}
F(x,y,\lambda) = y+\frac{\phi^*(\lambda)}{\lambda}x-\phi^*(\lambda).
\label{eq:fam_tan}
\end{equation}
and whose envelope is the desired arctic curve.

\subsection{Refined partition functions}

The most likely entry point $(0,\phi^*)$ will be obtained from the saddle-point analysis, by maximizing over $d$ the following quantity:
\begin{equation}
\frac{Z_{m_1,n_1,k,l}^{m_2,n_2}(\beta)
}{
	Z_{m_1,n_1,k}^{m_2,n_2}(\beta)
} = \sum_{d=0}^{m_2} \frac{
	Z_{m_1,n_1,k}^{m_2,n_2,d}(\beta)
}{
	Z_{m_1,n_1,k}^{m_2,n_2}(\beta)}
Y_{l,d}(\beta).
\label{eq:saddle_point}
\end{equation}

where
\begin{itemize}
	\item $Z_{m_1,n_1,k}^{m_2,n_2,d}(\beta)$ the refined partition function of the double Aztec rectangle with horizontal (resp. vertical) dimers having a weight $1$ (resp. $\beta$) and with two monomers centered at $(-1/4,1/4)$ and $(1/4,d-1/4)$, see Figure \ref{fig:tangent_method}(b). 
\end{itemize}
\begin{itemize}
	\item $Y_{l,d}(\beta)$ the weighted sum over paths between $(l,0)$ and $(0,d)$ using steps $(0,1),(-1,0)$ and $(-1,1)$, in the coordinate system $(X,Y)$, with respective weights $\beta,\beta,1$ and such that the last step is either $(-1,0)$ or $(-1,1)$. Hence, $Y_{l,d}(\beta)= D_{l-1,d-1}(\beta) + \beta\cdot D_{l-1,d}(\beta)$ with $D_{m,n}(\beta)$ given\footnote{The quantity $D_{m,n}(\beta)$ ($m,n\in \mathbb{N}$) gives the weighted enumeration of paths between $(m,-n)$ and $(0,0)$ made up of steps $(0,1),(-1,0)$ and $(-1,1)$ of respective weights $\beta,\beta,1$. In particular $D_{m,n}(1)$ is known as the Delannoy number.} by:
\begin{equation}
D_{m,n}(\beta)=\sum_{p=0}^{\min{(m,n)}} \beta^{m+n-2p}\binom{m+n-p}{p,m-p,n-p}.
\end{equation}
\end{itemize}

The computation will rely on the following result, which is a generalization of the results obtained in \cite{lai2016double_aztec_rect}. For the sake of clarity, the proof ---which is essentially graphical and exploits similar arguments as those presented in \cite{lai2016double_aztec_rect}--- is presented in Appendix \ref{App:DAR}.
\begin{theorem}
	The refined partition function $Z_{m_1,n_1,k}^{m_2,n_2,d}(\beta)$ is given by:
	\begin{equation}
	\begin{split}
	Z_{m_1,n_1,k}^{m_2,n_2,d}(\beta) = &\beta^{(n_1-m_1)(m_1+k)+d}{(1+\beta^2)}^{\binom{m_1+1}{2}+\binom{m_2}{2}}\times\\ &\sum_{r=0}^{d}\binom{m_2-r}{d-r}{(1+\beta^{-2})}^r T\big(\mathcal{H}^{(r)}_{n_1-m_1,m_2-k+1,m_1+k}\big),
	\end{split}
	\label{eq:refinedZ}
	\end{equation}
	where $T\big(\mathcal{H}^{(r)}_{n_1-m_1,m_2-k+1,m_1+k}\big)$ is the refined enumeration of lozenge tilings of a regular hexagon of side lengths $n_1-m_1$, $m_2-k+1$, $m_1+k$, $n_1-m_1$, $m_2-k+1$, $m_1+k$ in clockwise order with the horizontal sides of length $n_1-m_1$ and such that the unique left tile along the southeast side belongs to the $r$-th row, starting from the bottom.
\end{theorem}
Using (\ref{eq:refinedZ}) and the expression of the non-refined partition function of double Aztec rectangles (see \textbf{Theorem 4.1} of \cite{lai2016double_aztec_rect}):
\begin{equation}
Z_{m_1,n_1,k}^{m_2,n_2}(\beta) = \beta^{(n_1-m_1)(m_1+k)}{(1+\beta^2)}^{\binom{m_1+1}{2}+\binom{m_2+1}{2}} T\big(\mathcal{H}_{n_1-m_1,m_2-k+1,m_1+k}\big),
\label{eq:ZDAR}
\end{equation}
with $T\big(\mathcal{H}_{n_1-m_1,m_2-k+1,m_1+k}\big)$ the number of lozenge tilings of a regular hexagon of side lengths $n_1-m_1$, $m_2-k+1$, $m_1+k$, $n_1-m_1$, $m_2-k+1$, $m_1+k$, we deduce:
\begin{equation}
\frac{Z_{m_1,n_1,k}^{m_2,n_2,d}(\beta)}{Z_{m_1,n_1,k}^{m_2,n_2}(\beta)} = {\big(1+\beta^2\big)}^{-m_2}\beta^d \sum_{r=0}^{d}
\frac{
T\big(\mathcal{H}^{(r)}_{n_1-m_1,m_2-k+1,m_1+k}\big)
}{
T\big(\mathcal{H}_{n_1-m_1,m_2-k+1,m_1+k}\big)
}
\binom{m_2-r}{d-r}{\big(1+\beta^{-2}\big)}^{r}.
\end{equation}
We are now ready to extract the likeliest value $\phi^*(\lambda)$.
\subsection{Saddle-point analysis}
Let us now perform the saddle-point analysis of (\ref{eq:saddle_point}). One has:
\begin{equation}
\sum_{d=0}^{m_2} \frac{Z_{m_1,n_1,k}^{m_2,n_2,d}(\beta)}{Z_{m_1,n_1,k}^{m_2,n_2}(\beta)}Y_{l,d}(\beta)
\sim \int_{0}^{\sigma_2} d\phi \int_{0}^{\phi} d\rho \int_{0}^{\min{(\lambda,\phi)}}d\eta\, e^{n_2 S(\phi,\rho,\eta)},
\label{eq:saddle}
\end{equation}
with 
\begin{equation}
\begin{split}
&S(\phi,\rho,\eta)=\\
&-\sigma_2 \log(1+\beta^2) + \phi \log \beta +(\lambda+\phi-2\eta)\log \beta + \rho \log(1+\beta^{-2})+\mathcal{L}(\sigma_2-\rho)\\
 &-\mathcal{L}(\phi-\rho)-\mathcal{L}(\sigma_2-\phi)+\mathcal{L}(\lambda+\phi-\eta)-\mathcal{L}(\eta)-\mathcal{L}(\lambda-\eta)-\mathcal{L}(\phi-\eta)+S_{\sigma_1,\sigma_2,\kappa},
\end{split}
\end{equation}
where
\begin{equation}
\begin{split}
&\phi:=\frac{d}{n_2}~,~ \rho:=\frac{r}{n_2}~,~ \eta:=\frac{p}{n_2}~,~ \lambda=\frac{l}{n_2}~,~ \kappa=\frac{k}{n_2}~,~ \sigma_1=\frac{m_1}{n_2}~,~ \sigma_2=\frac{m_2}{n_2},\\
&\mathcal{L}(x)=x\log x
\end{split}
\end{equation}
and
\begin{equation}
S_{\sigma_1,\sigma_2,\kappa}(\rho) = \lim_{n_2\rightarrow+\infty}\frac{1}{n_2}\log \frac{T\big(\mathcal{H}^{(r)}_{n_1-m_1,m_2-k+1,m_1+k}\big)}{T\big(\mathcal{H}_{n_1-m_1,m_2-k+1,m_1+k}\big)}.
\end{equation}
The saddle-point equations read
\begin{subequations}\label{eq:ptselle}
	\begin{empheq}[left=\empheqlbrace]{align}
	&\partial_{\phi}S\vert_{\phi = \phi^*} = 0 \Leftrightarrow \beta^2(\sigma_2-\phi^*)(\lambda+\phi^*-\eta^*) = (\phi^*-\eta^*)(\phi^*-\rho^*), \label{eq:ptselle1}
	\\
	&\partial_{\rho}S\vert_{\rho = \rho^*} = 0 \Leftrightarrow S'_{\sigma_1,\sigma_2,\kappa}(\rho^*) = -\log\Big(\frac{(1+\beta^{-2})(\phi^*-\rho^*)}{\sigma_2-\rho^*}\Big), \label{eq:ptselle2}
	\\
	&\partial_{\eta}S\vert_{\eta = \eta^*} = 0 \Leftrightarrow (\lambda-\eta^*)(\phi^*-\eta^*) = \beta^2\eta^*(\lambda+\phi^*-\eta^*). \label{eq:ptselle3}
	\end{empheq}
\end{subequations}
From these equations, we wish to obtain $\phi^*(\lambda)$ in order to derive the family of tangent lines given in (\ref{eq:fam_tan}). From (\ref{eq:ptselle1}) and (\ref{eq:ptselle3}), we obtain:
\begin{equation}
\eta^* = \lambda \frac{\sigma_2-\phi^*}{\sigma_2-\rho^*}.
\label{eq:eta}
\end{equation}
The function $S_{\sigma_1,\sigma_2,\kappa}(\rho)$ can be computed explicitly based on the results given in \cite{colomo2016tangent} where it is shown that:
	\begin{equation}
	\frac{T\big(\mathcal{H}^{(r)}_{n_1-m_1,m_2-k+1,m_1+k}\big)}{T\big(\mathcal{H}_{n_1-m_1,m_2-k+1,m_1+k}\big)} = \binom{m_1+m_2-r}{m_2-k}\binom{n_1-m_1+r-1}{n_1-m_1-1}{\binom{n_1+m_2}{m_1+k}}^{-1}.
	\label{eq:ratio_T}
	\end{equation}

We obtain:
\begin{equation}
S_{\sigma_1,\sigma_2,\kappa}(\rho) = \mathcal{L}(\sigma_1+\sigma_2-\rho) - \mathcal{L}(\sigma_2-\kappa) - \mathcal{L}(\sigma_1+\kappa-\rho) + \mathcal{L}(1-\sigma_2+\rho) - \mathcal{L}(1-\sigma_2) - \mathcal{L}(\rho),
\end{equation}
from which we deduce:
\begin{equation}
S_{\sigma_1,\sigma_2,\kappa}'(\rho) = \log \frac{(\sigma_1+\kappa-\rho)(1-\sigma_2+\rho)
}{
\rho(\sigma_1+\sigma_2-\rho)}.
\end{equation}
To extract the likeliest value $\phi^*(\lambda)$, it is convenient to introduce the following new parametrisation:
\begin{equation}
z:=\frac{(1+\beta^{-2})(\phi^*-\rho^*)}{\sigma_2-\rho^*}.
\label{eq:z}
\end{equation}
The domain of definition of the parameter $z$ is included in the interval $[0,1+\beta^{-2}]$ since $0\leq \rho\leq \phi \leq \sigma_2 $. In addition, we must also have $\lambda\geq 0$, which requires $z\geq 1$.

From eq. (\ref{eq:ptselle2}), we obtain a quadratic equation for $\rho^*$ whose solutions are:
\begin{equation}
\begin{split}
\rho_{\pm}^* = &\frac{-\sigma_1-\sigma_2 + z(\kappa+\sigma_1+\sigma_2-1)}{2(z-1)}\\
&\pm \frac{\sqrt{-4z(z-1)(\sigma_2-1)(\sigma_1+\kappa) + 
		{\Big(\sigma_1+\sigma_2+z(1-\sigma_2-\sigma_1-\kappa)\Big)}^2	
	} 
}{
	2(z-1)
}.
\label{eq:rho}
\end{split}
\end{equation}
We must keep $\rho^*_{+}$ since $\rho_{-}^*(z)$ is negative for $z>1$ and thus not acceptable. 

Let us also notice that $\rho_{+}^*(z)$ is well defined when $z=1$ and takes the value $\rho_{+}(1) = \frac{(1-\sigma_2)(\sigma_1+\kappa)}{1-\kappa}$.

Combining (\ref{eq:eta}), (\ref{eq:z}) and (\ref{eq:rho}), we get :
\begin{subequations}\label{eq:rho_lambda}
	\begin{empheq}[left=\empheqlbrace]{align}
	&\phi^* = \frac{z(\sigma_2-\rho_{+}^*(z))}{1+\beta^{-2}} + \rho_{+}^*(z), \label{eq:phi}
	\\
	&\lambda = \frac{\phi^*(z)(\rho_{+}^*(z)-\sigma_2)(\phi^*(z)-\rho_{+}^*(z)+\beta^2(\phi^*(z)-\sigma_2))
	}{
		(\phi^*(z)-\rho_{+}^*(z))(\phi^*(z)-\sigma_2)(1+\beta^2).
	} \label{eq:lambda}
	\end{empheq}
\end{subequations}
Let us notice that $0\leq \rho^*_{+}\leq \phi^*$ as required by the saddle-point analysis. However, we must also ensure that $\rho^*<\sigma_2$, which requires:
\begin{equation}
\frac{\kappa z + \sigma_1 z}{\sigma_1+z} < \sigma_2 < 1
\label{eq:constraint1}
\end{equation}
In addition if $1-\sigma_1 < \kappa \leq 1$, we must also have $z<\min(\frac{\sigma_1}{\sigma_1+\kappa-1},1+\beta^{-2})$.  
Two additional constraints come from the fact that $0\leq \eta^*\leq\lambda$ and $0\leq \eta^*\leq\phi^*$, as imposed by equation (\ref{eq:saddle}). Both constraints are satisfied, as follows from an explicit computation. If some of the aforementioned constraints are not satisfied, it means that the saddle-point equations don't admit a local interior extremum and consequently the arctic curve has no contact point along the southwest boundary $\{(0,y)\}$ with $0<y<\sigma_2$. In this case, the displaced path will cross the domain at the point $(0,\sigma_2)$ and then reach the arctic curve tangentially.

\subsection{Family of tangent lines and arctic curve}

Taking into account (\ref{eq:phi}), (\ref{eq:lambda}) and (\ref{eq:rho}), one obtains the family of tangent lines:
\begin{equation}
\resizebox{\textwidth}{!}{% 
$
\begin{aligned}
&F(x,y,z) = y + \frac{z\big(1-\beta^2(z-1)\big)}{(1+\beta^2)(z-1)} x\\
&+ \frac{\sigma_1 \big(1-\beta^2(z-1)\big) (z-1) - \sigma_2 \big(1+\beta^2(z-1)\big) (z-1)
	+ \big(-1+\beta^2(z-1)\big)(z(\kappa-1)+\Psi_{\sigma_1,\sigma_2,\kappa}(z))
}{
2(1+\beta^2)(z-1)},
\end{aligned}
$
}
\end{equation}
with 
\begin{equation}
\resizebox{\textwidth}{!}{% 
	$
\Psi_{\sigma_1,\sigma_2,\kappa}(z) = \sqrt{
	{(\sigma_1+\sigma_2)}^2-2\big(
	\sigma_1-\sigma_2 + {\sigma_1}^2 + {\sigma_2}^2 + \kappa(2+\sigma_1-\sigma_2)
	\big)z
	+{(1+\kappa+\sigma_1-\sigma_2)}^2z^2
}
$
}
\end{equation}
From this, we deduce the parametric equations of the arctic curve:
\begin{equation}\label{eq:paramAC}
\resizebox{\textwidth}{!}{% 
	$
	\begin{split}
	 x(z) = &\fr{1}{2\lr{ 1+\beta^2{(z-1)}^2) }}\Bigg\{
	 \bigg(\chi(z) \left[\frac{ z (\kappa+\sigma_1-\sigma_2+1)^2- \left(\kappa
		(\sigma_1-\sigma_2+2)+\sigma_1^2+\sigma_1+(\sigma_2-1) \sigma_2\right)}{ \Psi(z)}+\kappa-1\right]\\
	&+ \beta^2 (z-1)(\sigma_1-\sigma_2) \bigg)(z-1)
	+\Psi(z)+(\kappa-1)z
	\Bigg\} \\
	 y(z) = &\fr{1}{2\Psi(z)(1+\beta^2)\lr{ 1+\beta^2{(z-1)}^2) }}\Bigg\{
	{\chi}^2(z)\Big((z-1)({\sigma_1}^2+{\sigma_2}^2)+2\kappa z\\
	& -2\sigma_1\sigma_2+z \sigma_1(\kappa+1)+\sigma_1\Psi(z)\Big)
	+\sigma_2\Bigg[{(1+\beta^2)}^2\Psi(z) -(1+\kappa)\beta^4z^3\\
	&+\beta^2z^2\Big(
	2(1+\beta^2)(\kappa+1) + 2\Psi(z) + \beta^2\Psi(z)
	\Big)
	-(1+\beta^2)z\Big(1+\kappa+\beta^2(1+\kappa)+2\beta^2\Psi(z)\Big)
	\Bigg]
	\Bigg\}
	\end{split}
$
}
\end{equation}

with 
\begin{subequations}\label{eq:Psi_chi}
	\begin{empheq}[left=\empheqlbrace]{align}
	& \Psi(z) = \Psi_{\sigma_1,\sigma_2,\kappa}(z) \\
	& \chi(z) = -1+\beta^2(z-1)
	\end{empheq}
\end{subequations}

%\subsubsection{Special cases}
The arctic curve given in (\ref{eq:paramAC}) depends on four parameters, three of them, namely $\sigma_1, \sigma_2, \kappa$, being related to the size of the domain and the last one, $\beta$, being the ratio between the weights attributed to a vertical domino and a horizontal domino. In this subsection, we give the expression of the arctic curve for the double Aztec rectangle in correspondence with the $6$V model with pDWBC, which is the case when $\sigma_1 = \sigma_2$ and $\kappa=0$, see Figure \ref{fig:from6VpDWBCtoDAD}.
\begin{itemize}
	\item $\kappa = 0, \sigma_1 = \sigma = \sigma_2$\newline
	In this case, we have:
	\begin{equation}
	\Psi(z) = \Psi_{\sigma,\sigma,0}(z) = \sqrt{z^2-4{\sigma}^2(z-1)} 
	\end{equation}
	We thus obtain:
	\begin{equation}\label{eq:param_eq}
		\begin{split}
		& x(z) = -\fr{1}{2}+\fr{2\sigma^2(1-z)+z+\beta^2{(z-1)}^2(z-2\sigma^2)}{2\sqrt{z^2-4{\sigma}^2(z-1)}\big(1+\beta^2{(z-1)}^2\big)}\\
		& y(z) = \sigma + \fr{{\sigma}^2(z-2){\big(1-\beta^2(z-1)\big)}^2
		}
		{
			(1+\beta^2)\big(1+\beta^2{(z-1)}^2\big)\sqrt{z^2-4{\sigma}^2(z-1)}}
		\end{split}
	\end{equation}
	\item $\kappa = 0, \sigma_1 = \sigma = \sigma_2, \beta=1$\newline
	This is a particular case of the previous point, obtained when horizontal and vertical dominoes have the same weight. As stated in the beginning of this section, we expect, in this case, the arctic curve to be identical to the one of the $6$V model with pDWBC and symmetric weights $a=1=b$ and $c=\sqrt{2}$. Making the change of variables $x\rightarrow -x$ and $z\rightarrow -\xi$, we obtain:
	\begin{equation}
	\begin{split}
	x(\xi)&=\frac{1}{2}+\frac{
		2{\sigma}^2\xi(\xi+1)+\xi(2+\xi(2+\xi))
	}
	{
		2(2+\xi(2+\xi))\sqrt{{\xi}^2+4{\sigma}^2(1+\xi)}
	},\\
	y(\xi)&={\sigma}-\frac{
		{\sigma}^2{(2+\xi)}^3
	}
	{
		2(2+\xi(2+\xi))\sqrt{{\xi}^2+4{\sigma}^2(1+\xi)}
	},
	\end{split}
	\end{equation} 
	which is exactly the equation obtained for the $6$V model with pDWBC at $\Delta=0$ and $t=1$, see (\ref{param_AC}). 
	
	\item $\kappa = 0, \sigma_1 = 1 = \sigma_2$\newline
	This is a special case for which $\Psi(z)=\vert z-2 \vert$:
	Let us recall that $1\leq z \leq 1+\beta^{-2}$. We thus obtain, for $1 \leq z \leq 2$:
	\begin{equation}\label{eq:param_eq1}
		\begin{split}
		& x(z) = -\fr{\beta^2{(z-1)}^2
		}{1+\beta^2{(z-1)}^2}\\
		& y(z) = \fr{\beta^2 z^2}
		{(1+\beta^2)\big(1+\beta^2{(z-1)}^2\big)}
		\end{split}
	\end{equation}
	while for $z>2$, we have:
	\begin{equation}\label{eq:param_eq2}
		\begin{split}
		& x(z) = -\fr{1}{
			1+\beta^2{(z-1)}^2}\\
		& y(z) =  \fr{2+\beta^2{(z-2)}^2+2\beta^4{(z-1)}^2
		}
		{(1+\beta^2)\big(1+\beta^2{(z-1)}^2\big)}
		\end{split}
	\end{equation}
	The analytic continuation is an ellipse whose algebraic equation is given by:
	\begin{equation}
	\beta^4{(1+x-y)}^2 + {(x+y)}^2+2\beta^2\Big(x+x^2+y^2-y\Big)=0
	\end{equation}
	for $0 \leq z \leq 2$ and 
	\begin{equation}
	\beta^4{(2+x-y)}^2 + {(-1+x+y)}^2+2\beta^2\Big(2+x+x^2+y^2-3y\Big)=0
	\end{equation}
	for $z>2$. 
	When $\beta = 1$, we recover two Arctic circles, centered respectively at $(-\fr{1}{2},\fr{1}{2})$ and $(-\fr{1}{2},\fr{3}{2})$.
\end{itemize}
 
\begin{figure}[h!]
	\centering
	\includegraphics[scale = 0.7, trim = 0 0 0 0]{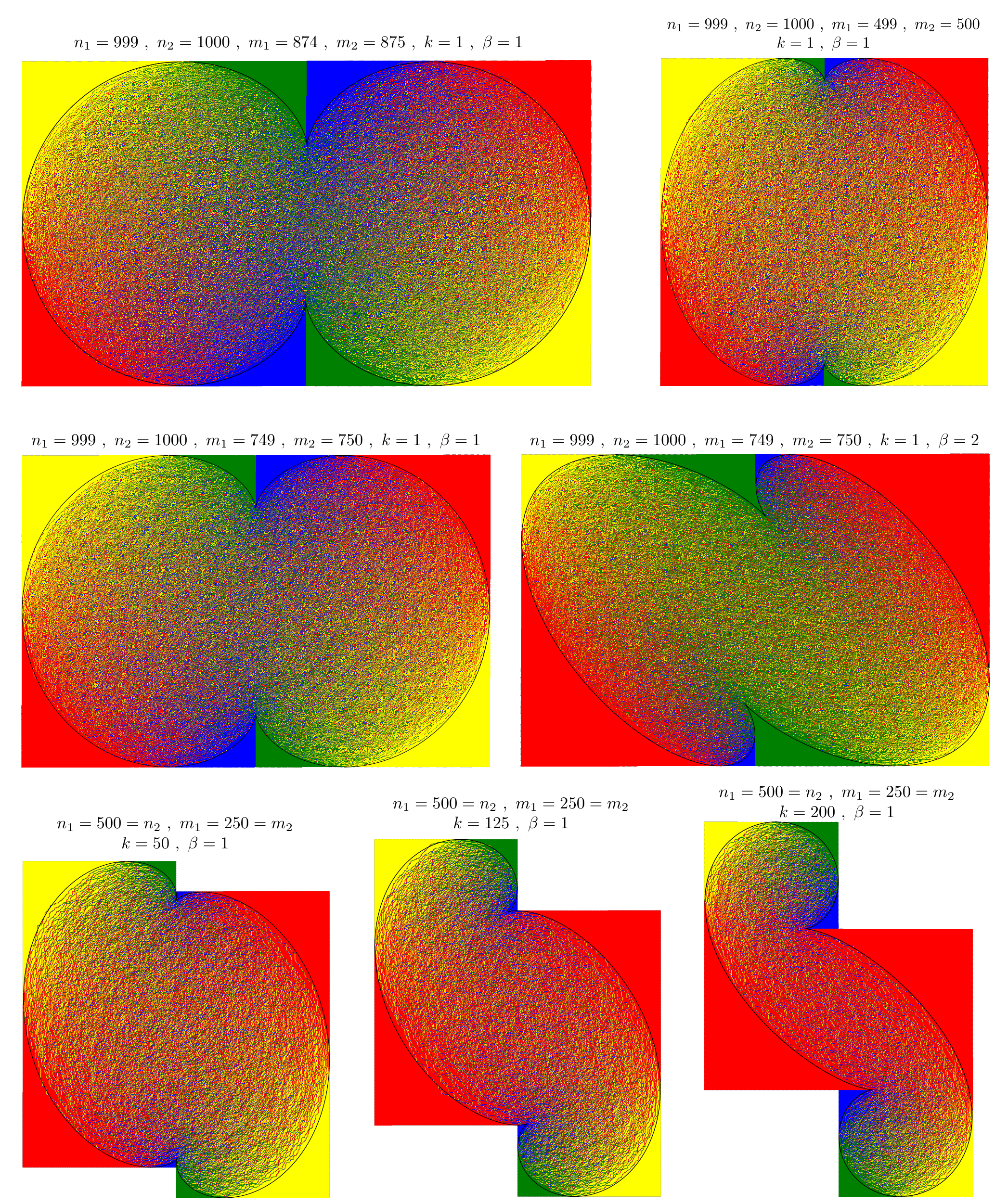}
	\caption{\small Configurations of the double Aztec rectangle model for several values of the parameters. Configurations were generated using the Janvresse algorithm \cite{janvresse2006shuffling}. The black curve was obtained from eq. (\ref{eq:paramAC}). }
	\label{fig:config_DAR}
\end{figure}

Figure \ref{fig:config_DAR} shows a few configurations generated using the Janvresse algorithm. Let us recall that the above computations give a portion of the arctic curve, the one for which the parameter $z$ satisfies the constraints mentioned previously, namely:
\begin{equation}
\frac{\kappa z + \sigma_1 z}{\sigma_1 + z} < \sigma_2, 
\label{eq:constraint}
\end{equation}
$ \forall z \in [1,1+\beta^{-2}]$ if $0 \leq \kappa \leq 1-\sigma_1$ and $ \forall z \in [1,\min (1+\beta^{-2}, \frac{\sigma_1}{\sigma_1 + \kappa -1}) ]$ if $\kappa > 1-\sigma_1$.
Since the l.h.s. of eq. (\ref{eq:constraint}) increases with $z$ ($z>0$), the arctic curve will have a contact point $(0,\phi^*)$ if $\sigma_2\geq \frac{\sigma_1 + \kappa}{1+\sigma_1}$. In particular, for $\sigma_1 = \sigma_2 \equiv \sigma $, this condition is met when $\sigma^2>\kappa$. The three bottom configurations shown in Figure \ref{fig:config_DAR} and corresponding to $\sigma = 1/2$ corroborate this fact. In this case, the arctic curve has a contact point $(0,\phi^*)$ provided $\kappa < 1/4$. Nevertheless, Figure \ref{fig:config_DAR} suggests that the parametric equations (\ref{eq:paramAC}) remain valid $\forall z \in \mathbb{R}$. Indeed, the black curve shown in Figure \ref{fig:config_DAR} was obtained from eq. (\ref{eq:paramAC}), relaxing the constraints on $z$. As for the $6$V model with pDWBC, not all the portions of the arctic curve can be recovered from eq. (\ref{eq:paramAC}); we took advantage of the symmetry of the domain to infer the remaining portions. 

\section{Conclusion}

Using the tangent method, we have derived the analytic expression of the arctic curve of the six-vertex ($6$V) model with partial domain wall boundary conditions (pDWBC) and for particular values of the weights ($a=1,b=1$ and $c=\sqrt{2}$). The computation relied on a LU decomposition of the determinant involved in the partition function and on techniques to deal with alternating sums of products of binomial coefficients. Large configurations generated with a Metropolis Monte-Carlo algorithm for several sizes of the domain are in very good agreement with the predicted arctic curve. \newline 
  
For the particular weights considered, configurations of the $6$V model with pDWBC are related to domino tilings of double Aztec rectangles, in the same way as the $6$V model with DWBC is connected to the Aztec diamond. We also computed the arctic curve of double Aztec rectangles using the tangent method. The computation was done for generic sizes of the domain and was confirmed by means of configurations generated with a generalized version of the shuffling algorithm. For particular sizes of the double Aztec rectangle, the arctic curve was found to be identical to that of the $6$V model with pDWBC and weights $a=1,b=1$ and $c=\sqrt{2}$. 

\JFadd{We conclude by giving a few research perspectives we think are promising. First, it should be possible to generalise the computation of the arctic curve of double Aztec rectangles to the case where the measure keeps track of the area under the paths. In this case, the arctic curve would be recovered as the envelope of a family of tangent geodesics\cite{di2019tangent3,di2019tangent4}. Secondly, it would be of clear interest to extend the computation of the arctic curve of the $6$V model with pDWBC to generic values of the weights (in the disordered regime for which an arctic phenomenon is found). Thirdly, it would be interesting to investigate the statistics of domino tilings of double Aztec rectangles as it was done for double Aztec diamonds and skew-Aztec rectangles\cite{adler2014double,adler2019singular}.} 

\section*{Acknowledgments}
BD and JFDK contributed equally to this work. The authors thank Timoteo Carletti \rev{and Filippo Colomo} for useful discussion\rev{s}. BD acknowledges the financial support of the Fonds Spéciaux de Recherche (FSR) of the Université catholique de Louvain, the Fonds de la Recherche Scientifique (FNRS) and the Fonds Wetenschappelijk Onderzoek—Vlaanderen (FWO) under EOS project no 30889451. JFDK is supported by a FNRS Aspirant Fellowship under the Grant FC38477. PR is a Senior Research Associate of FRS-FNRS. Part of the results were obtained using the computational resources provided by the ``Consortium des Equipements de Calcul Intensif'' (CECI), funded by the Fonds de la Recherche Scientifique de Belgique (F.R.S.-FNRS) under Grant No. 2.5020.11 and by the Walloon Region.
\appendix
\section{Technicalities of the 6V model with pDWBC \label{app_6VpDWBC}} 

In this appendix, we gather some technicalities in the computation of the arctic curve of the $6$V model with pDWBC at $\Delta=0$ and $t=1$. The core issue is the evaluation of sums involving factorials. There are several ways to deal with these sums. We will use generating functions and make use of the following lemmas, in the spirit of what is done in \cite{di2018tangent1}.

\begin{lemma} \label{lemma_sum_gen_fct} Let $f(x)$ and $g(x)$ be two generating functions. \JFadd{We denote by $f(x)|_{x^k} \equiv f_k$ the coefficient of $x^k$ in $f(x)$ and similarly for g(x).} We have
	\begin{equation}
	f(x)g(x)\big| _{x^k}=\sum_{l=0}^kf(x)\big| _{x^l} \, g(x)\big| _{x^{k-l}}.
	\label{eq_sum_gen_fct}
	\end{equation}
\end{lemma}
\begin{lemma} \label{lemma_gen_fct_binom}
	The binomial coefficient can be represented in four different ways by generating series or polynomial
	\begin{equation}
	\binom{n}{k}={(1+x)}^n\big| _{x^k}={(1+x)}^n\big| _{x^{n-k}} =\left.\frac{1}{{(1-x)}^{k+1}}\right\vert _{x^{n-k}} =\left.\frac{1}{{(1-x)}^{n-k+1}}\right\vert _{x^k}.\\
	\label{eq_gen_fct_bino}
	\end{equation}
\end{lemma}

\subsection{Homogeneous limit}\label{sec:App_Homogeneous}

In \cite{pronko2019pronko_pronko}, they show that in the free-fermion case ($\Delta=0$), the partition function of the vertically inhomogeneous $6$V model with pDWBC is
\begin{equation}
Z_{n,s}(t_1,\cdots,t_s)=\prod_{i=1}^s\frac{c_i}{1-t_i}\prod_{1\leq i < j \leq s}\frac{1+t_i t_j}{(1-t_it_j)(t_j-t_i)}\det_{1\leq i,j \leq s}\left[f_i(t_j)\right],
\label{det_Pronko_app}
\end{equation}
where the inhomogeneous weights are $a_j=1$, $t_j=b_j$ and $c_j=\sqrt{1+t_j^2}$ and 
\begin{equation}
f_i(t_j)=t_j^{i-1}-{t_j}^{n+s-i}.
\end{equation} 
Let us show how to take the homogeneous limit $t_i \to t=1$. When we successively take the limits $t_1 \to 1$, $t_2 \to 1$, $t_3 \to 1$, \textit{etc.} the prefactor develops poles of odd orders $\frac{1}{1-t_1}$, $\frac{1}{(1-t_2)^3}$, $\frac{1}{(1-t_3)^5}$, \textit{etc.}. We will show that these poles are exactly matched by zeros of equal orders, coming from the dominant contribution in the determinant once the corresponding column is Taylor expanded
\begin{equation}
f_i(t_j)=f_i(1)+f_i^{(1)}(1)(t_j-1)+\frac{f_i^{(2)}(1)}{2!}(t_j-1)^2+\cdots,
\end{equation}
where
\begin{equation}
f_i^{(m)}(1)=\frac{(i-1)!}{(i-1-m)!}-\frac{(n+s-i)!}{(n+s-i-m)!}.
\end{equation}
The (unusual) cancellation of  zeros of even order stems from the identity
\begin{equation}
\frac{f_i^{(m)}(1)}{m!} =\frac12 \sum_{k=1}^{m-1} (-1)^{k+1}{n+s-m-1-k \choose k} \frac{f_i^{(m-k)}(1)}{(m-k)!}
\label{eq:fm_even_linearcomb}
\end{equation}
that holds when $m$ is even. This relation follows from 
\begin{equation}
\begin{aligned}
&\sum_{k=0}^m (-1)^{k+1}{n+s-m-1-k \choose k}\left[{i-1 \choose m-k} - {n+s-i \choose m-k}\right]\\
&=\sum_{k=0}^m \left. \frac{1}{(1+x)^{n+s-m}}\right|_{x^k} \left.\left[ (1+x)^{n+s-i}-(1+x)^{i-1}\right] \right|_{x^{m-k}}\\
&= \left. \left[ \frac{1}{(1+x)^{i-m}}-\frac{1}{(1+x)^{n+s-m-i+1}}\right]\right|_{x^{m}}\\
&= (-1)^m\left\{{i-1 \choose m} -{n+s-i \choose m} \right\},
\end{aligned}
\end{equation}
that is valid for any $m$. When $m$ is even, the relation can be used to get \eqref{eq:fm_even_linearcomb}.

Identity \eqref{eq:fm_even_linearcomb} states that any $f^{(m)}_i(1)$ with even $m$ is a linear combination (with coefficient that do not depend on $i$) of $f^{(\ell)}_i(1)$ with $\ell<m$ odd. When $m$ is odd however, since $f^{(m)}_i(1)$ is a polynomial in $i$ of order $m$, it cannot be written as a linear combination of $f^{(\ell)}_i(1)$ with $\ell<m$. Using these properties, it is now easy to evaluate the homogeneous limit and one gets 
\begin{equation}
Z_{n,s}(1,\cdots,1)= 
-2^{s^2/2}{\prod_{k=1}^s\frac{1}{(2k-1)!}}
\det_{1\leq i,j \leq s}\left[
f_i^{(2j-1)}(1)
\right].
\label{eq:Z_hom}
\end{equation}

\subsection{Proof of the LU decomposition}\label{sec:App_LUdec}
The matrix we want to decompose has matrix elements ($s \le n$ are positive integers)
\be
B_{ij} = \frac{(s-i)!}{(s+1-i-2j)!} - \frac{(n+i-1)!}{(n+i-2j)!}, \qquad 1 \le i,j \le s.
\ee
The lower triangular matrix $L$ has been conjectured as well as its inverse, which is given by
\be
L^{-1}_{ij} = (-1)^{i+j} \, {i-1 \choose j-1} \, \frac{(n-s+2i-1)! \, (n-s+j-1)!}{(n-s+i+j-1)! \, (n-s+i-1)!}.
\ee
The proof of the conjecture reduces to show that $U \equiv L^{-1} B$ is upper triangular.
Using elementary operations, the elements of $U$ can be expressed as:
\begin{equation}
U_{k,l} = (2l-1)!{n-s+k-1 \choose n-s}^{-1}(C_{k,l}-D_{k,l})
\label{eq:U_kl},
\end{equation}
with
\begin{equation}
C_{k,l} = \sum_{j=0}^{k-1}{(-1)}^{k+j+1}{n+2k-s-1 \choose k-j-1}{n-s+j \choose n-s}{s-j-1 \choose 2l-1}
\label{eq:Ckl}
\end{equation}
and 
\begin{equation}
D_{k,l} = \sum_{j=0}^{k-1}{(-1)}^{k+j+1}{n+2k-s-1 \choose k-j-1}{n-s+j \choose n-s}{n+j \choose 2l-1}.
\end{equation}
We wish to show that $C_{k,l} = D_{k,l}$ for $k>l$. To this end, we will use the following representations of binomial coefficients:
\begin{equation}
\binom{n}{k}={(1+x)}^n\big| _{x^k}={(1+x)}^n\big| _{x^{n-k}} =\left.\frac{1}{{(1-x)}^{k+1}}\right\vert _{x^{n-k}} =\left.\frac{1}{{(1-x)}^{n-k+1}}\right\vert _{x^k}.\\
\label{eq_gen_fct_bino2}
\end{equation}
Let us first treat the term $C_{k,l}$. Using the above representations, we have:
\begin{equation}
\begin{aligned}
&{n+2k-s-1 \choose k-j-1} 
= \left. {{(1+x)}^{n-s+2k-1}}\right\vert _{x^{k-j-1}} 
= \left. \frac{{(1+x)}^{n-s+2k-1}}{x^{k-1}}\right\vert _{x^{-j}},\\
&{(-1)}^{j} \cdot {n-s+j \choose j} = \left. \frac{1}{{(1+y)}^{n-s+1}}\right\vert_{y^j},\\
&{s-j-1 \choose 2l-1} = \left. \frac{1}{{(1-z)}^{2l}} \right\vert_{z^{s-j-2l}} = \left. \frac{1}{{z^{s-2l}(1-z)}^{2l}} \right\vert_{z^{-j}}.
\end{aligned}
\end{equation}
Inserting the latter relations into equation (\ref{eq:Ckl}) leads to
\begin{equation}
\begin{aligned}
C_{k,l} &= {(-1)}^{k+1} \sum_{j=0}^{k-1} \left.
\frac{{(1+x)}^{n-s+2k-1}}{x^{k-1}}
\frac{1}{{(1+y)}^{n-s+1}}
\frac{1}{{z^{s-2l}(1-z)}^{2l}}
\right \vert_{{\big(\frac{y}{xz}\big)}^j}\\
&={(-1)}^{k+1} \sum_{j=0}^{k-1} {\big(\frac{xz}{y}\big)}^j
\left.
\frac{{(1+x)}^{n-s+2k-1}}{x^{k-1}}
\frac{1}{{(1+y)}^{n-s+1}}
\frac{1}{{z^{s-2l}(1-z)}^{2l}}
\right \vert_{{x^0y^0z^0}}\\
&={(-1)}^{k+1} \frac{1-{\big(\frac{xz}{y}\big)}^k}{1-\frac{xz}{y}}
\left.
\frac{{(1+x)}^{n-s+2k-1}}{x^{k-1}}
\frac{1}{{(1+y)}^{n-s+1}}
\frac{1}{{z^{s-2l}(1-z)}^{2l}}
\right \vert_{{x^0y^0z^0}}.
\end{aligned}
\end{equation}
Using the Cauchy theorem, the previous resut can be expressed as:
\begin{equation}
C_{k,l}
= {(-1)}^{k+1} \oint\limits \frac{1}{{(2\pi i)}^3} \frac{\mathrm{d}x\, \mathrm{d}y\, \mathrm{d}z}{ x\,y\,z}
\frac{1-{\big(\frac{xz}{y}\big)}^k}{1-\frac{xz}{y}}
\frac{{(1+x)}^{n-s+2k-1}}{x^{k-1}}
\frac{1}{{(1+y)}^{n-s+1}}
\frac{1}{{z^{s-2l}(1-z)}^{2l}},
\end{equation}

where the contours are positively oriented circles centered at the origin and of radius strictly smaller than $1$ so that we only pick up the residues at the origin. 
The term ${\big(\frac{xz}{y}\big)}^k$ brings no contribution since the corresponding residue at $x=0$ vanishes. Hence, we have:
\begin{equation}
C_{k,l}
= {(-1)}^{k+1} \oint\limits \frac{1}{{(2\pi i)}^3}\frac{\mathrm{d}x\, \mathrm{d}y\, \mathrm{d}z}{ x\,y\,z}
\frac{1}{1-\frac{xz}{y}}
\frac{{(1+x)}^{n-s+2k-1}}{x^{k-1}}
\frac{1}{{(1+y)}^{n-s+1}}
\frac{1}{{z^{s-2l}(1-z)}^{2l}}.
\end{equation}
Expressing the residue at $x=0$ in terms of that of $y/z$ leads to:
\begin{equation}
\begin{aligned}
C_{k,l}(y\rightarrow 0,z\rightarrow 0) &= {(-1)}^{k+1} \oint\limits \frac{1}{{(2\pi i)}^2} \frac{\mathrm{d}y\, \mathrm{d}z}{ y\,z}
\frac{{(1+y/z)}^{n-s+2k-1}}{{(y/z)}^{k-1}}
\frac{1}{{(1+y)}^{n-s+1}}
\frac{1}{{z^{s-2l}(1-z)}^{2l}}\\
&={(-1)}^{k+1} \oint\limits \frac{1}{{(2\pi i)}^2} \frac{\mathrm{d}y\, \mathrm{d}z}{ y\,z}
\frac{{(y+z)}^{n-s+2k-1}
}{
z^{n+k-2l}{(1-z)}^{2l}y^{k-1}{(1+y)}^{n-s+1}
},
\end{aligned}
\end{equation}
where the notation $C_{k,l}(z\rightarrow a)$ indicates that the we extract the residue at $z=a$. By Cauchy's theorem, $C_{k,l}(z\rightarrow 0) + C_{k,l}(z\rightarrow \infty) +C_{k,l}(z\rightarrow 1)=0$. Let us now observe that the residue at $z=\infty$ vanishes. To see this, we make the change\footnote{The computation of the residue at $z=\infty$ requires to consider a \textit{clockwise}-oriented contour whose interior englobes all the other residues. The change of variables $z=1/w$ transforms this contour into a \textit{counterclockwise} oriented contour the origin, excluding all the other poles.} of variables $\omega=1/z$:
\begin{equation}
\begin{aligned}
C_{k,l}(z\rightarrow \infty)&=-{(-1)}^{k+1} \oint\limits \frac{1}{{(2\pi i)}^2} \frac{\mathrm{d}y\, \mathrm{d}\omega}{ y \omega}
\frac{{(y+1/\omega)}^{n-s+2k-1}
}{
	{(1/\omega)}^{n+k-2l}{(1-1/\omega)}^{2l}y^{k-1}{(1+y)}^{n-s+1}
}\\
&={(-1)}^{k} \oint\limits \frac{1}{{(2\pi i)}^2} \frac{\mathrm{d}y\, \mathrm{d}\omega}{ y\,\omega}
\frac{{(1+y\omega)}^{n-s+2k-1}{\omega}^{s+1-k}
}{
	{(1-\omega)}^{2l}y^{k-1}{(1+y)}^{n-s+1}
}\\
&=0,
\end{aligned}
\end{equation}
the last equality being obtained since the expansion of the integrand around $\omega = 0$ is a polynomial of degree non-negative since $s+1>k$.
Hence, we deduce that $C_{k,l}(y\rightarrow 0, z\rightarrow 0) = -C_{k,l}(y\rightarrow 0, z\rightarrow 1)$.

Making the change of variables $z=1-u$ leads to:
\begin{equation}
C_{k,l}(u\rightarrow 0, y\rightarrow 0) = {(-1)}^{k+1} \oint\limits \frac{1}{{(2\pi i)}^2} \frac{\mathrm{d}y\, \mathrm{d}u}{ y^{k}\,u^{2l}}
\frac{{(1+y-u)}^{n-s+2k-1}
}{
	{(1-u)}^{n+k-2l+1}{(1+y)}^{n-s+1}
},
\end{equation}
where now the two contours are both positively oriented circles around $0$ and of radius smaller than $1$. Let us now make the change of variables\footnote{Under this change of variables, we have $\mathrm{d}y \mathrm{d}u = \frac{1-u}{{(1-v)}^2} \mathrm{d}u \mathrm{d}v$.} $y = \frac{(1-u)v}{1-v}$. This brings us to:
\begin{equation}
C_{k,l}(u\rightarrow 0, v\rightarrow 0) = {(-1)}^{k+1} \oint\limits \frac{1}{{(2\pi i)}^2} \frac{\mathrm{d}v\, \mathrm{d}u}{ v^{k}\,u^{2l}}
\frac{1
}{
	{(1-u)}^{s+1-2l}{(1-v)}^{k}{(1-uv)}^{n-s+1}
}.
\end{equation}
We can choose suitable contours for $u$ and $y$ to guarantee that the contour for $v$ is centered around $0$ and of radius smaller than $1$. 

We deduce:
\begin{equation}
C_{k,l}= {(-1)}^{k+1}
\left.
\frac{1}{{(1-u)}^{s+1-2l}{(1-v)}^{k}{(1-uv)}^{n-s+1}}
\right\vert_{v^{k-1}u^{2l-1}}.
\end{equation}
Let us investigate the powers in the denominator. We first notice that $n-s+1\geq 1$ and $k\geq 1$. If $s+1-2l<1$, then the coefficient $C_{k,l}$ is exactly $0$ since the last binomial coefficient appearing in (\ref{eq:Ckl}) vanishes (because $s-j-1<2l-1, \forall j=1\cdots k-1$ in this case). For $s-2l\geq 0$, using the fact that:
\begin{equation}
\frac{1}{{(1-x)}^{q+1}} = \sum_{j\geq 0} {q+j \choose j}x^j \quad \forall q\in\mathbb{N}_0,
\end{equation}
we obtain:
\begin{equation}
\begin{aligned}
C_{k,l} & = {(-1)}^{k+1}
\left.
\sum_{i,j,m\geq 0} 
{k-1+j \choose j}
{n-s+i \choose i}
{s-2l+m \choose m}
u^{i+m}v^{i+j}
\right\vert_{v^{k-1}u^{2l-1}}\\
&={(-1)}^{k+1}\sum_{i=0}^{\min (k-1,2l-1)}
{n-s+i \choose n-s}
{2k-2-i \choose k-1}
{s-1-i \choose 2l-1-i}.
\end{aligned}
\label{eq:Ckl_non_alternant}
\end{equation}
In the latter expression, the upper bound guarantees that the binomial coefficients have nonnegative integer entries and obey the definition\footnote{Throughout this analysis, we will always consider this definition.}:
\begin{equation}
{n \choose k} = 
\left\{
\begin{aligned}
&\frac{n!}{k!(n-k)!} &\text{ for } 0\leq k\leq n,\\
&0 &\text{otherwise}.
\end{aligned}
\right.
\end{equation}  
We observe that $C_{k,l}=0$ as soon as $s-2l<0$. 

Let us now proceed similarly for $D_{k,l}$. Using the fact that:
\begin{equation}
{n+j \choose 2l-1} = \left. \frac{1}{z^{n-2l+1}{(1-z)}^{2l}} \right\vert_{z^j},
\end{equation}
we obtain:
\begin{equation}
\begin{aligned}
D_{k,l} &= {(-1)}^{k+1} \sum_{j=0}^{k-1} \left.
\frac{{(1+x)}^{n-s+2k-1}}{x^{k-1}}
\frac{1}{{(1+y)}^{n-s+1}}
\frac{1}{{z^{n+1-2l}(1-z)}^{2l}}
\right \vert_{{\big(\frac{yz}{x}\big)}^j}\\
&={(-1)}^{k+1} \frac{1-{\big(\frac{x}{yz}\big)}^k}{1-\frac{x}{yz}}
\left.
\frac{{(1+x)}^{n-s+2k-1}}{x^{k-1}}
\frac{1}{{(1+y)}^{n-s+1}}
\frac{1}{{z^{n+1-2l}(1-z)}^{2l}}
\right \vert_{{x^0y^0z^0}}\\
&= {(-1)}^{k+1} \oint\limits \frac{1}{{(2\pi i)}^3}\frac{\mathrm{d}x\, \mathrm{d}y\, \mathrm{d}z}{ x\,y\,z}
\frac{1}{1-\frac{x}{yz}}
\frac{{(1+x)}^{n-s+2k-1}}{x^{k-1}}
\frac{1}{{(1+y)}^{n-s+1}}
\frac{1}{{z^{n+1-2l}(1-z)}^{2l}}\\
& ={(-1)}^{k+1} \oint\limits \frac{1}{{(2\pi i)}^2} \frac{\mathrm{d}y\, \mathrm{d}z}{ y\,z}
\frac{{(1+yz)}^{n-s+2k-1}
}{
	z^{n+k-2l}{(1-z)}^{2l}y^{k-1}{(1+y)}^{n-s+1}
}.
\end{aligned}
\end{equation}
As before, using Cauchy's theorem, we have $D_{k,l}(y\rightarrow 0, z\rightarrow 0) + D_{k,l}(y\rightarrow 0, z\rightarrow 1) + D_{k,l}(z\rightarrow \infty) = 0$. The residue at $z=\infty$ vanishes. Indeed, using the change of variables $z=1/\omega$, we have:
\begin{equation}
D_{k,l}(z\rightarrow \infty)={(-1)}^k \oint\limits \frac{1}{{(2\pi i)}^2} \frac{\mathrm{d}y\, \mathrm{d}\omega}{ y^k\,{\omega}^{k-s}}
\frac{{(\omega + y)}^{n+2k-s-1}
}{
{(1+y)}^{n-s+1}
} = 0,
\end{equation}
because $k\leq s$. Hence $D_{k,l}(z\rightarrow 0)= -D_{k,l}(z\rightarrow 1)$.
Under the change of variables $z=1-u$, we obtain:
\begin{equation}
D_{k,l}(y\rightarrow 0,u\rightarrow 0)={(-1)}^{k+1} \oint\limits \frac{1}{{(2\pi i)}^2} \frac{\mathrm{d}y\, \mathrm{d}u}{y^{k} u^{2l}}
\frac{{(1+y-uy)}^{n-s+2k-1}
}{
	{(1-u)}^{n+k-2l+1}{(1+y)}^{n-s+1}
}.
\end{equation}
In the above expression, the contours are positively oriented circles centered at the origin and of radius smaller than $1$.
Setting $y=\frac{1-v}{v(1-u)}$, one obtains\footnote{We have $1+y = \frac{1-uv}{v(1-u)}$, $1+y-uy = \frac{1}{v}$ and $\frac{\mathrm{d}y}{\mathrm{d}v} = -\frac{1}{v^2(1-u)}$}:
\begin{equation}
D_{k,l}(u\rightarrow 0, v\rightarrow 1) = {(-1)}^{k} \oint\limits \frac{1}{{(2\pi i)}^2} \frac{\mathrm{d}v\, \mathrm{d}u}{v^{k} u^{2l}}
\frac{1
}{
	{(1-u)}^{s-2l+1}{(1-v)}^{k}{(1-uv)}^{n-s+1}
},
\end{equation}
Let us note that we must pick the residues at $u=0$ and $v=1$. Indeed, we have $v=\frac{1}{1+y-uy}$ with $u$ and $y$ close to the origin. Let us set $v=1-t$ such that we get:
\begin{equation}
D_{k,l}(u\rightarrow 0, t\rightarrow 0) = {(-1)}^{k+1} \oint\limits \frac{1}{{(2\pi i)}^2} \frac{\mathrm{d}t\, \mathrm{d}u}{t^{k} u^{2l}}
\frac{1
}{
	{(1-u)}^{s-2l+1}{(1-t)}^{k}{(1-u(1-t))}^{n-s+1}
}.
\end{equation}
We now distinguish two cases depending on the value of $s-2l$.
\begin{itemize}
	\item $s-2l\geq 0$
	
	In this case, since we extract the residues at $u=0$ and $t=0$, we have:
	\begin{equation}
	\begin{aligned}
	D_{k,l} =  {(-1)}^{k+1} &\sum_{\substack{i\geq 0 \\ j\geq 0}}\left. {n-s+i \choose i}{s-2l+j \choose j} u^{i+j}{(1-t)}^{i-k}\right\vert_{u^{2l-1}t^{k-1}}\\
	=   {(-1)}^{k+1} &\sum_{i=0}^{k-1}\sum_{\substack{ j\geq 0 \\ m \geq 0}}\left. {n-s+i \choose i}{s-2l+j \choose j}{k-i-1+m \choose m}u^{i+j}t^m \right\vert_{u^{2l-1}t^{k-1}}\\
	+{(-1)}^{k+1} &\sum_{i=k+1}^{\infty}\sum_{j\geq 0}\left.{n-s+i \choose i}{s-2l+j \choose j}u^{i+j}\sum_{m=0}^{i-k} {(-1)}^{m} {i-k \choose m}t^m \right\vert_{u^{2l-1}t^{k-1}}\\
	={(-1)}^{k+1} &\sum_{i=0}^{\text{min}(k-1,2l-1)} {n-s+i \choose i}{s-1-i \choose 2l-1-i}{2k-i-2 \choose k-1} \\
	+&\sum_{i=2k-1}^{2l-1}{n-s+i \choose i}{s-1-i \choose 2l-1-i}{i-k \choose k-1}
	\end{aligned}
	\label{eq:Dkl_non_alternant_1}
	\end{equation}	
In the last sum, the summation starts at $i=2k-1$, otherwise the binomial coefficient ${i-k \choose k-1}$ vanishes and ends at $i=2l-1$ because ${s-1-i \choose 2l-1-i} = 0$ for $i>2l-1$. In particular, if $k>l$, thist last sum is empty and $D_{k,l}=C_{k,l}$, which implies $U_{k,l}=0$ as expected. 

	\item $s-2l< 0$
	
	In this case, a similar computation leads to:
	\begin{equation}
	\begin{aligned}
	D_{k,l} &= {(-1)}^{k+1} \left. \frac{ {(1-u)}^{2l-s-1} 
	}{{(1-t)}^k {(1-u(1-t))}^{n-s+1}
	}\right\vert_{u^{2l-1}t^{k-1}}\\
	& =  {(-1)}^{k+1} \sum_{j=0}^{2l-s-1}{2l-s-1 \choose j}{(-u)}^j \sum_{i\geq 0} \left. {n-s+i \choose i}{u}^i{(1-t)}^{i-k}\right\vert_{u^{2l-1}t^{k-1}}\\
	& = {(-1)}^{k+1} \sum_{j=0}^{2l-s-1} \sum_{i\geq k} {n-s+i \choose i} {2l-s-1 \choose j}  \sum_{m= 0}^{i-k}\left. {(-1)}^{j+m} {i-k\choose m} u^{i+j}t^m\right\vert_{u^{2l-1}t^{k-1}}\\
	&+ {(-1)}^{k+1} \sum_{j=0}^{2l-s-1} \sum_{i=0}^{k-1}{(-1)}^{j} {n-s+i \choose i} {2l-s-1 \choose j}  \sum_{m\geq 0}\left.  {k-i-1+m\choose m} u^{i+j}t^m\right\vert_{u^{2l-1}t^{k-1}}\\
	& = -\sum_{i= 2k-1}^{2l-1}{(-1)}^i {n-s+i \choose i} {2l-s-1 \choose 2l-1-i} {i-k \choose k-1}\\
	& \quad + {(-1)}^k \sum_{i=s}^{k-1}{(-1)}^i{n-s+i \choose i} {2l-s-1 \choose 2l-i-1}{2k-2-i \choose k-1}\\
	& =  - \sum_{i= 2k-1}^{2l-1}{(-1)}^i {n-s+i \choose i} {2l-s-1 \choose 2l-1-i} {i-k \choose k-1}.
	\end{aligned}
	\end{equation}
The last equality was obtained because $s>k-1$. When $k>l$, $D_{k,l}=0$ because the last sum is empty. It follows that $U_{k,l} = 0$ because the term $C_{k,l}$ also vanishes when $k>l$.		
\end{itemize}
Bringing all the pieces together, we have shown that $U_{k,l} = 0$ as $k>l$. 
For $k=l$, we have:
\begin{itemize}
	\item $s-2k\geq 0$
	
	From equations (\ref{eq:U_kl}), (\ref{eq:Ckl}), (\ref{eq:Ckl_non_alternant}) and (\ref{eq:Dkl_non_alternant_1}), we obtain:
	\begin{equation}
	\begin{aligned}
	U_{k,k} &= -{n-s+2k-1 \choose 2k-1}{s-2k \choose 0}{k-1 \choose k-1}\\
	&=-\frac{(n-s+2k-1)!(k-1)!
	}{(n-s+k-1)!
	},
	\end{aligned}
	\end{equation}
	
	which is exactly the result given in equation (\ref{decomp_LU}).
	
	\item $s-2k< 0$
	
	In this case, we have $C_{k,k}=0$ and $D_{k,l} = {n-s+2k-1 \choose 2k-1}$ and thus:
	\begin{equation}
	U_{k,k} = -\frac{(n-s+2k-1)!(k-1)!
	}{(n-s+k-1)!
	},
	\end{equation}
as expected.
\end{itemize}

\subsection{Computation of alternating sums in $\tilde{U}_{ss}$}\label{sec:App_alternating}
We now show how to get rid of the alternating sign in \eqref{U_ss}. We rewrite
\begin{equation}
\begin{aligned}
&\Tilde{U}_{ss}=\sum_{p=1}^{s}{(-1)}^{s+p}\binom{s-1}{p-1}\prod_{l=p}^{s-1}\frac{n+l}{n-s+l}\left[{(1+\xi)}^{s-p}-{(1+\xi)}^{n+p-1}\right]\\
%&=\sum_{p=1}^{s}    \frac{{(-1)}^{p+s}(s-1)!}{(p-1)!(s-p)!}\frac{(n+s-1)!}{(n+p-1)!}\frac{(n-s+p-1)!}{(n-1)!}\left[{(1+\xi)}^{s-p}-{(1+\xi)}^{n+p-1}\right]\\
%&={\binom{n-1}{s-1}}^{-1} \sum_{p=1}^{s}\frac{{(-1)}^{p+s}(n+s-1)!}{(n+p-1)!(s-p)!}\frac{(n-s+p-1)!}{(p-1)!(n-s)!} \left[{(1+\xi)}^{s-p}-{(1+\xi)}^{n+p-1}\right]\\
%&={\binom{n-1}{s-1}}^{-1}\sum_{p=1}^{s}{(-1)}^{p+s}\binom{n-s+p-1}{p-1}\binom{n+s-1}{n+p-1}\left[{(1+\xi)}^{s-p}-{(1+\xi)}^{n+p-1}\right]\\
%&=-{\binom{n-1}{s-1}}^{-1}\sum_{p=1}^{s}{(-1)}^{p}\binom{n-p}{s-p}\binom{n+s-1}{n+s-p}\left[{(1+\xi)}^{p-1}-{(1+\xi)}^{n+s-p}\right],
\end{aligned}
\label{eq_Uss}
\end{equation}
as $\tilde{U}_{ss} = (-1)^s {\binom{n-1}{s-1}}^{-1}( U_2-U_1)$, with
\begin{equation}
U_1\equiv \sum_{p=1}^{s}{(-1)}^{s-p}\binom{n-p}{s-p}\binom{n+s-1}{n+s-p}{(1+\xi)}^{p-1},
\label{termeU1}
\end{equation}
and
\begin{equation}
U_2\equiv \sum_{p=1}^{s}{(-1)}^{s-p}\binom{n-p}{s-p}\binom{n+s-1}{n+s-p}{(1+\xi)}^{n+s-p},
\label{termeU2_alt}
\end{equation}
and treat both terms separately. We start by $U_{1}$
\begin{equation}
\begin{aligned}
U_1&=\sum_{p=1}^{s}\left.\frac{1}{{(1+z)}^{n-s+1}}\right\vert_{z^{s-p}}\vertical{{(1+z)}^{n+s-1}}_{z^{p-1}}{(1+\xi)}^{p-1}\\
&=\sum_{p=1}^{s}\left.\frac{1}{{(1+z)}^{n-s+1}}\right\vert_{z^{s-p}}\vertical{{(1+(1+\xi)z)}^{n+s-1}}_{z^{p-1}}\\
&=\sum_{p=0}^{s-1}\left.\frac{1}{{(1+z)}^{n-s+1}}\right\vert_{z^{s-p-1}}  \vertical{{(1+(1+\xi)z)}^{n+s-1}}_{z^{p}}\\
%&=\vertical{\frac{{(1+(1+\xi)z)}^{n+s-1}}{{(1+z)}^{n-s+1}}}_{z^{s-1}}\\
&=\vertical{\frac{{(1+z+z \xi)}^{n+s-1}}{{(1+z)}^{n-s+1}}}_{z^{s-1}}\\
&=\vertical{\sum_{l=0}^{n+s-1}\binom{n+s-1}{l}{(z\xi)}^l{(1+z)}^{2s-2-l}}_{z^{s-1}}\\
&=\vertical{\sum_{l=0}^{s-1}\binom{n+s-1}{l}{(z\xi)}^l{(1+z)}^{2s-2-l}}_{z^{s-1}}\\
&=\vertical{
	\sum_{l=0}^{s-1}\binom{n+s-1}{l}z^l{\xi}^l\sum_{m=0}^{2s-2-l}\binom{2s-2-l}{m}z^m
}_{z^{s-1}}\\
&=\sum_{l=0}^{s-1}\binom{n+s-1}{l}\binom{2(s-1)-l}{s-1-l}{\xi}^l.
\end{aligned}
\label{termeU1_simpl}
\end{equation}
Similarly for $U_2$
\begin{equation}
\begin{aligned}
U_2&=\sum_{p=1}^{s}{(-1)}^{s-p}\binom{n-p}{s-p}\binom{n+s-1}{n+s-p}{(1+\xi)}^{n+s-p}\\
&={(1+\xi)}^{n+s-1}\sum_{p=1}^s\vertical{\frac{1}{{(1+z)}^{n-s+1}}}_{z^{s-p}}\vertical{{(1+z)}^{n+s-1}}_{z^{p-1}}{(1+\xi)}^{-(p-1)}\\
&={(1+\xi)}^{n+s-1}\sum_{p=1}^{s}\vertical{\frac{1}{{(1+z)}^{n-s+1}}}_{z^{s-p}}\vertical{
	{\lr{1+\frac{z}{1+\xi}}}^{n+s-1}}_{z^{p-1}
}\\
&=\vertical{\frac{{(1+z+\xi)}^{n+s-1}}{{(1+z)}^{n-s+1}}}_{z^{s-1}}\\
&=\vertical{\sum_{k=0}^{n+s-1}\binom{n+s-1}{k}{(1+z)}^{2(s-1)-k}{\xi}^k}
_{z^{s-1}}\\
&=\sum_{k=0}^{2(s-1)}
\binom{n+s-1}{k}{(1+z)}^{2(s-1)-k}{\xi}^k+\vertical{\sum_{k=2s-1}^{n+s-1}
	\binom{n+s-1}{k}\frac{1}{{(1+z)}^{k-2(s-1)}}{\xi}^k}_{z^{s-1}}
\\
&=\sum_{k=0}^{s-1}\binom{n+s-1}{k}\binom{2(s-1)-k}{s-1}{\xi}^k +(-1)^{s-1}\sum_{k=2s-1}^{n+s-1}\binom{n+s-1}{k}\binom{k-s}{s-1}{\xi}^k.
\label{termeU2}
\end{aligned}
\end{equation}
Hence, at the end of the day we get
\begin{equation}
\Tilde{U}_{ss}=-{\binom{n-1}{s-1}}^{-1}\sum_{k=2s-1}^{n+s-1}\binom{n+s-1}{k}\binom{k-s}{s-1}{\xi}^k.
\end{equation}

\section{Refined partition functions in double Aztec rectangles}\label{App:DAR}

\begin{figure}
	\centering
	\includegraphics[scale = 0.6]{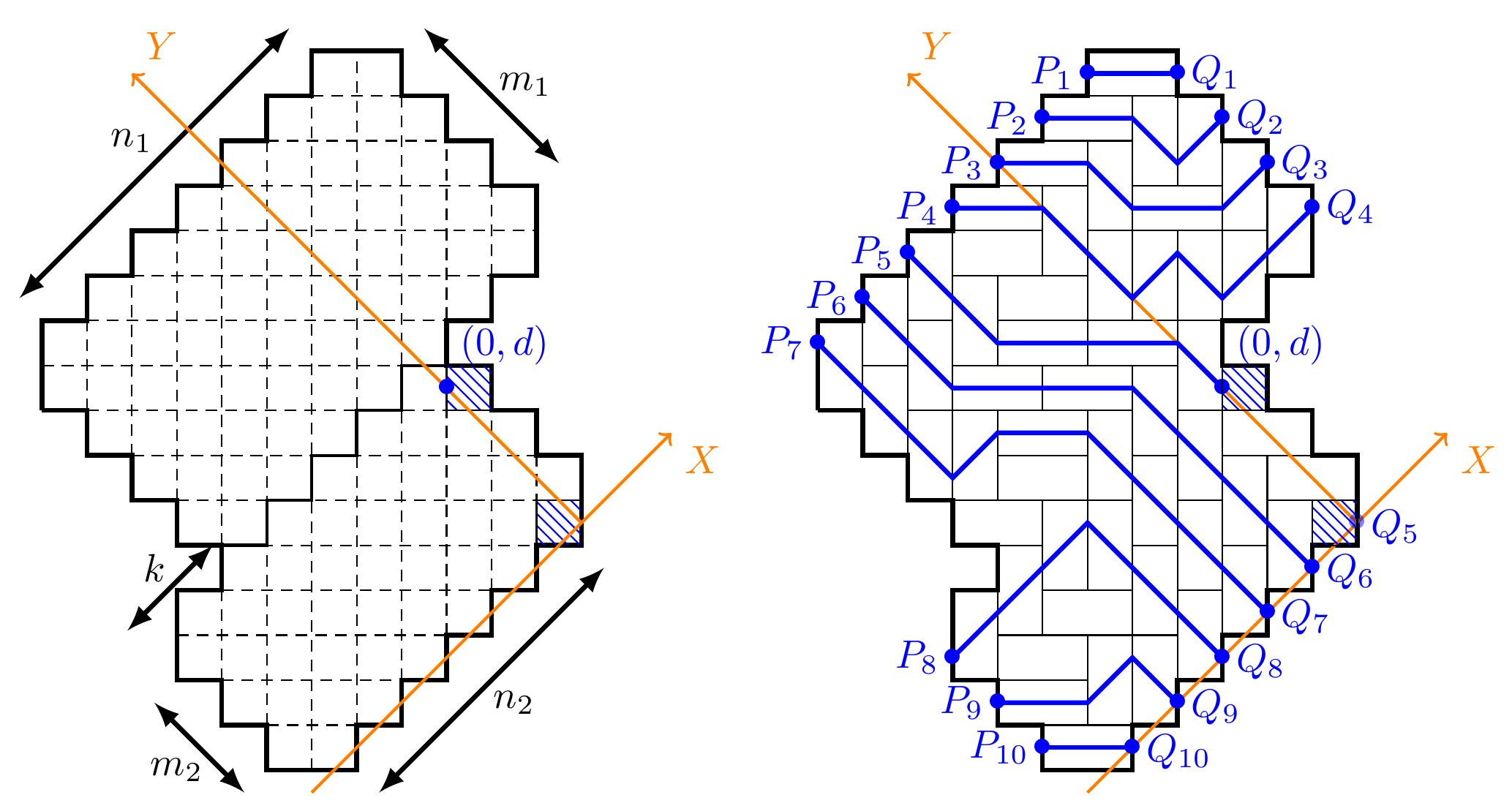}
	\caption{\JFadd{Left: the double Aztec rectangle $\mathcal{DR}_{4,7,2}^{3,6,3}$ corresponding to $m_1=4, n_1=7, m_2=3, n_2=6, k=2$ and $d=3$. Right: a domino tiling of $\mathcal{DR}_{4,7,2}^{3,6,3}$ with its equivalent description in terms of non-intersecting lattice paths.}}
	\label{fig:app_2refinedDAR}
\end{figure}

In this appendix, we extend the derivation of \cite{lai2016double_aztec_rect} and compute the $1$-refined partition function $Z_{m_1,n_1,k}^{m_2,n_2,d}(\beta)$ of the double Aztec rectangle $\mathcal{DR}_{m_1,n_1,k}^{m_2,n_2,d}$ obtained by inserting two monomers centered at $(-1/4,1/4)$ and $(1/4,d-1/4)$, see Figure \ref{fig:app_2refinedDAR}. \JFadd{We recall that $n_1-m_1=n_2-m_2$ and $k\leq \min(m_2,n_2-1)$.}

\begin{figure}
	\centering
	\begin{tikzpicture}[scale=0.5,circ/.style={shape=circle, inner sep=1.2pt, draw, fill=white, node contents=}]
	\draw[black] (-1,7)--(0,8);
	\draw[black] (-1,5)--(2,8);
	\draw[black] (-1,3)--(4,8);
	\draw[black] (-1,1)--(6,8);
	\draw[black] (-6,-6)--(8,8);
	\draw[black] (-5,-7)--(10,8);
	\draw[black] (-3,-7)--(12,8);
	\draw[black] (-1,-7)--(12,6);
	\draw[black] (1,-7)--(6,-2);
	\draw[black] (3,-7)--(6,-4);
	\draw[black] (10,0)--(12,2);
	\draw[black] (8,0)--(12,4);
	\draw[black] (12,0)--(13,1);
	\draw[black] (12,2)--(13,3);
	\draw[black] (12,4)--(13,5);
	\draw[black] (12,6)--(13,7);
	\draw[black] (-6,-2)--(-5,-1);
	\draw[black] (-6,-4)--(-3,-1);
	\draw[black] (-6,-4)--(-3,-7);
	\draw[black] (-6,-2)--(-1,-7);
	\draw[black] (-5,-1)--(1,-7);
	\draw[black] (-3,-1)--(3,-7);
	\draw[black] (-1,-1)--(5,-7);
	\draw[black] (-1,1)--(6,-6);
	\draw[black] (-1,3)--(6,-4);
	\draw[black] (-1,5)--(6,-2);
	\draw[black] (-1,7)--(6,0);
	\draw[black] (0,8)--(8,0);
	\draw[black] (2,8)--(10,0);
	\draw[black] (4,8)--(12,0);
	\draw[black] (6,8)--(13,1);
	\draw[black] (8,8)--(13,3);
	\draw[black] (10,8)--(13,5);
	\draw[black] (12,8)--(13,7);
	\draw[red] (5,-7)--(6,-6);
	\draw[red] (5,-5)--(6,-4);
	\draw[red] (5,-3)--(6,-2);
	\draw[black] (-6,-6)--(-5,-7);
	\draw[dashed,blue] (-1,0)--(13,0);
	\draw[blue] (13.5,0) node () {$d_1$};
	\draw[dashed,blue] (-6.5,-1)--(6,-1);
	\draw[blue] (6.5,-1) node () {$d_2$};
	\foreach \x in {0,...,6}
	%\draw[blue] (\x,1) node{$\bullet$};
	\draw node () at (2*\x, 0) [circ, label=below:{}];
	\foreach \x in {0,...,6}
	\draw node () at (2*\x, 2) [circ, label=below:{}];
	\foreach \x in {0,...,6}
	\draw node () at (2*\x, 4) [circ, label=below:{}];
	\foreach \x in {0,...,6}
	\draw node () at (2*\x, 6) [circ, label=below:{}];
	\foreach \x in {0,...,6}
	\draw node () at (2*\x, 8) [circ, label=below:{}];
	\foreach \x in {0,...,7}
	\draw[black] (2*\x-1,1) node{$\bullet$};
	\foreach \x in {0,...,7}
	\draw[black] (2*\x-1,3) node{$\bullet$};
	\foreach \x in {0,...,7}
	\draw[black] (2*\x-1,5) node{$\bullet$};
	\foreach \x in {0,...,7}
	\draw[black] (2*\x-1,7) node{$\bullet$};
	\foreach \x in {-2,...,3}
	\draw[black] (2*\x-1,-1) node{$\bullet$};
	\foreach \x in {-2,...,3}
	\draw[black] (2*\x-1,-3) node{$\bullet$};
	\foreach \x in {-2,...,3}
	\draw[black] (2*\x-1,-5) node{$\bullet$};
	\foreach \x in {-2,...,3}
	\draw[black] (2*\x-1,-7) node{$\bullet$};
	\foreach \x in {-3,...,3}
	\draw node () at (2*\x, -2) [circ, label=below:{}];
	\foreach \x in {-3,...,3}
	\draw node () at (2*\x, -4) [circ, label=below:{}];
	\foreach \x in {-3,...,3}
	\draw node () at (2*\x, -6) [circ, label=below:{}];
	\draw[<->,very thick, >=latex] (0,8.5)--(12,8.5);
	\draw (6,9) node () {$n_1$};
	\draw[<->,very thick, >=latex] (13.5,1)--(13.5,7);
	\draw (14.25,4) node () {$m_1$};
	\draw[<->,very thick, >=latex] (-6,-7.5)--(6,-7.5);
	\draw (0,-8) node () {$n_2$};
	\draw[<->,very thick, >=latex] (6.5,-6)--(6.5,-2);
	\draw (7.25,-4) node () {$m_2$};
	\draw[<->,very thick, >=latex] (-6,-0.5)--(0,-0.5);
	\draw (-3,0) node () {$k$};
	\end{tikzpicture}
	\caption{Dual graph of the double Aztec rectangle $\mathcal{DR}_{m_1,n_1,k}^{m_2,n_2}(\beta,\omega)$, for $m_1 = 4, n_1=7, m_2=3, n_2=6, k=3$. Red edges have a weight $\omega$, black edges oriented SW-NE carry a weight $\beta$ and the other edges (oriented along the SE-NW direction) have a weight $1$. \JFadd{The Aztec rectangle of size $(m_1,n_1)$ is located above the dashed horizontal line $d_2$ while the Aztec rectangle of size $(m_2,n_2)$ is located below this line.}} 
	\label{fig:Dual graph}
\end{figure}
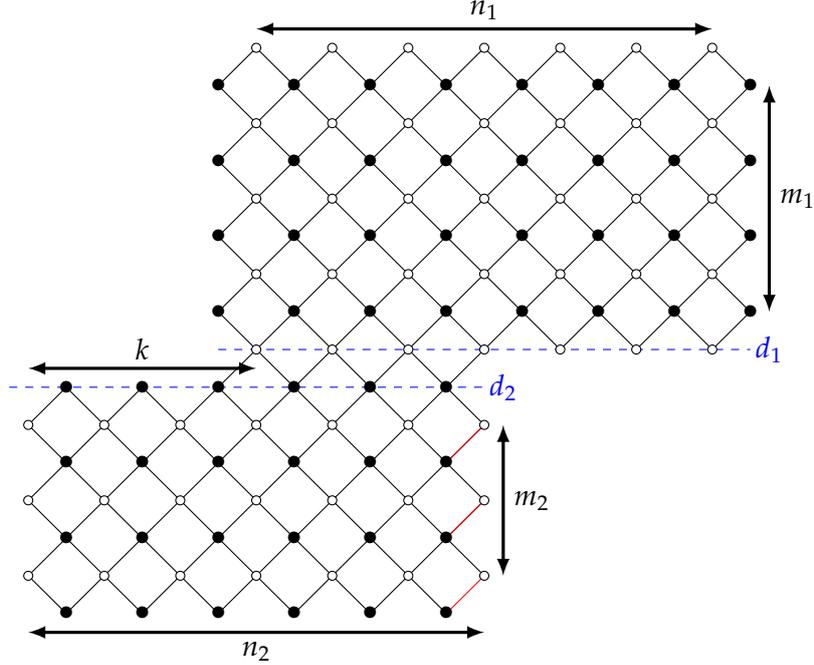

The goal of this appendix is to prove the following result:
\begin{theorem}\label{thm:1refinedZ}
	The $1$-refined partition function $Z_{m_1,n_1,k}^{m_2,n_2,d}(\beta)$ is given by:
	\begin{equation}
	\begin{aligned}
	Z_{m_1,n_1,k}^{m_2,n_2,d}(\beta) = &\beta^{(n_1-m_1)(m_1+k)+d}{(1+\beta^2)}^{\binom{m_1+1}{2}+\binom{m_2}{2}}\times\\ &\sum_{r=0}^{d}\binom{m_2-r}{d-r}{(1+\beta^{-2})}^r T\big(\mathcal{H}^{(r)}_{n_1-m_1,m_2-k+1,m_1+k}\big),
	\end{aligned}
	\label{eq:app_refinedZ}
	\end{equation}
	where $T\big(\mathcal{H}^{(r)}_{n_1-m_1,m_2-k+1,m_1+k}\big)$ is the refined enumeration of lozenge tilings of a regular hexagon of side lengths $n_1-m_1$, $m_2-k+1$, $m_1+k$, $n_1-m_1$, $m_2-k+1$, $m_1+k$ in clockwise order with the horizontal sides of length $n_1-m_1$ and such that the unique left tile along the southeast side belongs to the $r$-th row, starting from the bottom.
	\label{App:thm_refinedZ}
\end{theorem}
\JFadd{To prove Theorem \ref{App:thm_refinedZ}, it will be useful to use the bijection between domino tilings of double Aztec rectangles and perfect matchings of their dual graphs, see Figure \ref{fig:Dual graph}. As for dominoes, dimers of perfect matchings are assigned a weight $1$ or $\beta$ depending on their orientation. We consider here a slightly broader model where the $m_2$ dimers adjacent to the bottom right boundary have a weight $\omega$ instead of $\beta$, see Figure \ref{fig:Dual graph}. The interest of doing this is that the partition function $Z_{m_1,n_1,k}^{m_2,n_2}(\beta,\omega)$ of this broader model can be computed explicitly and is related to the $1$-refined partition functions $Z_{m_1,n_1,k}^{m_2,n_2,d}(\beta)$ through the following relation:}
\begin{equation}
Z_{m_1,n_1,k}^{m_2,n_2}(\beta,\omega)
=
\sum_{d=0}^{m_2}Z_{m_1,n_1,k}^{m_2,n_2,d}(\beta)\,\omega^d,
\label{eq:series_omega}
\end{equation}  
The remaining of this section will be devoted to the proof of the following lemma, which is a generalization of the results obtained in \cite{lai2016double_aztec_rect}.
\begin{lemma}
	\begin{equation}
	\begin{aligned}
	Z_{m_1,n_1,k}^{m_2,n_2}(\beta,\omega)=&\beta^{(n_1-m_1)(m_1+k)+d}{(1+\beta^2)}^{\binom{m_1+1}{2}+\binom{m_2}{2}} \times\\
	&\sum_{d=0}^{m_2}\left\{\sum_{r=0}^d\binom{m_2-r}{d-r}{(1+\beta^{-2})}^rT\big(\mathcal{H}^{(r)}_{n_1-m_1,m_2-k+1,m_1+k}\big)\right\}\omega^d,
	\end{aligned}
	\end{equation}
	\label{lemma:enumeration_omega}
\end{lemma}

From Lemma \ref{lemma:enumeration_omega} and (\ref{eq:series_omega}), one obtains the result stated in Theorem \ref{App:thm_refinedZ}. Lemma \ref{lemma:enumeration_omega} will be proved using the same techniques as the ones used in \cite{lai2016double_aztec_rect}, \JFadd{based themselves on the following preliminary lemmas that we recall for the sake of completeness.}

\subsection{Preliminary lemmas}

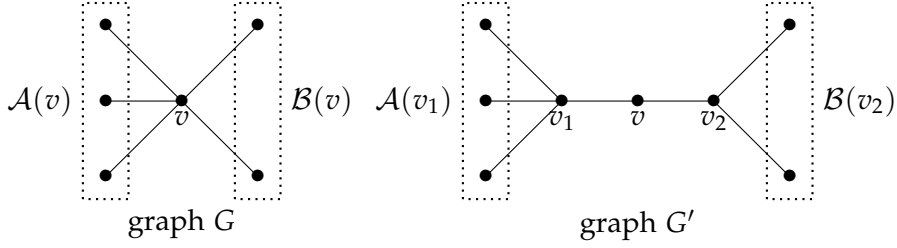
\begin{figure}
	\centering
	\begin{tikzpicture}
	\draw[black] (0,0) node{$\bullet$} node[below] () {$v$};
	\draw[black] (-1,0) node{$\bullet$};
	\draw[black] (-1,-1) node{$\bullet$};
	\draw[black] (-1,+1) node{$\bullet$};
	\draw[black] (1,-1) node{$\bullet$};
	\draw[black] (1,1) node{$\bullet$};
	\draw[black] (0,0) -- (-1,0);
	\draw[black] (0,0) -- (-1,1);
	\draw[black] (0,0) -- (-1,-1);
	\draw[black] (0,0) -- (1,1);
	\draw[black] (0,0) -- (1,-1);
	\draw[thick,dotted]     ($(-1.3,-1.3)$) rectangle ($(-0.7,1.3)$);
	\draw[thick,dotted]     ($(0.7,-1.3)$) rectangle ($(1.3,1.3)$);
	\draw[black] (-1.3,0) node[left] () {$\mathcal{A}(v)$};
	\draw[black] (1.3,0)  node[right] () {$\mathcal{B}(v)$};
	\draw[black] (0,-1.3) node[below] () {$\text{graph } G$};%
	\foreach \j in {6} {
		\draw[black] (\j,0) node{$\bullet$} node[below] () {$v$};
		\draw[black] (\j-1,0) node{$\bullet$} node[below] () {$v_1$};
		\draw[black] (\j+1,0) node{$\bullet$} node[below] () {$v_2$};
		\draw[black] (\j-2,+1) node{$\bullet$};
		\draw[black] (\j-2,0) node{$\bullet$};
		\draw[black] (\j-2,-1) node{$\bullet$};
		\draw[black] (\j+2,+1) node{$\bullet$};
		\draw[black] (\j+2,-1) node{$\bullet$};
		\draw[black] (\j-1,0) -- (\j-2,0);
		\draw[black] (\j-1,0) -- (\j-2,1);
		\draw[black] (\j-1,0) -- (\j-2,-1);
		\draw[black] (\j+1,0) -- (\j+2,1);
		\draw[black] (\j+1,0) -- (\j+2,-1);
		\draw[black] (\j,0) -- (\j+1,0);
		\draw[black] (\j,0) -- (\j-1,0);
		\draw[thick,dotted]     (\j-1-1.3,-1.3) rectangle (\j-1-0.7,1.3);
		\draw[thick,dotted]     (\j+1+0.7,-1.3) rectangle (\j+1+1.3,1.3);
		\draw[black] (\j-1-1.3,0) node[left] () {$\mathcal{A}(v_1)$};
		\draw[black] (\j+1+1.3,0)  node[right] () {$\mathcal{B}(v_2)$};
		\draw[black] (\j,-1.3) node[below] () {$\text{graph } G'$};}
	\end{tikzpicture}
	\caption{Vertex-splitting lemma.}
	\label{fig:Vertex-splitting lemma}
\end{figure}

\begin{lemma}
	(Vertex-splitting lemma, see Figure \ref{fig:Vertex-splitting lemma}) Let $G$ be a graph and $v$ one of its vertices. Let $\mathcal{N}(v)$ be the set of neighbors of $v$. Suppose $\mathcal{N}(v)=\mathcal{A}(v)\cup\mathcal{B}(v)$ with $\mathcal{A}(v)\cap\mathcal{B}(v) = \emptyset$. Let $G'$ be the new graph obtained by introducing two new vertices $v_1$ and $v_2$ such that $\mathcal{N}(v)=\{v_1,v_2\}$, $\mathcal{N}(v_1)=\mathcal{A}(v)\cup \{v\}$ and $\mathcal{N}(v_2)=\{v\}\cup \mathcal{B}(v)$. Then $T(G')=T(G)$ \JFadd{where $T(G)$ is the weighted enumeration of perfect matchings of the graph $G$}.
\end{lemma}

\begin{figure}
	\centering
	\begin{tikzpicture}[scale=0.5]
	\draw[black] (0,0) node{$\bullet$};
	\draw[black] (1,1) node{$\bullet$};
	\draw[black] (1,-1) node{$\bullet$};
	\draw[black] (2,0) node{$\bullet$};
	\draw[black] (1,-1)--(0,0)--(1,1)--(2,0);
	\draw[black] (2,0)--(1,-1);
	\draw[black] (0,0)--(-1,0);
	\draw[black] (2,0)--(3,0);
	\draw[black] (1,1)--(1,2);
	\draw[black] (1,-1)--(1,-2);
	\draw[black] (-1,0) node{$\bullet$} node[left] () {$A$};
	\draw[black] (3,0) node{$\bullet$} node[right] () {$B$};
	\draw[black] (1,2) node{$\bullet$} node[above] () {$C$};
	\draw[black] (1,-2) node{$\bullet$} node[below] () {$D$};
	\draw[black] (0.75,0.75) node[left] () {$e$};
	\draw[black] (1.25,0.75)  node[right] () {$f$};
	\draw[black] (0.75,-0.75) node[left] () {$g$};
	\draw[black] (1.25,-0.75) node[right] () {$h$};
	\draw[black] (1,-3) node[below] () {$\text{graph } G$};
	\foreach \j in {8} {
	\draw[black] (\j-1,0) node{$\bullet$} node[left] () {$A$};
	\draw[black] (\j+3,0) node{$\bullet$} node[right] () {$B$};
	\draw[black] (\j+1,2) node{$\bullet$} node[above] () {$C$};
	\draw[black] (\j+1,-2) node{$\bullet$} node[below] () {$D$};
	\draw[black] (\j-1,0)--(\j+1,2);
	\draw[black] (\j-1,0)--(\j+1,-2);
	\draw[black] (\j+3,0)--(\j+1,2);
	\draw[black] (\j+3,0)--(\j+1,-2);
	\draw[black] (\j+0.25,1+0.25) node[left] () {$h/\Delta$};
	\draw[black] (\j+2-0.25,1+0.25)  node[right] () {$f/\Delta$};
	\draw[black] (\j+0.25,-1-0.25) node[left] () {$g/\Delta$};
	\draw[black] (\j+2-0.25,-1-0.25) node[right] () {$e/\Delta$};
	\draw[black] (\j+1,-3) node[below] () {$\text{graph } G'$};
	}
	\end{tikzpicture}
	\caption{Spider lemma.}
	\label{fig:Spider_lemma}
\end{figure}
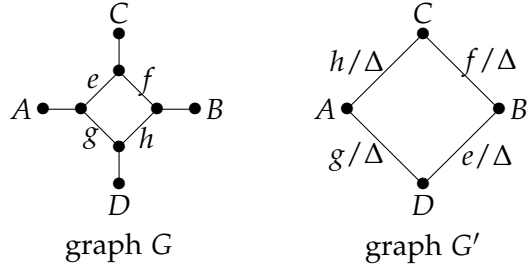

\begin{lemma}
	(Spider lemma, see Figure \ref{fig:Spider_lemma}.) Let $G$ be a graph containing the subgraph $K$ where the weights are indicated along the edges (Figure \ref{fig:Spider_lemma} Left). The four inner vertices have no other connection while vertices $A,B,C$ and $D$ can be connected to other nodes of graph $G$. If we replace the subgraph $K$ by the subgraph $K'$ (Figure \ref{fig:Spider_lemma} Right) with $\Delta=ef+gh$, then $T(G)=\Delta T(G')$.
\end{lemma}

\begin{lemma}
	(Star lemma) Let $G$ be a graph and $v$ one of its vertices. Let $G'$ be the new graph obtained by multiplying the weights of all the edges incident to $v$ by $x\neq 0$. Then $T(G)=\frac{1}{x}T(G')$.
\end{lemma}

\subsection{Proof of Theorem \ref{thm:1refinedZ}}
\begin{figure}
	\centering
	\includegraphics[width=\textwidth]{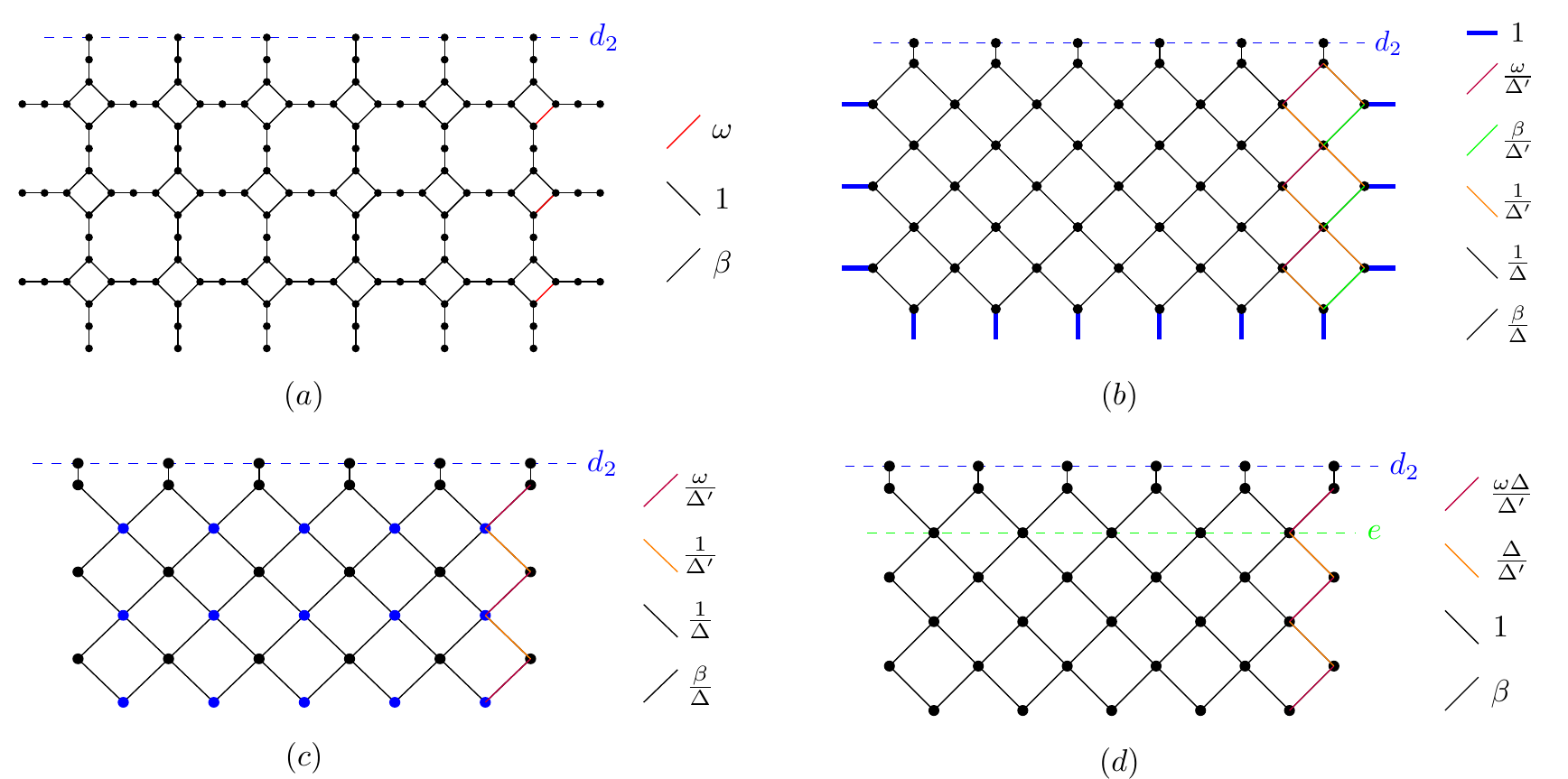}
	\caption{\JFadd{Steps to transform the dual graph of a double Aztec rectangle into the dual graph of regular hexagon. The vertex-splitting lemma (a), the spider lemma (b) and the star lemma are successively applied to all the vertices below the horizontal line $d_2$. Edge weights are indicated on the rigt. Applyig again those lemmas to the vertices below the horizontal line $e$, and so on, enables to relate the partition functions of double Aztec rectangles to those of regular hexagons. }}
	\label{fig:steps_lai}
\end{figure}

\JFadd{Applying successively the preliminary lemmas to the dual graph of the double Aztec rectangle will enable us to decompose $Z_{m_1,n_1,k}^{m_2,n_2}(\beta,\omega)$ as a sum involving refined enumerations of lozenge tilings of hexagons, for which exact expressions are known. Let us illustrate the procedure for the dual graph shown in Figure \ref{fig:Dual graph}. As stated in \cite{lai2016double_aztec_rect}, it is convenient to cut the dual graph into two parts ---one part, $G^{\text{up}}$, being above the horizontal line $d_1$, the other one, $G^{\text{down}}$, being below $d_2$--- and to apply consecutively the vertex-splitting, spider and star lemmas to each of the two parts. We give the details for the part below $d_2$ since the part above $d_1$ is treated similarly. Let use the notation $G^{\text{up}} \# G^{\text{down}}$ to indicate that the graphs $G^{\text{up}}$ and $G^{\text{down}}$ are connected through the nodes belonging to $d_2$ \cite{lai2016double_aztec_rect}.  The first step is the application of the \textit{vertex-splitting lemma} to all the nodes below $d_2$, see Figure \ref{fig:steps_lai}(a). This does not modify the partition function. In the second step, we use the spider lemma, see Figure \ref{fig:steps_lai}(b), leading to an overall multiplicative factor ${\Delta}^{m_2(n_2-1)}{\Delta'}^{\,m_2}$, with $\Delta = 1+\beta^2$ and $\Delta' = 1+\beta\,\omega$. We also remove the forced edges of degree $1$ along with the vertices attached to them. In the third step, we use the star lemma and multiply, by the factor $\Delta$, the weights of all the edges incident to the remaining $m_2(n_2-1)$ nodes (below $d_2$), see Figure \ref{fig:steps_lai}(c). This leads to an overall factor ${\big(\frac{1}{\Delta}\big)}^{m_2(n_2-1)}$. In summary, we obtain: 
\begin{equation}
\begin{aligned}
T( G^{\text{down}} \# G^{\text{up}} ) 
	&={\Big(\frac{1}{\Delta}\Big)}^{m_2(n_2-1)}{\Delta}^{m_2(n_2-1)}{\Delta'}^{\,m_2} \,T(G_{\text{new}} \# G^{\text{up}})\\
	&={\Delta'}^{\,m_2} T(G_{\text{new}} \# G^{\text{up}}),
\end{aligned}
\end{equation}
with $G_{\text{new}}$ the new graph obtained after the successive applications of the operations described above, see Figure \ref{fig:steps_lai}(d). }

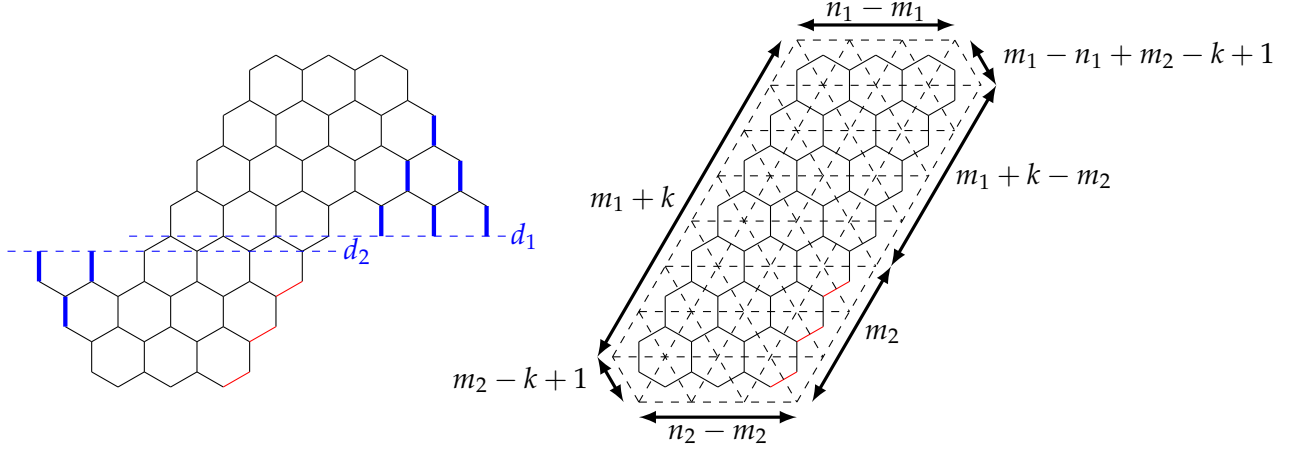
\begin{figure}
	\centering
\begin{tikzpicture}[scale=0.4]
%\foreach \i in {0,...,3} 
%\foreach \j in {0,...,3} {
	\foreach \a in {90,210,-30} \draw (0,0) -- +(cos{\a},sin{\a});
	\foreach \a in {90,210,-30} \draw (2*cos{30},0) -- +(cos{\a},sin{\a});
	\foreach \a in {90,-30} \draw (-2*cos{30},0) -- +(cos{\a},sin{\a});
	\foreach \a in {90} \draw (4*cos{30},0) -- +(cos{\a},sin{\a});
	\foreach \a in {210} \draw[red] (4*cos{30},0) -- +(cos{\a},sin{\a});
	% Ligne 2
	\foreach \a in {90,210,-30} \draw (-cos{30},1+sin{30}) -- +(cos{\a},sin{\a});
	\foreach \a in {90,210,-30} \draw (cos{30},1+sin{30}) -- +(cos{\a},sin{\a});
	\foreach \a in {90,210,-30} \draw (3*cos{30},1+sin{30}) -- +(cos{\a},sin{\a});
	\foreach \a in {90} \draw (5*cos{30},1+sin{30}) -- +(cos{\a},sin{\a});
	\foreach \a in {210} \draw[red] (5*cos{30},1+sin{30}) -- +(cos{\a},sin{\a});
	\foreach \a in {-30} \draw (-3*cos{30},1+sin{30}) -- +(cos{\a},sin{\a});
	\foreach \a in {90} \draw[blue,ultra thick] (-3*cos{30},1+sin{30}) -- +(cos{\a},sin{\a});
	% Ligne 3
	\foreach \a in {-30} \draw (-4*cos{30},2+2*sin{30}) -- +(cos{\a},sin{\a});
	\foreach \a in {210,-30} \draw (-2*cos{30},2+2*sin{30}) -- +(cos{\a},sin{\a});
	\foreach \a in {90,210,-30} \draw (0*cos{30},2+2*sin{30}) -- +(cos{\a},sin{\a});
	\foreach \a in {90,210,-30} \draw (2*cos{30},2+2*sin{30}) -- +(cos{\a},sin{\a});
	\foreach \a in {90,210,-30} \draw (4*cos{30},2+2*sin{30}) -- +(cos{\a},sin{\a});
	\foreach \a in {90} \draw (6*cos{30},2+2*sin{30}) -- +(cos{\a},sin{\a});
	\foreach \a in {210} \draw[red] (6*cos{30},2+2*sin{30}) -- +(cos{\a},sin{\a});
	\foreach \a in {90} \draw[blue,ultra thick] (-2*cos{30},2+2*sin{30}) -- +(cos{\a},sin{\a});
	\foreach \a in {90} \draw[blue,ultra thick] (-4*cos{30},2+2*sin{30}) -- +(cos{\a},sin{\a});
	% Ligne 4
	\foreach \a in {90,210,-30} \draw (cos{30},3+3*sin{30}) -- +(cos{\a},sin{\a});
	\foreach \a in {90,210,-30} \draw (3*cos{30},3+3*sin{30}) -- +(cos{\a},sin{\a});
	\foreach \a in {90,210,-30} \draw (5*cos{30},3+3*sin{30}) -- +(cos{\a},sin{\a});
	\foreach \a in {90,210} \draw (7*cos{30},3+3*sin{30}) -- +(cos{\a},sin{\a});
	\foreach \i in{1,2,3}{
	\foreach \a in {90} \draw[blue,ultra thick] (7*cos{30}+2*\i*cos{30},3+3*sin{30}) -- +(cos{\a},sin{\a});}
	% Ligne 5
	\foreach \a in {90,210,-30} \draw (2*cos{30},4+4*sin{30}) -- +(cos{\a},sin{\a});
	\foreach \a in {90,210,-30} \draw (4*cos{30},4+4*sin{30}) -- +(cos{\a},sin{\a});
	\foreach \a in {90,210,-30} \draw (6*cos{30},4+4*sin{30}) -- +(cos{\a},sin{\a});
	\foreach \a in {90,210,-30} \draw (8*cos{30},4+4*sin{30}) -- +(cos{\a},sin{\a});
	\foreach \a in {210,-30} \draw (10*cos{30},4+4*sin{30}) -- +(cos{\a},sin{\a});
	\foreach \a in {210,-30} \draw (12*cos{30},4+4*sin{30}) -- +(cos{\a},sin{\a});
	\foreach \a in {90} \draw[blue,ultra thick] (10*cos{30},4+4*sin{30}) -- +(cos{\a},sin{\a});
	\foreach \a in {90} \draw[blue,ultra thick] (12*cos{30},4+4*sin{30}) -- +(cos{\a},sin{\a});
	% Ligne 6
	\foreach \a in {90,210,-30} \draw (3*cos{30},5+5*sin{30}) -- +(cos{\a},sin{\a});
	\foreach \a in {90,210,-30} \draw (5*cos{30},5+5*sin{30}) -- +(cos{\a},sin{\a});
	\foreach \a in {90,210,-30} \draw (7*cos{30},5+5*sin{30}) -- +(cos{\a},sin{\a});
	\foreach \a in {90,210,-30} \draw (9*cos{30},5+5*sin{30}) -- +(cos{\a},sin{\a});
	\foreach \a in {210,-30} \draw (11*cos{30},5+5*sin{30}) -- +(cos{\a},sin{\a});
	\foreach \a in {90} \draw[blue,ultra thick] (11*cos{30},5+5*sin{30}) -- +(cos{\a},sin{\a});
	% Ligne 7
	\foreach \a in {90,210,-30} \draw (4*cos{30},6+6*sin{30}) -- +(cos{\a},sin{\a});
	\foreach \a in {90,210,-30} \draw (6*cos{30},6+6*sin{30}) -- +(cos{\a},sin{\a});
	\foreach \a in {90,210,-30} \draw (8*cos{30},6+6*sin{30}) -- +(cos{\a},sin{\a});
	\foreach \a in {90,210,-30} \draw (10*cos{30},6+6*sin{30}) -- +(cos{\a},sin{\a});
	% Ligne 8
	\foreach \a in {210,-30} \draw (5*cos{30},7+7*sin{30}) -- +(cos{\a},sin{\a});
	\foreach \a in {210,-30} \draw (7*cos{30},7+7*sin{30}) -- +(cos{\a},sin{\a});
	\foreach \a in {210,-30} \draw (9*cos{30},7+7*sin{30}) -- +(cos{\a},sin{\a});
%	\foreach \a in {0,120,-120} \draw (3*\i+3*cos{60},2*sin{60}*\j+sin{60}) -- +(\a:1);}
	\draw[dashed,blue] (-4.5,4)--(6.5,4);
	\draw[blue] (7,4) node () {$d_2$};
	\draw[dashed,blue] (-0.5,4+sin{30})--(12,4+sin{30});
	\draw[blue] (12.5,4+sin{30}) node () {$d_1$};
%\end{tikzpicture}
%\begin{tikzpicture}[scale=0.4]
\foreach \j in {18}{ 
\foreach \a in {90,210,-30} \draw (\j+0,0) -- +(cos{\a},sin{\a});
\foreach \a in {90,210,-30} \draw (\j+2*cos{30},0) -- +(cos{\a},sin{\a});
\foreach \a in {90,-30} \draw (\j-2*cos{30},0) -- +(cos{\a},sin{\a});
\foreach \a in {90} \draw (\j+4*cos{30},0) -- +(cos{\a},sin{\a});
\foreach \a in {210} \draw[red] (\j+4*cos{30},0) -- +(cos{\a},sin{\a});
% Ligne 2
\foreach \a in {90,210,-30} \draw (\j-cos{30},1+sin{30}) -- +(cos{\a},sin{\a});
\foreach \a in {90,210,-30} \draw (\j+cos{30},1+sin{30}) -- +(cos{\a},sin{\a});
\foreach \a in {90,210,-30} \draw (\j+3*cos{30},1+sin{30}) -- +(cos{\a},sin{\a});
\foreach \a in {90} \draw (\j+5*cos{30},1+sin{30}) -- +(cos{\a},sin{\a});
\foreach \a in {210} \draw[red] (\j+5*cos{30},1+sin{30}) -- +(cos{\a},sin{\a});
% Ligne 3
\foreach \a in {90,210,-30} \draw (\j+0*cos{30},2+2*sin{30}) -- +(cos{\a},sin{\a});
\foreach \a in {90,210,-30} \draw (\j+2*cos{30},2+2*sin{30}) -- +(cos{\a},sin{\a});
\foreach \a in {90,210,-30} \draw (\j+4*cos{30},2+2*sin{30}) -- +(cos{\a},sin{\a});
\foreach \a in {210} \draw[red] (\j+6*cos{30},2+2*sin{30}) -- +(cos{\a},sin{\a});
\foreach \a in {90} \draw (\j+6*cos{30},2+2*sin{30}) -- +(cos{\a},sin{\a});
% Ligne 4
\foreach \a in {90,210,-30} \draw (\j+cos{30},3+3*sin{30}) -- +(cos{\a},sin{\a});
\foreach \a in {90,210,-30} \draw (\j+3*cos{30},3+3*sin{30}) -- +(cos{\a},sin{\a});
\foreach \a in {90,210,-30} \draw (\j+5*cos{30},3+3*sin{30}) -- +(cos{\a},sin{\a});
\foreach \a in {90,210} \draw (\j+7*cos{30},3+3*sin{30}) -- +(cos{\a},sin{\a});
% Ligne 5
\foreach \a in {90,210,-30} \draw (\j+2*cos{30},4+4*sin{30}) -- +(cos{\a},sin{\a});
\foreach \a in {90,210,-30} \draw (\j+4*cos{30},4+4*sin{30}) -- +(cos{\a},sin{\a});
\foreach \a in {90,210,-30} \draw (\j+6*cos{30},4+4*sin{30}) -- +(cos{\a},sin{\a});
\foreach \a in {90,210} \draw (\j+8*cos{30},4+4*sin{30}) -- +(cos{\a},sin{\a});
% Ligne 6
\foreach \a in {90,210,-30} \draw (\j+3*cos{30},5+5*sin{30}) -- +(cos{\a},sin{\a});
\foreach \a in {90,210,-30} \draw (\j+5*cos{30},5+5*sin{30}) -- +(cos{\a},sin{\a});
\foreach \a in {90,210,-30} \draw (\j+7*cos{30},5+5*sin{30}) -- +(cos{\a},sin{\a});
\foreach \a in {90,210} \draw (\j+9*cos{30},5+5*sin{30}) -- +(cos{\a},sin{\a});
% Ligne 7
\foreach \a in {90,210,-30} \draw (\j+4*cos{30},6+6*sin{30}) -- +(cos{\a},sin{\a});
\foreach \a in {90,210,-30} \draw (\j+6*cos{30},6+6*sin{30}) -- +(cos{\a},sin{\a});
\foreach \a in {90,210,-30} \draw (\j+8*cos{30},6+6*sin{30}) -- +(cos{\a},sin{\a});
\foreach \a in {90,210} \draw (\j+10*cos{30},6+6*sin{30}) -- +(cos{\a},sin{\a});
% Ligne 8
\foreach \a in {210,-30} \draw (\j+5*cos{30},7+7*sin{30}) -- +(cos{\a},sin{\a});
\foreach \a in {210,-30} \draw (\j+7*cos{30},7+7*sin{30}) -- +(cos{\a},sin{\a});
\foreach \a in {210,-30} \draw (\j+9*cos{30},7+7*sin{30}) -- +(cos{\a},sin{\a});

\foreach \i in {0,...,6}{
\draw[dashed] (\j-3*cos{30}+\i*cos{30},\i+0.5+\i*sin{30})--(\j+5*cos{30}+\i*cos{30},\i+0.5+\i*sin{30});}
\draw[dashed] (\j-2*cos{30},-0.5-sin{30})--(\j+4*cos{30},-0.5-sin{30});
\draw[dashed] (\j+4*cos{30},7.5+7*sin{30})--(\j+10*cos{30},7.5+7*sin{30});
\draw[dashed] (\j-2*cos{30},-0.5-sin{30})--(\j-3*cos{30},0.5);
\draw[dashed] (\j-3*cos{30},0.5)--(\j+4*cos{30},7.5+7*sin{30});
\draw[dashed] (\j+10*cos{30},7.5+7*sin{30})--(\j+11*cos{30},7.5+4*sin{30});
\draw[dashed] (\j+4*cos{30},-0.5-sin{30})--(\j+11*cos{30},7.5+4*sin{30});
\draw[dashed] (\j-2*cos{30},2)--(\j+0*cos{30},-1);
\draw[dashed] (\j-1*cos{30},3.5)--(\j+2*cos{30},-1);
\draw[dashed] (\j+0*cos{30},5)--(\j+4*cos{30},-1);
\draw[dashed] (\j+1*cos{30},6.5)--(\j+5*cos{30},0.5);
\draw[dashed] (\j+2*cos{30},8)--(\j+6*cos{30},2);
\draw[dashed] (\j+3*cos{30},9.5)--(\j+7*cos{30},3.5);
\draw[dashed] (\j+4*cos{30},11)--(\j+8*cos{30},5);
\draw[dashed] (\j+6*cos{30},11)--(\j+9*cos{30},6.5);
\draw[dashed] (\j+8*cos{30},11)--(\j+10*cos{30},8);
\draw[<->,very thick, >=latex] (\j-2*cos{30},-1.5)--(\j+4*cos{30},-1.5);
\draw (\j+cos{30},-2) node () {$n_2-m_2$};
\draw[<->,very thick, >=latex] (\j+4*cos{30},11.5)--(\j+10*cos{30},11.5);
\draw (\j+7*cos{30},12) node () {$n_1-m_1$};
\draw[<->,very thick, >=latex] (\j-3*cos{30}-0.5,0.5)--(\j+4*cos{30}-0.5,11);
\draw (\j-2.25*cos{30},5.75) node () {$m_1+k$};
\draw[<->,very thick, >=latex] (\j-3*cos{30}-0.5,0.5)--(\j-2*cos{30}-0.5,-1);
\draw (\j-6.5*cos{30},-0.25) node () {$m_2-k+1$};
\draw[<->,very thick, >=latex] (\j+4*cos{30}+0.5,-1)--(\j+7*cos{30}+0.5,3.5);
\draw (\j+7.25*cos{30},1.25) node () {$m_2$};
\draw[<->,very thick, >=latex] (\j+7*cos{30}+0.5,3.5)--(\j+11*cos{30}+0.5,9.5);
\draw (\j+13*cos{30},6.5) node () {$m_1+k-m_2$};
\draw[<->,very thick, >=latex] (\j+11*cos{30}+0.5,9.5)--(\j+10*cos{30}+0.5,11);
\draw (\j+17*cos{30},10.5) node () {$m_1-n_1+m_2-k+1$};
\draw[dashed] (\j-2*cos{30},-1)--(\j+6*cos{30},7.5+7*sin{30});
\draw[dashed] (\j+0*cos{30},-1)--(\j+8*cos{30},7.5+7*sin{30});
\draw[dashed] (\j+2*cos{30},-1)--(\j+10*cos{30},7.5+7*sin{30});
}

\end{tikzpicture}
\caption{Applying successively the vertex-splitting, spider and star lemmas transforms the dual graph of a double Aztec rectangle into the dual graph of a regular $(n_1-m_1,m_2-k+1,m_1+k)$ hexagon (since by assumption $n_1-m_1 = n_2-m_2$). Blue edges (Left) are forced edges of weight $1$ that can be removed. Red edges (Right) have a weight $\omega \frac{\Delta}{\Delta'}$. }
\label{fig:refinedHexa}
\end{figure}

We repeat the process, considering now the part below the horizontal line $e$, see Figure \ref{fig:steps_lai}(d). \JFadd{Proceeding recursively} and doing the same for the part above $d_1$, we eventually get:
\begin{equation}
Z_{m_1,n_1,k}^{m_2,n_2}(\beta,\omega) = \Delta^{\binom{m_1+1}{2} + \binom{m_2}{2}}(\Delta')^{m_2}T\Big(\mathcal{H}_{n_1-m_1,m_2-k+1,m_1+k}\big(\omega \frac{\Delta}{\Delta'}\big)\Big),
\label{eq:refined}
\end{equation}
with $\Delta = 1+\beta^2$, $\Delta' = 1+\beta\,\omega$ and $T\Big(\mathcal{H}_{n_1-m_1,m_2-k+1,m_1+k}\big(\omega \frac{\Delta}{\Delta'}\big)\Big)$ the partition function of the regular hexagon with the $m_2$ edges along the bottom right side, starting from the bottom, having weight $\omega \frac{\Delta}{\Delta'}$, see Figure \ref{fig:refinedHexa}. This is a regular hexagon since we assume $n_1-m_1=n_2-m_2$.

Let us now develop the term $T\Big(\mathcal{H}_{n_1-m_1,m_2-k+1,m_1+k}\big(\omega \frac{\Delta}{\Delta'}\big)\Big)$, conditioning on the number of edges of weight $\omega \frac{\Delta}{\Delta'}$. Each tiling of a regular $(a,b,c)-$hexagon can be put in bijection with a set of $a$ non-intersecting lattice paths, each of these paths containing $c$ oblique steps SW-NE and $b$ oblique steps SE-NW. Hence, we have:
\begin{equation}
T\Big(\mathcal{H}_{n_1-m_1,m_2-k+1,m_1+k}\big(\omega \frac{\Delta}{\Delta'}\big)\Big) = \sum_{r=0}^{m_2}T\big(\mathcal{H}^{(r)}_{n_1-m_1,m_2-k+1,m_1+k}\big) \beta^{(n_1-m_1)\cdot(m_1+k)-r} {\Big(\omega \frac{\Delta}{\Delta'}\Big)}^{r},
\end{equation}
where $T\big(\mathcal{H}^{(r)}_{n_1-m_1,m_2-k+1,m_1+k}\big)$ gives the number of lozenge tilings of a regular $(n_1-m_1,m_2-k+1,m_1+k)$-hexagon such that the righmost path leaves the southeast boundary at position $r$ starting from the bottom. 
%using the fact that $\alpha=1$. 
Inserting this relation into (\ref{eq:refined}) leads to 
\begin{equation}
\resizebox{\textwidth}{!}{% 
$
\begin{aligned}
&Z_{m_1,n_1,k}^{m_2,n_2}(\beta,\omega)\\ 
&=\Delta^{\binom{m_1+1}{2} + \binom{m_2}{2}}\beta^{(n_1-m_1)(m_1+k)}\sum_{r=0}^{m_2}T\big(\mathcal{H}^{(r)}_{n_1-m_1,m_2-k+1,m_1+k}\big)\omega^r{\Big(\frac{\Delta}{\beta}\Big)}^r(\Delta')^{m_2-r}\\
&=\Delta^{\binom{m_1+1}{2} + \binom{m_2}{2}}\beta^{(n_1-m_1)(m_1+k)}\sum_{r=0}^{m_2}T\big(\mathcal{H}^{(r)}_{n_1-m_1,m_2-k+1,m_1+k}\big)\omega^r{\Big(\frac{\Delta}{\beta}\Big)}^r
\sum_{j=0}^{m_2-r}\binom{m_2-r}{j}{(\beta\omega)}^j\\
&=\Delta^{\binom{m_1+1}{2} + \binom{m_2}{2}}\beta^{(n_1-m_1)(m_1+k)}\sum_{r=0}^{m_2}T\big(\mathcal{H}^{(r)}_{n_1-m_1,m_2-k+1,m_1+k}\big){\Big(\frac{\Delta}{\beta}\Big)}^r
\sum_{d=r}^{m_2}\binom{m_2-r}{d-r}\beta^{d-r}\omega^d\\
&=\Delta^{\binom{m_1+1}{2} + \binom{m_2}{2}}\beta^{(n_1-m_1)(m_1+k)}\sum_{d=0}^{m_2}\left\{
\beta^d\sum_{r=0}^d T\big(\mathcal{H}^{(r)}_{n_1-m_1,m_2-k+1,m_1+k}\big) \binom{m_2-r}{d-r}{\Big(\frac{\Delta}{\beta^2}\Big)}^r
\right\}\omega^d
\end{aligned}
$
}
\end{equation}
This proves Lemma \ref{lemma:enumeration_omega}. Extracting the coefficient in front of $\omega^d$ leads to the proof of the Theorem \ref{App:thm_refinedZ}.

\section{Simulation of the $6$V model with pDWBC}\label{App:simu}

In this appendix, we briefly explain the algorithm used to generate configurations of the $6$ vertex model with pDWBC. Configurations of the $6$V model with pDWBC were obtained using a Markov chain Monte-Carlo algorithm in the same spirit as the ones discussed in \cite{allison20056V_MonteCarlo, lyberg20186V_varietyBC, lyberg2017density_6VDWBC}. Starting with an initial configuration $\mathcal{C}_0$, the idea is to generate a sequence of successive configurations $\mathcal{C}_1, \mathcal{C}_2,\cdots$ such that the probability $P_N(\mathcal{C}_0 \rightarrow \mathcal{C})$ of observing the configuration $\mathcal{C}$ after $N$ steps converges to the Gibbs measure
\begin{equation}
	\pi(C) = \fr{
	\omega(\mathcal{C})
	}{
	Z
	},
\end{equation}
as $N\rightarrow + \infty$, independently of the choice of the initial configuration $\mathcal{C}_0$ as long as it satisifies the ice rule. The Gibbs measure is expressed as the ratio between the weight $\omega(\mathcal{C})$ associated to the configuration $\mathcal{C}$ and the partition function $Z$. We use the description in terms of non-intersecting lattice paths. At each iteration, we select randomly a plaquette whose center is $(i,j)$ and, if allowed by the ice rule, we apply a local move among the $4$ possibilities \begin{tikzpicture} \draw[dotted](0,0) rectangle(0.3,0.3);\draw (0,0)--(0,0.3)--(0.3,0.3); \end{tikzpicture} $\rightarrow$  \begin{tikzpicture} \draw[dotted](0,0) rectangle(0.3,0.3);\draw(0.3,0.3)--(0.3,0)--(0,0); \end{tikzpicture} , \begin{tikzpicture} \draw[dotted](0,0) rectangle(0.3,0.3);\draw(0.3,0.3)--(0.3,0)--(0,0); \end{tikzpicture} $\rightarrow$ \begin{tikzpicture} \draw[dotted](0,0) rectangle(0.3,0.3);\draw (0,0)--(0,0.3)--(0.3,0.3); \end{tikzpicture} , \begin{tikzpicture} \draw[dotted](0,0.3)--(0,0)--(0.3,0)--(0.3,0.3) ;\draw(0,0)--(0.3,0)--(0.3,0.3); \end{tikzpicture}  $\rightarrow$ \begin{tikzpicture} \draw[dotted](0,0.3)--(0,0)--(0.3,0)--(0.3,0.3) ;\draw(0,0.3)--(0,0); \end{tikzpicture} , \begin{tikzpicture} \draw[dotted](0,0.3)--(0,0)--(0.3,0)--(0.3,0.3) ;\draw(0,0.3)--(0,0); \end{tikzpicture} $\rightarrow$ \begin{tikzpicture} \draw[dotted](0,0.3)--(0,0)--(0.3,0)--(0.3,0.3) ;\draw(0,0)--(0.3,0)--(0.3,0.3); \end{tikzpicture} .
Notice that the last two moves only imply vertices belonging to the upper boundary. The new configuration $\mathcal{C}'$ is then accepted with a Metropolis acceptance probability given by \cite{metropolis1953metrop_algo}:
\begin{equation}
A(\mathcal{C}\rightarrow \mathcal{C'})=\min\left(1,\frac{W_{i,j}(\mathcal{C}')}{W_{i,j}(\mathcal{C})}\right).
\label{eq_prob_metropolis}
\end{equation}
where $\mathcal{C}$ is the older configuration and $W_{i,j}(\mathcal{C})$ is the product of the weights of the $4$ vertices attached to the corners of the plaquette centered at $(i,j)$ in the configuration $\mathcal{C}$. This procedure defines a finite\footnote{The number of possible states corresponds to the number of distinct configurations which is finite for a finite domain.} Markov chain since the probability $p(\mathcal{C} \rightarrow \mathcal{C'})$ to move from a configuration $\mathcal{C}$ to a configuration $\mathcal{C'}$ only depends on the current configuration $\mathcal{C}$. This Markov chain is moreover ergodic since every configuration is accessible from any other one, after a finite number of local moves. Those properties guarantee that $\lim_{N\rightarrow + \infty} P(\mathcal{C}_0\rightarrow \mathcal{C}) = P(\mathcal{C})$ with $P$ a unique distribution \cite{doeblin1938ergo_stationary} satisfying:
\begin{equation}
P(\mathcal{C}) = \sum_{\mathcal{C'}} P(\mathcal{C'}) p(\mathcal{C} \rightarrow \mathcal{C'}).
\label{eq:Doeblin}
\end{equation}
Using the fact that $\sum_{\mathcal{C'}} p(\mathcal{C} \rightarrow \mathcal{C'})=1$, equation (\ref{eq:Doeblin}) is equivalent to the global balance condition:
\begin{equation}
\sum_{\mathcal{C'}\neq \mathcal{C}} P(\mathcal{C}) p(\mathcal{C} \rightarrow \mathcal{C'}) = 
\sum_{\mathcal{C'}\neq \mathcal{C}} P(\mathcal{C'}) p(\mathcal{C'} \rightarrow \mathcal{C}).
\end{equation}
The latter equation is itself satisfied if the detailed balanced condition is met, that is:
\begin{equation}
P(\mathcal{C}) p(\mathcal{C} \rightarrow \mathcal{C'}) = P(\mathcal{C'}) p(\mathcal{C'} \rightarrow \mathcal{C}).
\end{equation}
One can check that the detailed balance condition is satisfied if we choose $p(\mathcal{C} \rightarrow \mathcal{C'}) = \frac{1}{N_p} A(\mathcal{C}\rightarrow \mathcal{C'})$, with $N_p$, a constant, corresponding to the number of plaquettes of any configuration.

\begin{figure}[h!]
	\centering
	\begin{tikzpicture}[scale=0.8]
	\clip (-0.8,-1.7) rectangle (5.1,3);
	\draw[step=0.75cm,color=black, thick] (0,0) grid (3.75,2.25);
	\draw[step=0.75cm,color=black , thick] (-0.75,0) grid (0,0);
	\draw[step=0.75cm,color=black, thick] (-0.75,0.75) grid (0,0.75);
	\draw[step=0.75cm,color=black, thick] (-0.75,1.5) grid (0,1.5);
	\draw[step=0.75cm,color=black, thick] (-0.75,2.25) grid (0,2.25);
	\draw[step=0.75cm,color=black, thick] (3.75,0) grid (4.5,0);
	\draw[step=0.75cm,color=black, thick] (3.75,0.75) grid (4.5,0.75);
	\draw[step=0.75cm,color=black, thick] (3.75,1.5) grid (4.5,1.5);
	\draw[step=0.75cm,color=black, thick] (3.75,2.25) grid (4.5,2.25);
	\draw[step=0.75cm,color=black, thick] (0,-0.75) -- (0,0);
	\draw[step=0.75cm,color=black, thick] (0.75,-0.75) -- (0.75,0);
	\draw[step=0.75cm,color=black, thick] (1.5,-0.75) -- (1.5,0);
	\draw[step=0.75cm,color=black, thick] (2.25,-0.75) -- (2.25,0);
	\draw[step=0.75cm,color=black, thick] (3,-0.75) -- (3,0);
	\draw[step=0.75cm,color=black, thick] (3.75,-0.75) -- (3.75,0);
	\draw[step=0.75cm,color=black, thick] (0,2.25) -- (0,3);
	\draw[step=0.75cm,color=black, thick] (0.75,2.25) -- (0.75,3);
	\draw[step=0.75cm,color=black, thick] (1.5,2.25) -- (1.5,3);
	\draw[step=0.75cm,color=black, thick] (2.25,2.25) -- (2.25,3);
	\draw[step=0.75cm,color=black, thick] (3,2.25) -- (3,3);
	\draw[step=0.75cm,color=black, thick] (3.75,2.25) -- (3.75,3);
	\draw[fill = blue, opacity=0.3] (0, 0) rectangle (0.75, 0.75);
	\draw[fill = blue, opacity=0.3] (1.5, 0) rectangle (2.25, 0.75);
	\draw[fill = blue, opacity=0.3] (3, 0) rectangle (3.75, 0.75);
	\draw[fill = blue, opacity=0.3] (0, 1.5) rectangle (0.75, 2.25);
	\draw[fill = blue, opacity=0.3] (1.5, 1.5) rectangle (2.25, 2.25);
	\draw[fill = blue, opacity=0.3] (3, 1.5) rectangle (3.75, 2.25);
	\draw[fill = red, opacity=0.3] (0, 0.75) rectangle (0.75, 1.5);
	\draw[fill = red, opacity=0.3] (1.5, 0.75) rectangle (2.25, 1.5);
	\draw[fill = red, opacity=0.3] (3, 0.75) rectangle (3.75, 1.5);
	\draw[fill = red, opacity=0.3] (0, 2.25) rectangle (0.75, 3);
	\draw[fill = red, opacity=0.3] (1.5, 2.25) rectangle (2.25, 3);
	\draw[fill = red, opacity=0.3] (3, 2.25) rectangle (3.75, 3);
	\draw[fill = green, opacity=0.3] (0.75, 0.75) rectangle (1.5, 1.5);
	\draw[fill = green, opacity=0.3] (2.25, 0.75) rectangle (3, 1.5);
	%\draw[fill = green, opacity=0.3] (3.75, 0.75) rectangle (4.5, 1.5);
	\draw[fill = green, opacity=0.3] (0.75, 2.25) rectangle (1.5, 3);
	\draw[fill = green, opacity=0.3] (2.25, 2.25) rectangle (3, 3);
	%\draw[fill = green, opacity=0.3] (3.75, 2.25) rectangle (4.5, 3);
	\draw[fill = orange, opacity=0.3] (0.75, 0) rectangle (1.5, 0.75);
	\draw[fill = orange, opacity=0.3] (2.25, 0) rectangle (3, 0.75);
	%\draw[fill = orange, opacity=0.3] (3.75, 0) rectangle (4.5, 0.75);
	\draw[fill = orange, opacity=0.3] (0.75, 1.5) rectangle (1.5, 2.25);
	\draw[fill = orange, opacity=0.3] (2.25, 1.5) rectangle (3, 2.25);
	%\draw[fill = orange, opacity=0.3] (3.75, 1.5) rectangle (4.5, 2.25);
	\draw[line width=0.6mm,color=black] (-0.75,0) -- (2.25,0)--(2.25,0.75-0.1)--(2.25+0.1,0.75)--(3,0.75)--(3,1.5)--(3.75,1.5)--(3.75,3);
	\draw[line width=0.6mm,color=black] (-0.75,0.75) -- (1.5+0.65,0.75)--(2.25,1.5-0.65)--(2.25,3);
	\draw[line width=0.6mm,color=black] (-0.75,1.5) --(0.75,1.5)--(0.75,2.25)--(1.5,2.25)--(1.5,3);
	\draw[line width=0.6mm,color=black] (-0.75,2.25)--(0,2.25)--(0,3);
	\draw (1.875,-1) node(B) {$\mathcal{C}$};
	\end{tikzpicture}
	\begin{tikzpicture}[scale=0.8]
	\clip (-0.8,-1.7) rectangle (5.1,3);
	\draw[step=0.75cm,color=black, thick] (0,0) grid (3.75,2.25);
	\draw[step=0.75cm,color=black , thick] (-0.75,0) grid (0,0);
	\draw[step=0.75cm,color=black, thick] (-0.75,0.75) grid (0,0.75);
	\draw[step=0.75cm,color=black, thick] (-0.75,1.5) grid (0,1.5);
	\draw[step=0.75cm,color=black, thick] (-0.75,2.25) grid (0,2.25);
	\draw[step=0.75cm,color=black, thick] (3.75,0) grid (4.5,0);
	\draw[step=0.75cm,color=black, thick] (3.75,0.75) grid (4.5,0.75);
	\draw[step=0.75cm,color=black, thick] (3.75,1.5) grid (4.5,1.5);
	\draw[step=0.75cm,color=black, thick] (3.75,2.25) grid (4.5,2.25);
	\draw[step=0.75cm,color=black, thick] (0,-0.75) -- (0,0);
	\draw[step=0.75cm,color=black, thick] (0.75,-0.75) -- (0.75,0);
	\draw[step=0.75cm,color=black, thick] (1.5,-0.75) -- (1.5,0);
	\draw[step=0.75cm,color=black, thick] (2.25,-0.75) -- (2.25,0);
	\draw[step=0.75cm,color=black, thick] (3,-0.75) -- (3,0);
	\draw[step=0.75cm,color=black, thick] (3.75,-0.75) -- (3.75,0);
	\draw[step=0.75cm,color=black, thick] (0,2.25) -- (0,3);
	\draw[step=0.75cm,color=black, thick] (0.75,2.25) -- (0.75,3);
	\draw[step=0.75cm,color=black, thick] (1.5,2.25) -- (1.5,3);
	\draw[step=0.75cm,color=black, thick] (2.25,2.25) -- (2.25,3);
	\draw[step=0.75cm,color=black, thick] (3,2.25) -- (3,3);
	\draw[step=0.75cm,color=black, thick] (3.75,2.25) -- (3.75,3);
	\draw[fill = blue, opacity=0.3] (0, 0) rectangle (0.75, 0.75);
	\draw[fill = blue, opacity=0.3] (1.5, 0) rectangle (2.25, 0.75);
	\draw[fill = blue, opacity=0.3] (3, 0) rectangle (3.75, 0.75);
	\draw[fill = blue, opacity=0.3] (0, 1.5) rectangle (0.75, 2.25);
	\draw[fill = blue, opacity=0.3] (1.5, 1.5) rectangle (2.25, 2.25);
	\draw[fill = blue, opacity=0.3] (3, 1.5) rectangle (3.75, 2.25);
	\draw[fill = red, opacity=0.3] (0, 0.75) rectangle (0.75, 1.5);
	\draw[fill = red, opacity=0.3] (1.5, 0.75) rectangle (2.25, 1.5);
	\draw[fill = red, opacity=0.3] (3, 0.75) rectangle (3.75, 1.5);
	\draw[fill = red, opacity=0.3] (0, 2.25) rectangle (0.75, 3);
	\draw[fill = red, opacity=0.3] (1.5, 2.25) rectangle (2.25, 3);
	\draw[fill = red, opacity=0.3] (3, 2.25) rectangle (3.75, 3);
	\draw[fill = green, opacity=0.3] (0.75, 0.75) rectangle (1.5, 1.5);
	\draw[fill = green, opacity=0.3] (2.25, 0.75) rectangle (3, 1.5);
	%\draw[fill = green, opacity=0.3] (3.75, 0.75) rectangle (4.5, 1.5);
	\draw[fill = green, opacity=0.3] (0.75, 2.25) rectangle (1.5, 3);
	\draw[fill = green, opacity=0.3] (2.25, 2.25) rectangle (3, 3);
	%\draw[fill = green, opacity=0.3] (3.75, 2.25) rectangle (4.5, 3);
	\draw[fill = orange, opacity=0.3] (0.75, 0) rectangle (1.5, 0.75);
	\draw[fill = orange, opacity=0.3] (2.25, 0) rectangle (3, 0.75);
	%\draw[fill = orange, opacity=0.3] (3.75, 0) rectangle (4.5, 0.75);
	\draw[fill = orange, opacity=0.3] (0.75, 1.5) rectangle (1.5, 2.25);
	\draw[fill = orange, opacity=0.3] (2.25, 1.5) rectangle (3, 2.25);
	%\draw[fill = orange, opacity=0.3] (3.75, 1.5) rectangle (4.5, 2.25);
	\draw[line width=0.6mm,color=black] (-0.75,0) -- (3,0)--(3,1.5)--(3.75,1.5)--(3.75,3);
	\draw[line width=0.6mm,color=black] (-0.75,0.75) -- (2.25,0.75)--(2.25,3);
	\draw[line width=0.6mm,color=black] (-0.75,1.5) --(1.5,1.5)--(1.5,3);
	\draw[line width=0.6mm,color=black] (-0.75,2.25)--(0,2.25)--(0,3);
	\draw (1.875,-1) node(B) {$\mathcal{C'}$};
	\end{tikzpicture}
	\vspace{-0.5cm}
	\caption{\JFadd{Subdivision of the lattice into $4$ sublattices. Local moves are applied simultaneously to all plaquettes of the same sublattice with a Metropolis acceptance probability $A(\mathcal{C}\rightarrow \mathcal{C'})$, which, for the example shown (two local moves, on the same sublattice, are performed), is given by $A(\mathcal{C}\rightarrow \mathcal{C'}) = \text{min}(1,\frac{b^2}{c^2})\cdot \text{min}(1,\frac{b^2}{a^2})$.}}
	\label{fig:sublattices}
\end{figure}
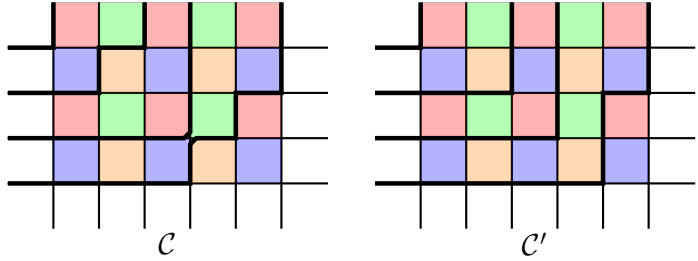

Although the algorithm guarantees the convergence to the desired distribution, in practice we must interrupt the algorithm after a finite number  $N$  of iterations, such that $P_N(\mathcal{C}_0 \rightarrow \mathcal{C}_N)\approx \pi(\mathcal{C})$. We decided to stop the algorithm when the arctic curve is stabilized, as done in \cite{allison20056V_MonteCarlo}. The drawback of this algorithm is that the expected number $N$ of required steps increases rapidly with the size of the lattice. One way to speed up the algorithm is by using the parallel computing capabilities of modern computers. Indeed a local move applied to a plaquette $(i,j)$ has only an impact on its nearest plaquettes \cite{keating2018sim_tiling_TW}. Hence we can subdivide the lattice into $4$ sublattices such that within a sublattice all the plaquettes are surronded by plaquettes belonging to the other sublattices, see Figure \ref{fig:sublattices}. Then, at each iteration, \JFadd{one of the four sublattices is randomly chosen (with a probability $1/4$)} and the local moves are applied simultaneously to \textit{all} the plaquettes within this sublattice. This enables to take advantage of parallelization techniques \cite{keating2018sim_tiling_TW}. 

\JFadd{For each plaquette, if a local move is allowed by the ice rule, it will be accepted with the Metropolis acceptance probability given by (\ref{eq_prob_metropolis}), which ensures that the detailed balanced condition is met. Indeed, let us suppose that the configuration $\mathcal{C'}$ can be obtained from the configuration $\mathcal{C}$ after $n$ local moves, involving $n$ plaquettes belonging to the same sublattice. Then, we have:
	\begin{equation}
	\fr{ P(\mathcal{C'}) }{ P(\mathcal{C}) } = \frac{1}{4} \prod_{\text{plaq}=1}^n\fr{ W_{\text{plaq} }(\mathcal{C'}) }{ W_{\text{plaq}}(\mathcal{C}) },
	\end{equation}
	where the product runs over the set of plaquettes $p$ and the prefactor $\frac{1}{4}$ is the probability to select the appropriate sublattice. At the same time, since the local moves are performed independently from each other, we have:
	\begin{equation}
	p(\mathcal{C} \rightarrow \mathcal{C'}) = \prod_{\text{plaq}=1}^n \min\left(1,\frac{W_{\text{plaq}}(\mathcal{C}')}{W_{\text{plaq}}(\mathcal{C})}\right),
	\end{equation} 
	from which we deduce that $\fr{p(\mathcal{C} \rightarrow \mathcal{C'})}{p(\mathcal{C'} \rightarrow \mathcal{C})} = \fr{P(\mathcal{C'})}{P(\mathcal{C})}$.}
	
\begin{figure}
	\centering
	\includegraphics[width=\textwidth]{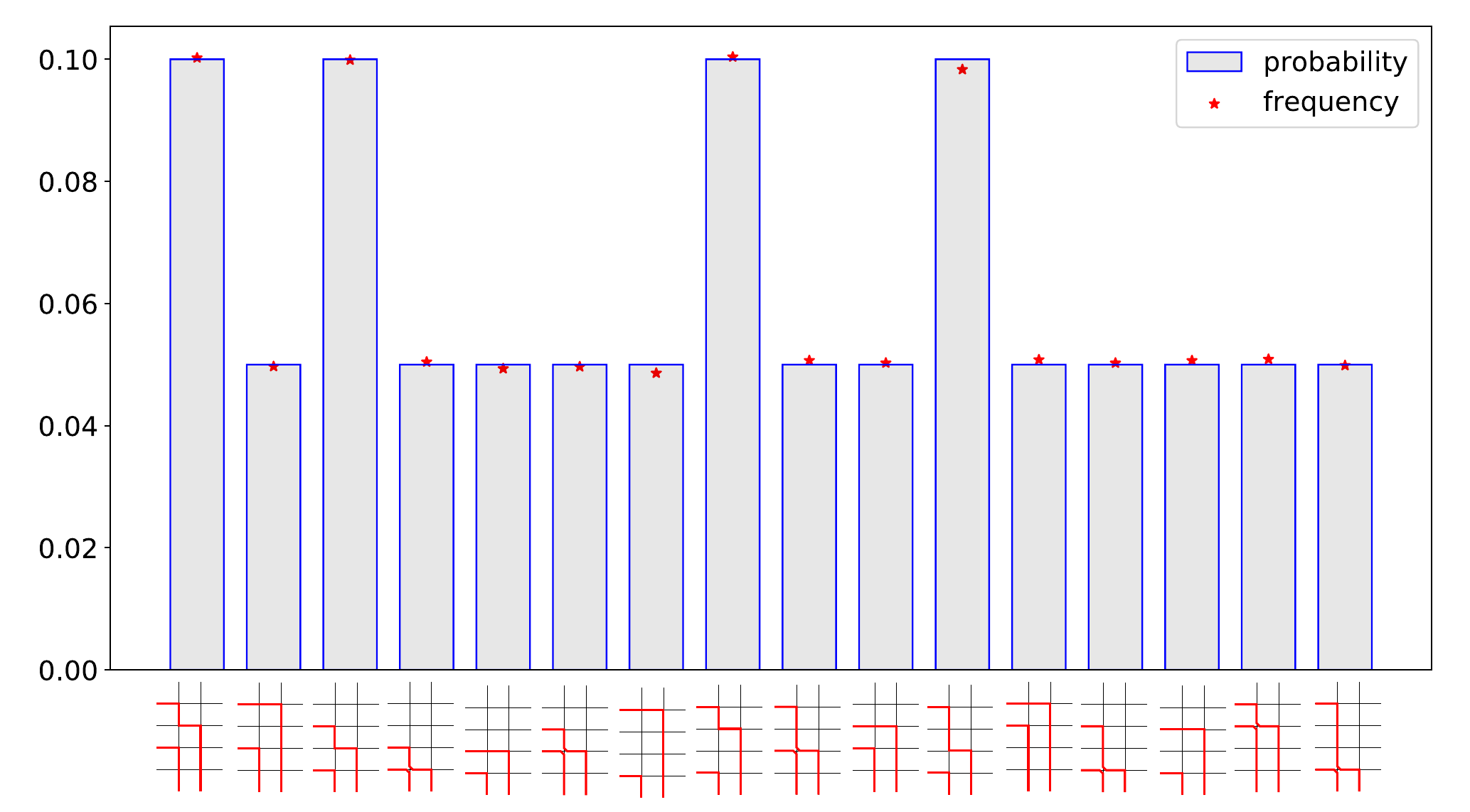}
	\caption{\JFadd{Observed frequencies for the $6$V model with pDWBC for $s=2, n=4$ and weights $a=1=b$ and $c=\sqrt{2}$, compared with the theoretical probabilities. Averages were performed over $100\,000$ configurations sampled after $100$ operations.}}
	\label{fig:freq}
\end{figure}

\JFadd{To check the validity of the algorithm, we ran the algorithm a large number of times for a small domain corresponding to $s=2$ and $n=4$ and compared the observed frequencies to the theoretical probabilities, computed from eq. (\ref{Z_hom}), see Figure \ref{fig:freq}.}

\cleardoublepage
\phantomsection

\addcontentsline{toc}{section}{References}

\bibliographystyle{mybibstyleJL_v2}
\bibliography{biblio}

\end{document}